\documentclass[11pt,a4paper]{article}
\usepackage{placeins}
\usepackage{graphicx}
\usepackage{xcolor}
\usepackage{float}
\usepackage{afterpage}
\usepackage{amssymb,amsmath}
\usepackage{multirow,booktabs}
\usepackage{cite}
\usepackage[colorlinks=true, linkcolor=black!50!blue, urlcolor=blue, citecolor=blue, anchorcolor=blue]{hyperref}
\usepackage[font=small,labelfont=bf,labelsep=period]{caption}
\usepackage[a4paper,top=3cm,bottom=2.5cm,left=2.5cm,right=2.5cm,bindingoffset=0mm]{geometry}
\usepackage{tabularx}
\newcolumntype{C}[1]{>{\centering\arraybackslash}p{#1}}

\newcommand{\be}{\begin{equation}}
\newcommand{\ee}{\end{equation}}
\newcommand{\bea}{\begin{eqnarray}}
\newcommand{\eea}{\end{eqnarray}}
\newcommand{\bi}{\begin{itemize}}
\newcommand{\ei}{\end{itemize}}
\newcommand{\ben}{\begin{enumerate}}
\newcommand{\een}{\end{enumerate}}
\newcommand{\la}{\left\langle}
\newcommand{\ra}{\right\rangle}
\newcommand{\lc}{\left[}
\newcommand{\rc}{\right]}
\newcommand{\lp}{\left(}
\newcommand{\rp}{\right)}

\def\frac#1#2{{{#1}\over {#2}}}
\def\gsim{\gtrsim}
\def\lsim{\lesssim}
\newcommand{\mrexp}{\mathrm{exp}}

\newcommand{\art}{\mathrm{art}}
\newcommand{\rep}{\mathrm{rep}}

\newcommand{\draft}[1]{}

\def\beq{\begin{equation}}
\def\eeq{\end{equation}}

\def\({\left(}
\def\){\right)}
\def\[{\left[}
\def\]{\right]}
\let\originalleft\left
\let\originalright\right
\renewcommand{\left}{\mathopen{}\mathclose\bgroup\originalleft}
\renewcommand{\right}{\aftergroup\egroup\originalright}

\numberwithin{equation}{section}
\numberwithin{figure}{section}
\numberwithin{table}{section}

\let\oldsubsection\subsection
\renewcommand\subsection[2][\subsectiontoc]{%
  \def\subsectiontoc{#2}%
  \oldsubsection[#1]{\boldmath #2}%
}

\let\oldsubsubsection\subsubsection
\renewcommand\subsubsection[2][\subsubsectiontoc]{%
  \def\subsubsectiontoc{#2}%
  \oldsubsubsection[#1]{\boldmath #2}%
}

\usepackage{xcolor}
\usepackage[utf8]{inputenc}
\usepackage[T1]{fontenc}
\usepackage[numbers, elide, sort&compress]{natbib}
\usepackage{lmodern, amsmath, amssymb, xcolor, a4wide, graphicx, datetime, rotating, adjustbox, array}

\makeatletter\g@addto@macro\bfseries{\boldmath}\makeatother

\def\equationautorefname~#1\null{Eq.\,(#1)\null}
\newcommand{\fullref}[2]{\hyperref[#2]{#1\,\ref*{#2}}}
\makeatletter

\newcommand{\rcite}[1]{\hyper@@link[cite]{}{cite.#1}{Ref.\,\cite*{#1}}}
\makeatother
\newcommand*{\eqsref}[2]{\hyperref[#1]{Eqs.\,(\ref*{#1}--}\hyperref[#2]{\ref*{#2})}}

\allowdisplaybreaks[4]

\makeatletter
\newcommand{\parenbar}{\mathpalette\p@renb@r}
\def\p@renb@r#1#2{%
	\vbox{%
		\ifx#1\scriptscriptstyle\dimen@.7em\dimen@ii.20em\else%
		\ifx#1\scriptstyle	\dimen@.8em\dimen@ii.25em\else%
					\dimen@.9em\dimen@ii.35em\fi\fi%
		\offinterlineskip%
		\ialign{%
			\hfill##\hfill\cr
			\vbox{\hrule width\dimen@ii height .35pt}\cr
			\noalign{\vskip-.35ex}%
			\hbox to\dimen@{$\mathchar300\hfil\mathchar301$}\cr
			\noalign{\vskip-.35ex}%
			$#1#2$\cr%
		}%
	}%
}%
\makeatother

\newcommand*{\tmp}[4]{\ensuremath{%
	{#4%
	\ifx\empty#3\empty\ifx\empty#1\empty\else^{#1}\fi\else^{#1(#3)}\fi%
	\ifx\empty#2\empty\else_{#2}\fi}%
}}
\newcommand*{\cc }[4][]{\tmp{#2}{#3}{#4}{#1{C}}}
\newcommand*{\ccc}[4][]{\tmp{#2}{#3}{#4}{#1{c}}}

\newcommand*{\qq }[4][]{\tmp{#2}{#3}{#4}{#1{O}}}

\let\Re\undefined

\DeclareMathOperator{\Re}{Re}

\newcommand{\hc}[1]{{}^\ddagger #1}

\newcommand*{\sw}{s_W}
\newcommand*{\cw}{c_W}

\newcommand{\FDF}{(\varphi^\dagger i\!\!\overleftrightarrow{D}_\mu\varphi)}
\newcommand{\FDFI}{(\varphi^\dagger i\!\!\overleftrightarrow{D}^I_\mu\varphi)}

\newcommand{\rien}[1]{}

\newcounter{affiliation}

\renewcommand{\kappa}{\sigma}

\begin{document}
\newgeometry{top=1.5cm,bottom=1.5cm,left=2.5cm,right=2.5cm,bindingoffset=0mm}
\begin{titlepage}
\thispagestyle{empty}
\noindent
\begin{flushright}
OUTP-18-07P \\
Nikhef-2018-058 \\
 CP3-19-02
\end{flushright}
\vspace{0.7cm}
\begin{center}
  {\LARGE \bf\boldmath A Monte Carlo
    global analysis of the \\[0.16cm]
    Standard Model Effective Field Theory:\\[0.33cm]
    the top quark sector}
\vspace{1.3cm}

Nathan~P.~Hartland,$^{1,2}$
Fabio Maltoni,$^{3,4}$
Emanuele R. Nocera,$^{2,5}$
Juan~Rojo,$^{1,2}$\\[0.1cm]
Emma Slade,$^{6}$
Eleni Vryonidou,$^7$ and
Cen Zhang$^8$

\vspace{0.7cm}
{\it \small
~$^1$ Department of Physics and Astronomy, Vrije Universiteit Amsterdam,\\ NL-1081 HV Amsterdam, The Netherlands\\[0.1cm]
 ~$^2$  Nikhef Theory Group, Science Park 105, 1098 XG Amsterdam, The Netherlands\\[0.1cm]
~$^3$ Centre for Cosmology, Particle Physics and Phenomenology (CP3),\\
  Universit\'e Catholique de Louvain, B-1348 Louvain-la-Neuve, Belgium\\[0.1cm]
~$^4$
Dipartimento di Fisica e Astronomia, Universit\`a di Bologna \\and INFN, Sezione di Bologna, 
via Irnerio 46, 40126 Bologna, Italy\\[0.1cm]
~$^5$ The Higgs Centre for Theoretical Physics,\\
  University of Edinburgh, JCMB, KB, Mayfield Rd, Edinburgh EH9 3FD, Scotland\\[0.1cm]
~$^6$ Rudolf Peierls Centre for Theoretical Physics, University of Oxford, \\
  Clarendon Laboratory, Parks Road, Oxford OX1 3PU, United Kingdom\\[0.1cm]
~$^7$ Theoretical Physics Department, CERN, CH-1211 Geneva, Switzerland\\[0.05cm]
~$^8$ Institute of High Energy Physics, and School of Physical Sciences,\\
University of Chinese Academy of Sciences, Beijing 100049, China

}

\vspace{0.7cm}

{\bf \large Abstract}

\end{center}

We present a novel framework for carrying out global analyses of the
Standard Model Effective Field Theory (SMEFT) at dimension-six: {SMEFiT}.
This approach is based on the Monte Carlo replica method for deriving a
faithful estimate of the experimental and theoretical uncertainties and enables
one to construct the probability distribution in the space of the SMEFT degrees
of freedom.
As a proof of concept of the SMEFiT methodology,
we present a first study of the constraints on the SMEFT
provided by top quark production measurements from the LHC.
Our analysis includes more than 30 independent measurements from 10 different 
processes at $\sqrt{s}=8$ and 13 TeV such as inclusive $t\bar{t}$ and 
single-top production and the associated
production of top quarks with weak vector bosons and the Higgs boson.
State-of-the-art theoretical calculations are adopted both for
the Standard Model and for the SMEFT contributions, where in the latter case
NLO QCD corrections are included for the majority of processes.
We derive bounds for the 34 degrees of freedom relevant for
the interpretation of the LHC top quark data and compare these bounds
with previously reported constraints.
Our study illustrates the significant potential of LHC precision measurements 
to constrain physics beyond the Standard Model in a model-independent way, and
paves the way towards a global analysis of the SMEFT.

\vspace{0.8cm}

\end{titlepage}

\restoregeometry

\tableofcontents

\section{Introduction}

The Large Hadron Collider (LHC) is pursuing an extensive
program of direct searches for physics beyond the
Standard Model (BSM) by exploiting its unique reach in energy.
Whilst these searches have not yet returned any convincing evidence for BSM 
physics, only a small fraction of the final LHC dataset has been analysed so 
far, and ample room for surprises remains.
A complementary approach to the searches for direct production
of new particles is that of {\it indirect} BSM searches,
where precise measurements of total cross-sections and differential
distributions are compared to Standard Model (SM) predictions
with the hope to uncover glimpses of BSM dynamics in the interactions between 
SM particles.
For instance, if new particles are too heavy to be directly produced at the LHC,
they could still leave imprints in the kinematical distributions of the SM 
particles via interference or virtual effects.

A powerful framework to identify, constrain, and parametrise potential
deviations with respect to the SM predictions in a model-independent way is
the Standard Model Effective Field Theory
(SMEFT)~\cite{Weinberg:1979sa,Buchmuller:1985jz,Grzadkowski:2010es}.
In this framework, the effects of BSM dynamics at high scales $E\simeq \Lambda$ 
are parametrised for $E\ll \Lambda$ in terms of higher-dimensional (irrelevant)
operators built up from the SM fields and respecting symmetries such as gauge
and Lorentz symmetry. 
This approach is robust and general, since one can construct non-redundant 
bases of independent operators at any given mass dimension ($\hbar=c=1$) that 
can then be systematically matched to explicit ultraviolet-complete scenarios 
for their interpretation at any order in $1/\Lambda$.
 
Analysing experimental data in the SMEFT framework is non trivial;
even restricting oneself to operators that conserve
baryon and lepton number~\cite{Grzadkowski:2010es}, one ends
up with $N_{\rm op}=59$ operators at dimension six for one generation,
growing to more than 2000 in absence of flavour assumptions.
This implies that global and model-independent SMEFT analyses need to explore 
a complicated parameter space with a large number of degenerate (``flat'') 
directions and local minima.

In this context, the wealth of precision measurements presented by
the LHC collaborations in recent  years, together with
the significant progress in the corresponding theoretical calculations
and modelling of collider processes, has motivated many
groups to pursue (partial) SMEFT analyses of the LHC data~\cite{Alioli:2017jdo,Englert:2015hrx,Ellis:2014jta,
Alioli:2017nzr,Alioli:2018ljm,Alte:2017pme,AguilarSaavedra:2018nen,Castro:2016jjv,deBlas:2016ojx,
Ellis:2018gqa,Ellis:2014dva,Buckley:2016cfg,Butter:2016cvz,Azatov:2015oxa,Biekotter:2018rhp,
Schulze:2016qas,Cirigliano:2016nyn,Alioli:2017ces} 
complemented often with input from lower-energy experiments
such as the LEP electroweak precision tests.
In these fits, constraints on the SMEFT operators can be provided
not only by ``traditional'' processes such as electroweak gauge boson and 
Higgs production, but also by other high-$p_T$ processes such as
jet and top quark production.
Interestingly, even when only considering electroweak
processes, these constraints are comparable or even superior to those provided 
by LEP~\cite{Franceschini:2017xkh,Alioli:2017jdo}.
Indeed, SMEFT corrections often grow quadratically with the energy
and thus directly benefit from the large kinematic reach, up to several
TeV, provided by present and future LHC measurements.

From the methodological point of view, a global fit of the SMEFT
from LHC measurements requires combining state-of-the-art theoretical
calculations (in the SM and in the SMEFT) with a wide variety of experimental 
cross-sections and distributions. 
This should be accomplished by means of a robust statistical analysis 
allowing for the reliable estimation of all sources of uncertainty and for the
minimisation of procedural and theoretical biases.
SMEFT fits therefore represent, conceptually, a similar problem to  
that arising in the global QCD analysis of the quark and gluon structure of 
the proton in terms of parton distribution functions (PDFs)~\cite{Gao:2017yyd,
Butterworth:2015oua,Rojo:2015acz}.
By exploiting these conceptual similarities, in this work we develop a novel
strategy for global SMEFT analyses inspired by the NNPDF framework, successfully
applied to the determination of the parton distributions of the proton~\cite{DelDebbio:2004xtd,DelDebbio:2007ee,Ball:2008by,Ball:2009mk,Ball:2010de,Ball:2011mu,Ball:2012cx,Ball:2014uwa,Ball:2016neh,Nocera:2014gqa,Ball:2017nwa} and of 
hadron fragmentation functions~\cite{Bertone:2017tyb,Bertone:2018ecm}.
This approach, which we denote by SMEFiT, combines the generation of 
Monte Carlo (MC) replicas, to estimate and propagate uncertainties, 
with cross-validation to prevent over-fitting.

As a proof of concept of the SMEFiT methodology, we apply it here for the first 
time to the detailed study of top quark production at the LHC in the SMEFT 
framework at dimension six.
The top quark, the only fermion with an $\mathcal{O}\lp 1\rp$
Yukawa coupling, plays a privileged role in most BSM scenarios aiming
to explain the origin of electroweak symmetry breaking and stabilise
the weak scale.
From the experimental data point of view, a global SMEFT analysis of top quark
production at the LHC is motivated by the large number of precision
measurements at $\sqrt{s}=7, 8$ and 13 TeV that have become
available recently. 
This data includes total rates and differential
distributions in inclusive $t\bar{t}$ and single-top production, associated
production of top quarks with vector bosons and the Higgs boson, 
and helicity fractions in top quark decay.
The wealth of data collected by the LHC is mirrored by the advancements on the
theoretical side, where  significant progress in higher order calculations in 
the top quark sector has been achieved. 
This is true both from the SM point of view, with the calculation of NNLO QCD 
and NLO electroweak corrections for inclusive top quark pair and single top 
production, as well as from the SMEFT side.
In the latter case, LO calculations are now automatised
in codes such as {\tt MadGraph5\_aMC@NLO}~\cite{Alwall:2014hca}
within a framework agreed within the LHC Top
WG~\cite{AguilarSaavedra:2018nen}, 
and NLO QCD corrections have been presented for a continuously growing
number of processes. 

Several SMEFT analyses of the top quark sector have been presented
based on either hadron collider~\cite{Maltoni:2016yxb,Degrande:2018fog,
Chala:2018agk,Durieux:2014xla,AguilarSaavedra:2010zi,DHondt:2018cww} 
or lepton collider~\cite{Durieux:2018ggn} processes, in the latter case also 
considering the sensitivity of future machines
such as the International Linear Collider (ILC).
The top quark sector of the SMEFT has been in particular studied
by the {\tt TopFitter} collaboration~\cite{Buckley:2015nca,Buckley:2015lku,
Buckley:2016cfg}.
Our analysis exhibits several improvements as compared to the available studies, allowing
us to assess the impact of several important aspects in the fit.
First, we include a broader range of input experimental measurements
from different processes, which allow us to constrain a larger number
of SMEFT operators.
Second, we include the NLO QCD corrections to the SMEFT
contributions. This entails an improved accuracy and a reduction of the theory systematic errors.
Third, we always compute both the leading linear ($\mathcal{O}(\Lambda^{-2})$) and the 
subleading quadratic  ($\mathcal{O}(\Lambda^{-4})$) contributions to the SMEFT predictions, so
that effects of including or not the quadratic terms can be systematically studied. 
Fourth, our methodology avoids any assumption about the specific profile
of the $\chi^2$ function and in particular we do not rely on any quadratic
approximation for error propagation.

By exploiting the SMEFiT methodology, here we derive the probability
distribution in the space of SMEFT Wilson coefficients that follows from all
available top quark production cross-sections.
We study the impact of individual processes on the SMEFT parameter space
and the role of higher order corrections, such as NLO QCD and the SMEFT $\mathcal{O}(\Lambda^{-4})$ 
corrections.
In general, we find that that higher order effects are non-negligible and can significantly affect the results. 
We also quantify the correlations between the operators, and compare the 
bounds derived here with previous constraints reported in the literature.
Our analysis illustrates the significant potential of LHC precision 
measurements to constrain, and possibly identify, BSM physics in a 
model-independent way.

The outline of this paper is the following.
In Sect.~\ref{sec:smefttop} we summarise the SMEFT description of the top 
quark sector at dimension six and introduce our choice of operator basis for
the fit.
In Sect.~\ref{sec:settings_expdata} we describe the experimental measurements 
of top quark production at the LHC which are used to constrain the SMEFT 
operators and the settings of the corresponding theoretical calculations of
the SM and SMEFT cross-sections.
The SMEFiT methodology is presented in Sect.~\ref{sec:fitsettings}, where it
is validated by means of closure tests.
The main results of this work are presented in Sect.~\ref{sec:results},
where we determine the confidence level intervals for the coefficients of the 
$N_{\rm op}=34$ SMEFT operators and their correlations, and compare them with
the bounds reported in previous studies.
In Sect.~\ref{sec:summary} we summarise our main conclusions
and outline possible directions
for generalising our analysis to other processes.

\section{The SMEFT in the top quark sector}
\label{sec:smefttop}

In this section we describe the theoretical formalism that will be adopted in
this work to interpret the LHC top quark production data within the SMEFT
framework.
First, we provide an introduction to the SMEFT, focusing on those
operators that affect the description of the top quark sector.
Then, we define the degrees of freedom that are more relevant to studying top 
quark production at the LHC.
Operators that do not involve top quarks and their constraints are also briefly
discussed.
We finally describe our theory calculations at NLO QCD accuracy, and comment
on some additional aspects of the SMEFT formalism relevant for this study.

\subsection{The SMEFT framework}

Let us begin by reviewing the SMEFT
formalism~\cite{Weinberg:1978kz,Buchmuller:1985jz},
with emphasis on its description of the top quark sector.
As mentioned in the introduction, the effects of new heavy BSM particles with
typical mass scale $M\simeq \Lambda$ can under general conditions be
parametrised at lower energies $E\ll \Lambda$ in a model-independent way in
terms of a basis of higher-dimensional operators constructed from the SM fields
and their symmetries.
The resulting effective Lagrangian then admits the following power expansion
\be
\label{eq:smeftlagrangian}
\mathcal{L}_{\rm SMEFT}=\mathcal{L}_{\rm SM} + \sum_i^{N_{d6}} \frac{c_i}{\Lambda^2}\mathcal{O}_i^{(6)} +
\sum_j^{N_{d8}} \frac{b_j}{\Lambda^4}\mathcal{O}_j^{(8)} + \ldots \, ,
\ee
where $\mathcal{L}_{\rm SM}$ is the SM Lagrangian, and
$\{\mathcal{O}_i^{(6)}\}$ and $\{\mathcal{O}_j^{(8)}\}$ stand for
the elements of the
operator basis of mass-dimension $d=6$ and $d=8$,
respectively.
Operators with $d=5$ and $d=7$, which violate lepton and/or baryon number
conservation~\cite{Degrande:2012wf,Kobach:2016ami}, are not considered here.

Whilst the choice of operator basis used in Eq.~(\ref{eq:smeftlagrangian}) is
not unique, it is possible to relate the results obtained in different
bases~\cite{Falkowski:2015wza}.
In this work we adopt the Warsaw basis for
$\{\mathcal{O}_i^{(6)}\}$~\cite{Grzadkowski:2010es}, and neglect effects
arising from operators with mass dimension $d\ge 8$.

For specific UV completions, the Wilson coefficients $\{c_i\}$ in
Eq.~(\ref{eq:smeftlagrangian}) can be evaluated in terms of the parameters of
the BSM theory, such as its coupling constants and masses.
However, in a
bottom-up approach, they are {\it a priori} free parameters and they need to be
constrained from experimental data.
In general, the effects of the dimension-6 SMEFT operators in a given 
observable, such as cross-sections at the LHC, differential distributions, 
or other pseudo-observables, can be written as follows:
\be
\label{eq:smeftXsecInt}
\sigma=\sigma_{\rm SM} + \sum_i^{N_{d6}}
\kappa_i \frac{c_i}{\Lambda^2} +
\sum_{i,j}^{N_{d6}}  \widetilde{\kappa}_{ij} \frac{c_ic_j}{\Lambda^4}  \, ,
\ee
where $\sigma_{\rm SM}$ indicates
the SM prediction and the Wilson coefficients $c_i$ are considered to be real 
for simplicity.  

In Eq.~(\ref{eq:smeftXsecInt}), the second term arises from operators 
interfering with the SM amplitude.
The resulting $\mathcal{O}\lp\Lambda^{-2}\rp$ corrections to the SM 
cross-sections represent formally the dominant correction, though in many cases 
they can be subleading for different reasons.
The third term in Eq.~(\ref{eq:smeftXsecInt}),
representing $\mathcal{O}\lp\Lambda^{-4}\rp$ effects, arises from the
squared amplitudes of the SMEFT operators, irrespectively of whether or not the
dimension-6 operators interfere with the SM diagrams.
In principle, this
second term may not need to be included, depending on if the truncation at
$\mathcal{O}\lp\Lambda^{-2}\rp$ order is done at the Lagrangian or the cross
section level, but in practice there are often
valid reasons to include them in the calculation.
We will
discuss in more details the impact of
these $\mathcal{O}\lp\Lambda^{-4}\rp$
corrections at the end of this section.

An important aspect of any SMEFT analysis is the need to include
all relevant operators that contribute
to the processes whose data is used as input
to the fit.
Only in this way can the SMEFT retain its
model and basis independence.
However, unless specific scenarios are adopted, the number of
non-redundant operators $N_{d6}$ becomes unfeasibly large:
59 for one generation of fermions \cite{Grzadkowski:2010es} and 2499 for
three~\cite{Alonso:2013hga}.
This implies that a global SMEFT fit, even if
restricted to dimension-6 operators,
will have to explore a huge
parameter space with potentially a large number of flat (degenerate) directions.

Due to the above consideration, in this work we follow closely the strategy
documented in the LHC Top Quark Working Group 
note~\cite{AguilarSaavedra:2018nen}.
In particular, we adopt the Minimal
Flavour Violation (MFV) hypothesis~\cite{DAmbrosio:2002vsn} in the quark
sector as the baseline scenario.
We further assume that the
Cabibbo-Kobayashi-Maskawa (CKM) matrix is diagonal, and that the Yukawa
couplings are nonzero only for the top and bottom quarks.
In other words, we
impose a $U(2)_q\times U(2)_u \times U(2)_d$ flavour symmetry among the first
two generations.
In addition, we restrict ourselves to the CP-even operators only, and focus on
those operators that induce modifications in the interactions of the top quark
with other SM fields.
As we will now show, under
the above assumptions, we will explore the parameter space
associated to the $N_{\rm op}=34$ linear combinations of dimension-6 operators
that are relevant for the description of the top quark sector.
Following Ref.~\cite{AguilarSaavedra:2018nen}, we will then define the specific
degrees of freedom relevant for the interpretation of top quark measurements.

\subsection{The top quark sector of the SMEFT}
\label{sec:SMEFTtop}

Given the scope of this study, we will consider here only those dimension-6 
operators that affect
the production and decay of top quarks at the LHC through the modifications of
their couplings to other SM fields.
Following Ref.~\cite{AguilarSaavedra:2018nen}, we adopt the  Warsaw
basis~\cite{Grzadkowski:2010es} of non-redundant, gauge-invariant dimension-six
operators, and then we define the specific degrees of freedom relevant for each
measurement.
These degrees of freedom are linear combinations of the Warsaw-basis
operator coefficients, which appear in the interference with SM amplitudes,
and in interactions with physical fields after electroweak symmetry breaking.
These combinations are then aligned with physically relevant directions of the
SMEFT parameter space.
They represent the maximal information that can be extracted from measuring a
certain process.
The rationale for using them in a global fit instead of the basis operator 
coefficients directly is that they
may reduce the number of relevant parameters and unconstrained combinations.

Since we only consider here those operators which contain at least one top quark
under the assumed flavour symmetries, we are implicitly assuming that
other operators affecting the considered processes are well
constrained from measurements of other processes that do not involve top quarks.
This assumption may not always be justified, but it is helpful for a better 
understanding of the top quark sector, and also for setting up the scope
of this work.
Without this assumption, it is likely that one would have to resort to a much 
more global analysis, including all currently available data, which goes beyond
the scope of the present analysis.
We will discuss explicitly how in our case this assumption is justified in the 
next subsection.

We are now ready to define the relevant degrees of freedom that will be used 
in this analysis in terms of the dimension-6 operators of the Warsaw basis.
The complete set of degrees of freedom can be found in 
Ref.~\cite{AguilarSaavedra:2018nen}, and for completeness we collect in 
Appendix~\ref{app:notation} the definitions and conventions
that will be adopted in the following.
To begin with,
concerning the operators involving four heavy quarks
(that is, either a right-handed top $t$, or a right-handed bottom $b$,
or a left-handed top-bottom doublet $Q$), we define the following degrees of
freedom:
\begin{equation}
	\begin{aligned}
	\ccc{1}{QQ}{} &\equiv 2\cc{1}{qq}{3333}-\frac{2}{3}\cc{3}{qq}{3333}, \\
	\ccc{8}{QQ}{} &\equiv 8\cc{3}{qq}{3333},	\\
	\end{aligned}
	\qquad
	\begin{aligned}
	\ccc{1}{Qt}{} &\equiv  \cc{1}{qu}{3333},	\\
	\ccc{8}{Qt}{} &\equiv  \cc{8}{qu}{3333},	\\
	\ccc{1}{Qb}{} &\equiv  \cc{1}{qd}{3333},	\\
	\ccc{8}{Qb}{} &\equiv  \cc{8}{qd}{3333},	\\
	\end{aligned}
	\qquad
	\begin{aligned}
	\ccc{1}{tt}{} &\equiv  \cc{1}{uu}{3333},	\\
	\end{aligned}
	\qquad
	\begin{aligned}
	\ccc{1}{tb}{} &\equiv  \cc{1}{ud}{3333},	\\
	\ccc{8}{tb}{} &\equiv  \cc{8}{ud}{3333},
	\end{aligned}
        \label{eq:DoF4hq}
\end{equation}
and in addition we also have
\begin{equation}
	\ccc{1}{QtQb}{}	\equiv	\Re\{\cc{1}{quqd}{3333}\},	\qquad
	\ccc{8}{QtQb}{}	\equiv	\Re\{\cc{8}{quqd}{3333}\} \, ,
\end{equation}
in terms of the Warsaw-basis operators listed in
Eq.~(\ref{eq:FourQuarkOp}).
We note that the imaginary parts of the last two operators
are CP-odd, and therefore are not included here since we restrict
ourselves to CP-conserving operators.

Note that in Eq.~(\ref{eq:DoF4hq}) all four flavour indices in these degrees of
freedom correspond to the third (heavy) quark generation.
For example, the degree of freedom labelled as
$c_{QQ}^1$ in Eq.~(\ref{eq:DoF4hq}) is constructed
from a linear combination of  the $\mathcal{O}_{qq}^{1(ijkl)}$
and $\mathcal{O}_{qq}^{3(ijkl)}$ operators
for which $i=j=k=l=3$.
From Eq.~(\ref{eq:DoF4hq}) we see that
within the specific flavour assumptions adopted here
there are 11 operators involving four heavy quarks.
These operators can only be constrained from processes involving four
heavy quarks in the final state, such as four-top quark production
or $t\bar{t}b\bar{b}$ production, as we will discuss below in
Sect.~\ref{sec:overview}.

Concerning the dimension-6 operators of the Warsaw basis
involving two light quarks and two heavy quarks, see the list
in Eq.~(\ref{eq:TwoQuarkOp}), we first note that operators involving a
light-quark scalar or tensor current are vetoed by the flavour assumptions
adopted here.
On the other hand, vector-like interactions such as $\bar LL\bar LL$, $\bar
LL\bar RR$, and $\bar RR\bar RR$ type operators are allowed by our flavour
scenario.
We can therefore define the following degrees of freedom in terms of
two-light-two-heavy operators:
\begin{gather}
\begin{aligned}
	\ccc{1,1}{Qq}{}&\equiv
		\cc{1}{qq}{ii33}
		+\frac{1}{6}\cc{1}{qq}{i33i}
		+\frac{1}{2}\cc{3}{qq}{i33i},
	\\
	\ccc{3,1}{Qq}{}&\equiv
		\cc{3}{qq}{ii33}
		+\frac{1}{6}(\cc{1}{qq}{i33i}-\cc{3}{qq}{i33i}),
	\\
	\ccc{1,8}{Qq}{}&\equiv
		\cc{1}{qq}{i33i}+3\cc{3}{qq}{i33i},
	\\
	\ccc{3,8}{Qq}{}&\equiv
		\cc{1}{qq}{i33i}-\cc{3}{qq}{i33i},
\end{aligned}
\qquad
\begin{aligned}
	\ccc{1}{tu}{} &\equiv \cc{}{uu}{ii33} +\frac{1}{3} \cc{}{uu}{i33i},\\
	\ccc{8}{tu}{} &\equiv 2 \cc{}{uu}{i33i},	\\
	\ccc{1}{td}{} &\equiv \cc{1}{ud}{33ii},		\\
	\ccc{8}{td}{} &\equiv \cc{8}{ud}{33ii},		\\
\end{aligned}
\qquad
\begin{aligned}
	\ccc{1}{tq}{}	&\equiv\cc{1}{qu}{ii33},		\\
	\ccc{1}{Qu}{}	&\equiv\cc{1}{qu}{33ii},		\\
	\ccc{1}{Qd}{}	&\equiv\cc{1}{qd}{33ii},		\\
	\ccc{8}{tq}{}	&\equiv\cc{8}{qu}{ii33},		\\
	\ccc{8}{Qu}{}	&\equiv\cc{8}{qu}{33ii},		\\
	\ccc{8}{Qd}{}	&\equiv\cc{8}{qd}{33ii},		\\
\end{aligned}
\label{14dofs}
\end{gather}
where $i$ corresponds to a light quark index, that is,
it is either $1$ or $2$, and recall that the first two generations
are massless and thus exhibit an SU(2) flavour symmetry.
For these degrees of freedom involving two heavy quarks and two light
quarks, we therefore end up with 14 independent coefficients.
These degrees of freedom can be constrained by processes such as inclusive
$t\bar{t}$, through the quark-antiquark component of the initial state, as well
as by $t\bar{t}$ production in association with gauge vector bosons.
The SU(2) triplet degrees of freedom can also be constrained by single top
processes.

Finally, we need to take into account the degrees of freedom involving operators
built from two heavy quarks and bosonic fields, including the Higgs field, 
namely those listed in Eq.~(\ref{eq:TwoQuarkOp}).
For these operators, the following combinations are defined:
\begin{gather}
	\begin{aligned}
	\ccc{}{t\varphi}{}	&\equiv \Re\{\cc{}{u\varphi}{33}\},	\\
	\end{aligned}
	\qquad
	\begin{aligned}
	\ccc{-}{\varphi Q}{}	&\equiv \cc{1}{\varphi q}{33}-\cc{3}{\varphi q}{33},	\\
	\ccc{3}{\varphi Q}{}	&\equiv \cc{3}{\varphi q}{33},	\\
	\ccc{}{\varphi t}{}	&\equiv \cc{}{\varphi u}{33},	\\
	\ccc{}{\varphi tb}{}	&\equiv \Re\{\cc{}{\varphi ud}{33}\},	\\
	\end{aligned}
	\qquad
	\begin{aligned}
	\ccc{}{tW}{}		&\equiv \Re\{\cc{}{uW}{33}\} ,	\\
	\ccc{}{tZ}{}		&\equiv \Re\{-\sw\cc{}{uB}{33}+\cw\cc{}{uW}{33}\},\\
	\ccc{}{bW}{}		&\equiv \Re\{\cc{}{dW}{33}\} ,	\\
	\ccc{}{tG}{}		&\equiv \Re\{\cc{}{uG}{33}\} .
	\end{aligned}
      \label{eq:bosonicDoF}
\end{gather}
We see for example that the $\ccc{}{tZ}{}$ degree of freedom is a combination
of the $\mathcal{O}_{uB}^{ij}$ and $\mathcal{O}_{uW}^{ij}$ operators with
$i=j=3$, weighted by the sine and the cosine of the Weinberg angle
respectively.
Since we account here only for CP-conserving effects,
the imaginary parts of the last five coefficients, being CP-odd, will not be
included.

Note that there are two additional degrees of freedom that fall into the same
category (two heavy quark fields plus bosonic fields), but they are not 
independent from those defined above.
First of all, we have the combination of $\qq{1,3}{\varphi q}{33}$ operators
that modifies the SM coupling of the $b$ quark to the $Z$ boson, defined as
\be
\label{cplusphiQ}
\ccc{+}{\varphi Q}{} \equiv \cc{3}{\varphi q}{33}+\cc{1}{\varphi q}{33} 
= 
\ccc{-}{\varphi Q}{}+2\ccc{3}{\varphi Q}{}\,,
\ee
as well as the combination of operators that affects the electromagnetic dipole
of the top quark, defined as
\be
\ccc{}{tA}{} \equiv \Re\{\cw \cc{}{uB}{33}+\sw \cc{}{uW}{33}\} 
= 
(\ccc{}{tW}{}-\cw \ccc{}{tZ}{})/\sw \, .
\ee
These two degrees of freedom, $\ccc{+}{\varphi Q}{}$
and $\ccc{}{tA}{}$,
are useful for instance in the interpretation
of processes such as $Z\to b\bar b$ and $t\bar t\gamma$.
Since they can be simply written as linear combinations of other
degrees of freedom, we will not discuss them further in this work.

Taking stock, in total we have 9 CP-conserving degrees of freedom constructed 
from operators that involve two heavy quarks and gauge and Higgs bosonic fields.
Operators involving gauge boson fields can be constrained either by
single top production, if they modify the charged current coupling, or
by the associated production of top quark pairs and single tops with
electroweak bosons, i.e.~processes such as $t\bar{t}V$ and $tV$, if they
modify only the neutral current couplings. The degree of freedom $c_{tG}$ will
enter at leading-order in top pair and top pair associated production and $tW$,
and at NLO in $t/s-$channel single top production.
The degree of freedom
$c_{t\varphi}$, on the other hand, can only be constrained from
the associated production of a top quark pair with a Higgs boson,
as we will discuss in the next section.
Fortunately in this case, the first cross-section measurements for $t\bar{t}H$
production have recently become available.

Putting everything together, in total our fitting basis will be composed of
$N_{\rm op}=34$ independent degrees of freedom constructed from the dimension-6 
SMEFT operators relevant for the description of the top quark sector: 
$N_{\rm op}=11$ four-heavy-quark operators, $N_{\rm op}=14$ two-heavy-two-light 
quark operators, and
$N_{\rm op}=9$ operators involving two heavy quarks and bosonic fields.
In Table~\ref{eq:summaryOperators} we summarise the
definition of these 34 degrees of freedom in terms of the
SMEFT operators in the Warsaw basis, as well as the internal notation
that we will use in the following to refer to them.
As in the above discussion,
the degrees of freedom are divided into the three relevant classes:
four-heavy-quark operators, two-heavy-two-light-quark operators,
and operators that couple two heavy quarks to gauge and Higgs bosonic fields.
We will discuss in the next section (see, in particular, 
Table~\ref{table:operatorprocess}) which
of these operators are constrained by each of the LHC
top quark measurements included in the analysis.

\begin{table}[p]
  \centering
  \small
   \renewcommand{\arraystretch}{1.39}
  \begin{tabular}{c|c|c|c}
Class  &  Notation   &   Degree of Freedom   & Operator Definition \\
    \toprule
\multirow{9}{*}{$QQQQ$}&   {\tt OQQ1} &$\ccc{1}{QQ}{}$&$2\cc{1}{qq}{3333}-\frac{2}{3}\cc{3}{qq}{3333}$ \\
    &    {\tt OQQ8} &$\ccc{8}{QQ}{}$&$8\cc{3}{qq}{3333}$\\
    &    {\tt OQt1} &$\ccc{1}{Qt}{}$&$\cc{1}{qu}{3333}$\\
    &    {\tt OQt8} &$\ccc{8}{Qt}{}$&$\cc{8}{qu}{3333}$\\
    &    {\tt OQb1} &$\ccc{1}{Qb}{}$&$\cc{1}{qd}{3333}$\\
    &    {\tt OQb8} &$\ccc{8}{Qb}{}$&$\cc{8}{qd}{3333}$\\
    &    {\tt Ott1} &$\ccc{1}{tt}{}$&$\cc{}{uu}{3333}$  \\
    &    {\tt Otb1} &$\ccc{1}{tb}{}$&$\cc{1}{ud}{3333}$\\
    &    {\tt Otb8} &$\ccc{8}{tb}{}$&$\cc{8}{ud}{3333}$ \\
    &    {\tt OQtQb1} &$\ccc{1}{QtQb}{}$&$\cc{1}{quqd}{3333}$ \\
    &    {\tt OQtQb8} &$\ccc{8}{QtQb}{}$&$\cc{8}{quqd}{3333}$ \\
        \midrule      
 \multirow{14}{*}{$QQqq$}   &    {\tt  O81qq }  &  	$\ccc{1,8}{Qq}{}$      & $\cc{1}{qq}{i33i}+3\cc{3}{qq}{i33i}$     \\
    &    {\tt O11qq }  &  $\ccc{1,1}{Qq}{} $     & $\cc{1}{qq}{ii33}+\frac{1}{6}\cc{1}{qq}{i33i}+\frac{1}{2}\cc{3}{qq}{i33i} $   \\
    &    {\tt O83qq }  &  $\ccc{3,8}{Qq}{} $     &  $\cc{1}{qq}{i33i}-\cc{3}{qq}{i33i} $   \\
    &    {\tt  O13qq }  &  $\ccc{3,1}{Qq}{} $     & 	$\cc{3}{qq}{ii33}+\frac{1}{6}(\cc{1}{qq}{i33i}-\cc{3}{qq}{i33i}) $   \\
    &    {\tt O8qt }  &   $\ccc{8}{tq}{}$     & $ \cc{8}{qu}{ii33}   $ \\
    &    {\tt O1qt }  & $\ccc{1}{tq}{} $      &  $  \cc{1}{qu}{ii33} $\\
    &    {\tt O8ut }  &  $ \ccc{8}{tu}{}$     &  $2\cc{}{uu}{i33i}$  \\
    &    {\tt O1ut }  &  $\ccc{1}{tu}{} $     &  $ \cc{}{uu}{ii33} +\frac{1}{3} \cc{}{uu}{i33i} $ \\
    &    {\tt O8qu }  & 	$\ccc{8}{Qu}{} $      & $  \cc{8}{qu}{33ii}$\\
    &    {\tt O1qu }  & 	$\ccc{1}{Qu}{} $      & $  \cc{1}{qu}{33ii}$  \\
    &    {\tt O8dt }  &   $	\ccc{8}{td}{} $    & $\cc{8}{ud}{33ii}$ \\
    &    {\tt O1dt }  &  $	\ccc{1}{td}{}$      &  $ \cc{1}{ud}{33ii}$ \\
    &    {\tt O8qd }  &  $\ccc{8}{Qd}{}  $    &  $ \cc{8}{qd}{33ii}$ \\
    &    {\tt O1qd}   & $\ccc{1}{Qd}{}  $     &  $ \cc{1}{qd}{33ii}$\\	
        \midrule
 \multirow{9}{*}{$QQ+V,G,\varphi$}   &     {\tt OtG }    &  $c_{tG}$   &   $\Re\{\cc{}{uG}{33}\}$   \\
    &    {\tt  OtW}  &  $\ccc{}{tW}{}$		& $\Re\{\cc{}{uW}{33}\} $                   \\
    &    {\tt ObW} &$\ccc{}{bW}{}$ & $\Re\{\cc{}{dW}{33}\}$\\
    &        {\tt OtZ} &$\ccc{}{tZ}{}$ & $\Re\{-\sw\cc{}{uB}{33}+\cw\cc{}{uW}{33}\}$ \\
    &    {\tt Off} &$\ccc{}{\varphi tb}{}$ &$\Re\{\cc{}{\varphi ud}{33}\}$ \\
    &    {\tt  Ofq3}  & $\ccc{3}{\varphi Q}{}$  &   \cc{3}{\varphi q}{33}  \\
    &	{\tt OpQM} &$\ccc{-}{\varphi Q}{}$ & $\cc{1}{\varphi q}{33}-\cc{3}{\varphi q}{33}$\\ 
   &	{\tt Opt} &$\ccc{}{\varphi t}{}$ &$\cc{}{\varphi u}{33}$ \\
   &     {\tt Otp} & $\ccc{}{t\varphi}{}$&$\Re\{\cc{}{u\varphi}{33}\}$ \\
        \bottomrule
  \end{tabular}
  \caption{\small The notation that we will use to denote the results
    of the fits presented in this work.
    In each case, we indicate the internal notation for the degree of freedom
    and the corresponding definition in terms of the
    operators in the Warsaw basis.
    The degrees of freedom are divided into three classes:
    four-heavy-quark operators ($QQQQ$), two-heavy-two-light-quark operators ($QQqq$),
    and operators that couple two heavy quarks to gauge and Higgs bosonic fields
    ($QQ+V,G,\varphi$).
    \label{eq:summaryOperators}
}
\end{table}


Note that some of the degrees of freedom
defined in Table~\ref{eq:summaryOperators}  have  already
been studied in the context of SMEFT fits of the top quark sector, see
{\it e.g.}~\cite{Buckley:2015lku,Zhang:2017mls,Englert:2017dev}
and references therein.
For instance, the chromomagnetic operator
$c_{tG}$ was constrained to be within  $[-1.3,1.2]$
for $\Lambda=1$ TeV at the 68\% confidence level in the
analysis of~\cite{Buckley:2015lku}.
However, so far the simultaneous determination of the complete set of
degrees of freedom of Table~\ref{eq:summaryOperators}
has never been carried out.
In most cases, existing fits consider only either varying one operator at a time
or marginalising over a smaller subset of operators, and the bounds derived in
this way can differ significantly from those derived in a more global analysis.
We will come back in Sect.~\ref{sec:results} to the comparison of the results
of our analysis with previous studies in the literature.

The SMEFT degrees of freedom defined in Table~\ref{eq:summaryOperators} have
been implemented at LO in the UFO model {\tt dim6top} as discussed
in~\cite{AguilarSaavedra:2018nen}. Results obtained with {\tt dim6top} have been
benchmarked with the independent UFO implementation
available in the {\tt SMEFTsim} package~\cite{Brivio:2017btx}.
In this work the {\tt dim6top} model is complemented with the necessary 
counter-terms to enable NLO computations.  
The UFO model has been interfaced to {\tt MadGraph5\_aMC@NLO} to compute
the $\mathcal{O}\lp \Lambda^{-2}\rp$ and  $\mathcal{O}\lp \Lambda^{-4}\rp$
SMEFT corrections to the relevant SM cross-sections as indicated in 
Eq.~(\ref{eq:smeftXsecInt}).

\subsection{Operators not involving top quarks}
\label{eq:nontopops}

In this work, we follow Ref.~\cite{AguilarSaavedra:2018nen} and only include
those operators that explicitly modify the  couplings of the top quark
with the other SM fields.
We therefore assume that other relevant
operators are well constrained by processes that do not involve top quarks.
In the following, we give
a brief overview of these operators and discuss how they are constrained.

Firstly, the operator
\begin{equation}
O_{G}=f^{ABC}G_\mu^{A\nu}G_\nu^{B\rho}G_\rho^{C\mu},
\end{equation}
enters $t\bar t (V/H)$ production through a modification of the triple gluon 
coupling.
Although it has been suggested that $t\bar t$ production can possibly
constrain it due to its non-interference with the SM in di-jet 
production~\cite{Cho:1994yu},
it has been shown recently that this operator is most tightly constrained by
multi-jet production measurements~\cite{Krauss:2016ely,Hirschi:2018etq}.
The bounds on the coefficient $c_G$ are found to lie within
$[-0.04,0.04]$ TeV$^{-2}$, beyond the sensitivity of top quark pair production
or associated production.

\newcommand{\OO}{\ensuremath{\mathcal{O}}}
\newcommand{\sss}{\scriptscriptstyle}
\newcommand{\Op}[1]{\OO_{\sss #1}}

Secondly, operators involving a modification of the electroweak gauge-boson 
couplings to light fermions are in principle relevant to the interpretation
of the single top, $tZ$ and $ttZ$ measurements.
Most of these operators are however reasonably well constrained by electroweak 
precision observables.
In the Warsaw basis, and under the assumed flavour structure, they are
$\Op{\phi q}^{(1)}$,
$\Op{\phi q}^{\sss(3)}$, $\Op{\phi u}$, $\Op{\phi d}$,
$\Op{\phi l}^{\sss(3)}$, $\Op{\phi l}^{\sss(1)}$, $\Op{\phi e}$, $\Op{ll}^{\sss
(3)}$, $\Op{\varphi WB}$ and $\Op{\varphi D}$.
Among these 10 operators, 8 degrees of freedom are stringently constrained
by electroweak observables \cite{Falkowski:2014tna}, while two flat directions
remain which are constrained only by diboson production processes,
as discussed in
Refs.~\cite{Grojean:2006nn,Alonso:2013hga,Brivio:2017bnu} for example.
The two flat directions can be conveniently parametrised with
\cite{Hagiwara:1993ck}:
\begin{eqnarray}
\Op{HW}&=&(D^\mu\phi)^\dagger\tau_{\sss I}(D^{\nu}\phi)W^{\sss I}_{\mu\nu},\\
\Op{HB}&=&(D^\mu\phi)^\dagger(D^{\nu}\phi)B_{\mu\nu}.
\end{eqnarray}
which are linear combinations of Warsaw basis operators. Together with another
basis operator $O_W$, they form the full set of operators that modify the
triple-gauge-boson couplings (TGC).
While in principle these couplings would enter the $tZ$
process considered in this analysis (as well as measurements of $t\gamma$ that 
we do not include), they are well constrained from
diboson production at LEP2 and the LHC\@.
An interesting question is whether
processes like $tZ$ production could enhance the sensitivity to the anomalous
TGCs, as the diagrams of this process in the SM display large cancellations
among each other as required by unitarity, which are then spoiled by anomalous
TGC leading to enhanced cross sections at large energy.
The study performed in
Ref.~\cite{Degrande:2018fog} shows that while the effect is indeed present,
it is not significant enough to compete with the sensitivities provided by 
diboson production.
Therefore neglecting these operators in the associated production of single 
top quarks is well justified.

Another operator that potentially would need to be taken into account is
associated to the modification of the $Zbb$ coupling, characterized by the
coefficient of
\begin{flalign}
	\qq{}{\varphi b}{}
	=\FDF (\bar{b}\gamma^\mu b) \, ,
\end{flalign}
as well as by the degree of freedom \ccc{+}{\varphi Q}{} defined
in Eq.~(\ref{cplusphiQ}).
These are constrained
by the decay rate of $Z\to b\bar b$ and the forward-backward asymmetry of
$e^+e^-\to b\bar b$ at the $Z$ pole measured by LEP\@.
The corresponding constraints on the two
coefficients are below $\mathcal{O}(0.1)~\mathrm{TeV}^{-2}$.
Therefore $C_{\varphi b}$ can be safely ignored in this analysis.
Conversely, the constraint on $\ccc{+}{\varphi Q}{}$
is in principle relevant as it is a linear combination of
$\ccc{-}{\varphi Q}{}$ and $\ccc{3}{\varphi Q}{}$, which enter this fit.
However, in this work we choose not to include this information, because
our goal is to quantify the direct constraints provided
by top quark measurements.

Another operator that one might have to consider is the Higgs-gluon operator
\begin{flalign}
	O_{\varphi G}=\varphi^\dagger\varphi G_{\mu\nu}^A G^{A\mu\nu} \, ,
\end{flalign}
which enters in the $t\bar tH$ production process.
While this operator is already tightly constrained by Higgs production in gluon 
fusion, $gg\to H$, this process is also affected by exactly the same top quark 
degrees of freedom that enter $t\bar tH$, namely $\ccc{}{t\varphi}{}$
and $\ccc{}{tG}{}$.
Therefore in principle the marginalised limit should be
derived by combining $gg\to h$ together with the other top quark measurements,
and fitting simultaneously $C_{\varphi G}$ and the relevant top quark
operators.
Such a combined fit to both the $t\bar tH$ and $gg\to H$ processes  
has been studied in~\cite{Maltoni:2016yxb}, showing that, within its 
marginalised bound, $C_{\varphi G}$ does have an impact on the $t\bar tH$ rate, 
but it is not very significant.
Therefore in this work we include the data on the $t\bar tH$ cross-sections
but not the $gg\to H$ ones,
since the latter will only fix the value of $C_{\varphi G}$ without
affecting the description of the $t\bar tH$ process too much.
We should keep in mind that it would be
possible to improve upon this by explicitly including
the $gg\to H$ cross-sections to our experimental inputs.

Another interesting operator is the following:
\begin{flalign}
	O_{bG}=\left( \bar Q\sigma^{\mu\nu}T^Ab \right)\phi G^A_{\mu\nu} \, ,
\end{flalign}
which, in addition to all four-fermion operators involving two light quarks 
and two right-handed $b$ quarks, will affect the description of 
$t\bar{t}b\bar{b}$ production.
Ref.~\cite{Bramante:2014hua} has reported $\mathcal{O}(1)~\mathrm{TeV}^{-2}$
bounds on $C_{bG}$ from the analysis of $pp\to b\bar b$ production.  \
The other four-fermion operators will enter the same process, and it is unlikely
that $t\bar{t}b\bar{b}$ production will provide an even stronger constraint.

Finally, two-lepton-two-top-quark operators, such as $(\bar t\gamma^\mu
t)(\bar e\gamma_\mu e)$ and $(\bar Q\gamma^\mu \tau^I Q)(\bar l\gamma_\mu
\tau^I l)$, could in principle affect the description of
$t\bar tZ$ and $tZ$ production as well as the
measurement of the $W$-helicity fractions in top quark decay, if the decays
of $W$ and $Z$ bosons are taken into account.
However, depending on the details of the analysis, the inclusion of these
operators requires a reinterpretation of the experimental measurements, since
for example the extrapolation from the fiducial to the total phase space could
be affected by SMEFT effects.
We thus postpone the inclusion of these operators to future studies.

We emphasize that the decoupling of the operators that do not involve
top quarks from the interpretation of top quark measurements at the LHC
is in principle only an approximation.
Not all of these operators are currently strictly constrained
by other processes, but they could be dealt with either by possible improvements
in the future, or by extending our fit by including additional measurements.
However, until the most complete global fit can be performed, approximations
like the ones we adopt here are always useful, because they allow us to focus 
on a certain sector
of SMEFT, to study a certain type of processes, and to obtain an intuitive
understanding of the underlying physics.
Fortunately, as we can see from the discussions above, for a study focused on
the interpretation of top quark measurements, the assumption of the decoupling 
of non-top operators is in general already very good, and future improvements 
can be envisioned.
We therefore expect our results to be robust and not significantly affected by
the possible inclusion of SMEFT operators that do not involve top quarks.

\subsection{NLO QCD effects in the SMEFT calculation}
\label{sec:nlosmeft}

In Eq.~(\ref{eq:smeftXsecInt}), the coefficients $\kappa_i$ and
$\widetilde{\kappa}_{ij}$ can be evaluated at either leading order in both the
QCD and electroweak couplings, or by also including higher-order perturbative
corrections.
Given the high precision of available top quark measurements, particularly from 
the LHC Run II, as well as the further improvements expected at Run II and 
during the High-Luminosity (HL) LHC, it is important to take into account the 
NLO QCD corrections to SMEFT effects.
This is necessary for a number of reasons, including:
\begin{itemize}
\item QCD corrections to total rates are often quite large, especially
  for processes that are proportional to $\alpha_s$ at the Born level.
  Taking them into account  results in general
  in an improvement of the bounds on the SMEFT Wilson
  coefficients.
  Additionally, NLO QCD corrections also reduce the theoretical
  uncertainties from scale variations, which is helpful in
  discriminating between different BSM scenarios.
\item QCD corrections can distort the distributions of key observables.
  Given that the interpretation of
  differential distributions plays an important role in
  SMEFT global fits, providing reliable predictions for them
  is crucial.
  For instance, it is shown in
  Ref.~\cite{Zhang:2016omx} that in the presence of a deviation
  from the SM, missing QCD corrections to certain differential distributions
  could lead us to make incorrect conclusions on the nature of BSM physics.
\item The experimental sensitivity to SM
  deviations can be improved by using the most accurate SMEFT
  predictions and by optimizing the experimental strategies in
  a top-down way.
  However, the large QCD corrections at the LHC make this improvement
  unrealistic without consistently taking into account NLO predictions.
\end{itemize}

Motivated by the above considerations, theoretical calculations in the
SMEFT at NLO in the QCD perturbative expansion have started to appear, in many
cases matched with parton showers.
For instance,
computations including higher dimensional operators
for top-related processes have been presented in
\cite{Degrande:2014tta,Franzosi:2015osa,Zhang:2016omx,Bylund:2016phk,
Maltoni:2016yxb,Degrande:2018fog,Durieux:2018tev},
and those for Higgs and electroweak processes in
\cite{Artoisenet:2013puc,Maltoni:2013sma,Demartin:2014fia,Demartin:2015uha,
Mimasu:2015nqa,Degrande:2016dqg,Alioli:2018ljm}.
In particular, based on the machinery of automatic computations in {\tt
MadGraph5\_aMC@NLO}, a systematic framework for including higher-dimensional
operators at NLO has been established through the studies of
Refs.~\cite{Degrande:2014tta,Franzosi:2015osa,Zhang:2016omx,Bylund:2016phk,
Maltoni:2016yxb,Degrande:2018fog,Durieux:2018tev}, and the implementation of
the full set of dimension-6 SMEFT operators is currently being studied.

A novel feature of the present work, as compared
to previous SMEFT studies of the top quark sector,
is that we will exploit this framework and
include the theoretical predictions at NLO in QCD
whenever possible.
This allows us to obtain the currently most accurate bounds on the coefficients
of the SMEFT operators affecting the top quark couplings.
Furthermore, by switching on and off the NLO QCD corrections
in the fit, we  can understand better the
importance of the higher-order corrections in the
SMEFT calculation when constraining different operators.
In the following, we briefly explain which corrections will be included.
As we will discuss in Sect.~\ref{sec:settings_expdata}, the SM calculations
are always performed using the highest perturbative order
available for each process.

Of all the degrees of freedom relevant in this work, the operators involving
only two fermion fields have been fully automated already in this framework,
it is therefore possible to straightforwardly evaluate their
associated NLO QCD corrections.
Four-fermion operators are being studied, and their complete implementation is 
expected to be publicly available in a short timescale.
In this work we will include the NLO QCD corrections
to the four-fermion operators only in the inclusive single top and top-pair
production processes, which are the most accurately measured processes.

One practical difficulty in obtaining stable numerical results at NLO
is that  the residual
  uncertainties arising from the
  numerical Monte Carlo integration of the cross-sections,
the so-called ``MC errors'', for the interference terms $\kappa_i$ and
in the cross terms
$\widetilde\kappa_{ij}, i\neq j$ can be large.
These $\kappa$ terms are obtained by
sampling the parameter space spanned by the full set of relevant degrees of
freedom, and computing the total cross section (or other observables)
iteratively.
The results are fitted to the most general quadratic function.
Therefore the MC error on interference/cross terms can be large, in
particular when these terms are suppressed (see discussions in
Ref.~\cite{Fichet:2016iuo}).
Examples of suppressed interference contributions
will be discussed in the next subsection.
Given the large number of SMEFT
operators relevant for the description
of top quark measurements, a full simulation at NLO QCD would be
very time consuming.

In this work, we adopt the following strategy:
\begin{itemize}
\item For $t\bar t$ and single top production, the
  experimental measurements exhibit the highest precision.
  We therefore use the full NLO simulation.
  This is done by sampling the parameter space following~\cite{Fichet:2016iuo}.
  For each point, we generate $8\times 10^5$
  events, and estimate the corresponding MC errors
  for each observable included in the fit.
  These MC uncertainties can be taken into account when
  constructing the $\chi^2$ function, as discussed in 
  Sect.~\ref{sec:fitsettings}.
\item For associated production processes, the measurements are less
  accurate.
  We generate the full LO predictions using
  the implementation provided in~\cite{AguilarSaavedra:2018nen}.
  We then apply $K$-factors from previous calculations of
  $t\bar tZ$, $t\bar tH$ and $tZj$ production, wherever available
  \cite{Bylund:2016phk,Maltoni:2016yxb,Degrande:2018fog}.
  For contributions or processes that have not been previously calculated
  (e.g.~contributions from the four-fermion operators, and $t\bar tb\bar b$ and
  $tt\bar t\bar t$ processes), we simply apply the SM $K$-factor.
 
\item The $W$-helicity in top quark decay is available at
  NLO in the form of analytical results~\cite{Zhang:2014rja}.
\end{itemize}
In Sect.~\ref{sec:results} we will assess the stability of our results
with respect to the inclusion or not of NLO QCD corrections
to the SMEFT dimension-6 effects.

Finally, the $t\bar tt\bar t$ process is a special one, because the dominant
contributions can come from $\mathcal{O}(c^4/\Lambda^8)$ terms, i.e.~it goes
beyond the parametrisation of Eq.~(\ref{eq:smeftXsecInt}).  
This is due to diagrams with two insertions of $qqtt$ operators at the amplitude
level, which leads to a rapid growth of the cross section as a function of 
energy, potentially causing problems with SMEFT validity \cite{Zhang:2017mls}.  
In this work we only keep the terms up
to $\mathcal{O}(\Lambda^{-4})$ from one insertion of the operators.  
This represents a good approximation if the coefficients
of $qqtt$ operators are 
constrained to be within the order of a few TeV$^{-2}$ at most.  

Beyond this limit, $\mathcal{O}(\Lambda^{-8})$ terms dominate, and our
predictions are not accurate, but they do give a lower bound of the true SMEFT
contribution.  Given that in practice, only the upper bound of this measurement
is useful for constraining operators, our approximation will always lead to
conservative result.  
In addition, to avoid possible EFT validity problems due to
the particular energy-growth behavior of this process, we impose a hard cut of
2 TeV on the four-top invariant mass in our prediction.  
Once again this leads to a lower bound of the true SMEFT contribution.  
When compared with the upper bound of the full cross section measurement 
without the cut, it gives a conservative bound.  
In Sect.~\ref{sec:highenergy} we will discuss the
dependence of our final result on this cut.

\subsection{General discussion}
\label{sec:generaldiscussion}

To complete this section,  we briefly discuss some additional
aspects of the SMEFT framework relevant to
the present analysis of the top quark sector.

\paragraph{RG running and mixing.}
 In general, the SMEFT operators $\{ \mathcal{O}_i^{(6)}\}$ in
 Eq.~(\ref{eq:smeftlagrangian}) will run with the
 scale and thus the coefficients $\{c_i\}$ will depend on the typical momentum
 transfer of the process.
 This dependence can be evaluated using renormalisation group (RG)
 equations~\cite{Jenkins:2013zja,Alonso:2013hga,Jenkins:2013wua}, but here 
 since we
 focus on processes with a similar energy scale, $E\simeq m_t$, we will
 not include these operator running effects.
In any case, the inclusion of NLO QCD
 corrections will reduce this scale dependence \cite{Maltoni:2016yxb,Deutschmann:2017qum},
  even for the case of differential distributions where significantly 
  different scales are involved. In \cite{Maltoni:2016yxb}, RGE effects are 
  calculated for $ttH$ production over a scale ranging from 
  150 GeV to 2 TeV. It is found that RGE effects remain below the 10\% level.  
   We expect a similar behaviour for other top production processes also at the
    differential level and therefore we can for this first study safely ignore RGE effects.

\paragraph{Energy enhancements.}
One important feature of Eq.~(\ref{eq:smeftXsecInt}) is that certain SMEFT
operators will induce a growth of the cross-sections $\sigma(E)$ with the
energy $E$~\cite{Alioli:2017jdo}.
This is a consequence of the fact that, in four space-time dimensions, field
theories involving operators with mass dimensions $d>4$ exhibit a strong
sensitivity to the UV cut-off of the theory $\Lambda$.
In other words, the coefficients $\{\kappa_i\}$ in Eq.~(\ref{eq:smeftXsecInt})
will include terms that grow quadratically with $E$.
Restricting ourselves to the $\mathcal{O}\lp\Lambda^{-2}\rp$ corrections,
Eq.~(\ref{eq:smeftXsecInt}) can be written schematically as:
\be
\label{eq:smeftlagrangianXsec}
\sigma(E)=\sigma_{\rm SM}(E)\left[1 + \sum_i^{N_{d6}}
\omega_i \frac{c_i
	v^2
}{\Lambda^2} +
\sum_i^{N_{d6}} \widetilde{\omega}_i \frac{c_i E^2}{\Lambda^2} +
\mathcal{O}\lp\Lambda^{-4}\rp\right]  \, ,
\ee
where $v$ is the Higgs boson vacuum expectation value (vev).
The coefficients $\{\omega_i\}$ and $\{\widetilde{\omega}_i\}$
are process-dependent and arise from the $\{\kappa_i\}$ coefficients in
Eq.~(\ref{eq:smeftXsecInt}), once we separate the energy-growing
contributions.

Whether a dimension-6 operator leads to a nonzero value
of the energy-growing coefficient $\tilde\omega_i$ depends on
many factors.
For example, four-fermion operators in $t\bar t$ and single top processes can 
interfere with the SM amplitude with the same helicity configurations
without any additional suppression, and therefore their contributions are
proportional to $E^2/\Lambda^2$ by simple power counting.
On the contrary, operators involving only two fermions contribute as
$v^2/\Lambda^2$: the current-current operators like $O_{\varphi Q}^{(1)}$ always
enter with two powers of the Higgs vev, while the dipole operators come with
one power of the Higgs vev, but the flip of the fermion chirality leads to an 
additional suppression factor $m_t$ or $m_b$ upon interfering with the SM.

The associated production channels are on the other hand more complicated.
There, even two-fermion operators can lead to $E^2/\Lambda^2$ contributions.
This is because already in the SM, each Feynman diagram could lead to 
energy-growing terms, but overall cancellations occur among the leading 
contributions in different diagrams as required by unitarity.
With higher-dimensional operators instead, even a $\mathcal{O}(v^2/\Lambda^2)$ 
change in one diagram could spoil the cancellation and lead to 
$\mathcal{O}(E^2/\Lambda^2)$ modification of the total rate.

Finally, the specific observable under consideration also matters.
The interference between SM and SMEFT
amplitudes with different helicities are suppressed by mass factors, but
this suppression can be lifted by considering the decay products of the particle
with different helicities~\cite{Panico:2017frx,Azatov:2017kzw}.
Note that
suppression due to different helicities only applies to the interference term.
As a result, at large energy the SMEFT contribution could be dominated by the
$\mathcal{O}(\Lambda^{-4})$ terms.
This is actually one of the reasons to
include the quadratic contributions from the operators.
A similar situation
occurs for instance in diboson production at the LHC,
see the discussion in~\cite{Falkowski:2016cxu}.

The schematic decomposition in
Eq.~(\ref{eq:smeftlagrangianXsec}) indicates that the effects of those
operators for which $\widetilde{\omega}_i\ne 0$
will be enhanced with the energy $E$ of the process.
In turn, this will lead to an increased sensitivity of the experimental
measurements at high energies with the values of the corresponding $c_i$
coefficients.
This property can be uniquely exploited at the LHC, where multiple processes
probe the TeV region.
It has been shown that in some cases the enhancement due to energy dependence in
LHC processes already leads to a sensitivity competitive with respect to the LEP
measurements~\cite{Farina:2016rws,Greljo:2017vvb,Falkowski:2017pss}.

It is therefore an interesting question whether the same happens for top quark
measurements, and which operators can benefit from the high energy reach
provided by the LHC\@.
In this work we are going to study the impact of high energy measurements by
comparing results with and without high mass bins in differential distributions,
for instance in the invariant mass distribution in $t\bar{t}$ production.
This will be discussed in Sect.~\ref{sec:highenergy}.

\paragraph{Validity of the SMEFT.}
Following up on the previous discussion,
while it is useful to make use of the large energy transfer in the observed
events, one has to pay special attention to remain in the region
$E\ll \Lambda$.
Otherwise, the whole validity of the SMEFT power expansion would be
questionable, and it would become impossible to interpret the resulting
constraints within any explicit BSM model.

To make sure that SMEFT analyses remain in their validity region,
it has been proposed in Ref.~\cite{Contino:2016jqw} and recommended in the
Top LHC WG EFT note~\cite{AguilarSaavedra:2018nen} that a
kinematic cut $E_{\rm cut}$ should be imposed
to events being analysed, as an upper bound of the energy transfer, so that
the condition
\begin{flalign}
	E<E_{\rm cut}<\Lambda
\end{flalign}
is always guaranteed.
Given that $\Lambda$ is the scale of the BSM dynamics and is
model-dependent, different values of $E_{\rm cut}$ should be used, and results
should be derived for each of these values.
While the input data used in this work is not provided with such explicit cut,
for specific distributions it is possible to remove bins with scale higher than
a given value of $E_{\rm cut}$.
A strong dependence of the final results on high mass bins would imply that the
sensitivity is dominated by the high energy events, and that the constraints can
only be interpreted for those BSM models where $\Lambda$ lies above the scale of
the largest bin used in the fit.

Among the input data used in this work, only the $m_{t\bar t}$ distribution
in $t\bar t$ production extends to energy scales above 1 TeV.
Therefore, in Sect.~\ref{sec:highenergy} we will study the dependence of our
results on the high mass bins in the $m_{t\bar{t}}$ distributions of ATLAS and
CMS\@.

\paragraph{Quadratic dimension-6 contributions.}
The last term in Eq.~(\ref{eq:smeftXsecInt}) arises from the squares of
dimension-six amplitudes. At order $\mathcal{O}(\Lambda^{-4})$, they are
formally subleading contributions, so one can decide to include them without
modifying the accuracy of the prediction of the central value.
Indeed, there are good reasons why it could be worth including them in the
analysis.
To begin with, in BSM models with relatively large couplings the quadratic
dimension-6 terms can become dominant without exiting the realm of validity of
the EFT, see for instance
Refs.~\cite{Domenech:2012ai,Biekoetter:2014jwa,Biekotter:2016ecg,Contino:2016jqw}.
Therefore including the quadratic terms allows the results of a SMEFT analysis
to be interpreted in the context of these scenarios.

Furthermore, the interference terms in Eq.~(\ref{eq:smeftXsecInt}) are often
suppressed, so that the leading contributions arise from quadratic dimension-6
terms.
In these cases,
one relies on the quadratic terms to extract meaningful bounds from the measurements.
As an extreme case, the SM amplitude may not interfere with the SMEFT amplitude
at all, because of different helicity and colour structures or different $CP$
parity.
As an illustration,
in our analysis $\ccc{}{\varphi tb}{}$ and $\ccc{}{bW}{}$
cannot interfere with the SM in the limit of $m_b\to 0$, so they can
only be constrained once $\mathcal{O}( \Lambda^{-4})$ terms are included in the
fit.
Similarly, several of the $qqtt$ operators do not interfere with SM due to
their color singlet interaction, though this is slightly lifted once NLO
corrections are added.

It is also possible that, while the interference term exists, it does not lead
to an energy growth behaviour.
We have already mentioned that some operators
cannot lead to a nonzero $\tilde\omega_i$ at the order
$\mathcal{O}(\Lambda^{-2})$.
One would then expect that the dominant
sensitivity for these operators comes from $\mathcal{O}(\Lambda^{-4})$ contributions
at large energies.
For instance, it has been observed that in diboson production
at the LHC, the helicity selection rule \cite{Azatov:2016sqh} leads to an energy
suppression in the interference term, so the sensitivity to TGC couplings is 
dominated by $\mathcal{O}(\Lambda^{-4})$ terms \cite{Falkowski:2016cxu}.
Finally, a suppression of the interference term could be simply accidental.
This has been observed for weak dipole operators in $t\bar tZ$ and $t\bar
t\gamma$ production processes \cite{Bylund:2016phk}, and is relevant also in
the present analysis.

Here we will follow the recommendations
of Ref.~\cite{AguilarSaavedra:2018nen} and repeat the
analysis with and without including the quadratic SMEFT contributions.
This
comparison will then tell us where quadratic dimension-6 contributions
are subleading and where the truncation at dimension-6 at the cross section level can
be a good approximation.
In addition, under certain assumptions, the $\mathcal{O}(\Lambda^{-4})$
corrections could also provide an estimate of how reliable are the SMEFT fit
results with respect to higher orders in the effective theory parameter
expansion.
Note, however, that for the degrees of freedom whose
contributions at $\mathcal{O}(\Lambda^{-2})$ are extremely suppressed, such as
$\ccc{}{\varphi tb}{}$ and $\ccc{}{bW}{}$, and for the color-singlet four-fermion
interactions $\ccc{1,1}{Qq}{}$, $\ccc{1}{tu}{}$, $\ccc{1}{td}{}$,
$\ccc{1}{tq}{}$, $\ccc{1}{Qu}{}$, and $\ccc{1}{Qd}{}$, our numerical approach
would lead to large MC errors on the interference, and therefore the
resulting bounds from a $\mathcal{O}(\Lambda^{-2})$ fit will be at most
qualitative.

\section{Experimental data and theoretical calculations}
\label{sec:settings_expdata}

In this section we describe the experimental measurements of top quark
production at the LHC which will be used to constrain the SMEFT operators
related to the top sector.
For each dataset we discuss its main features, the information that it
provides on the SMEFT effects, and the treatment of experimental
uncertainties.
We also describe the settings of the theoretical calculations of the SM and
SMEFT contributions to the cross-sections that are used for each process.
Finally, we summarise the main features of our choice of fitting basis in terms
of the sensitivity of each of the input LHC processes to the individual
operators.

\subsection{Top quark production at the LHC}

In the present analysis, we will constrain the top quark sector of the SMEFT
by using experimental measurements from the LHC Run I at $\sqrt{s}=8$ TeV 
and from Run II at $\sqrt{s}=13$ TeV.
We do not consider previous, less precise data at $\sqrt{s}=7$ TeV nor data from
the Tevatron.
The measurements of top quark production at the LHC 7 TeV
  are superseded by the more precise 8 and 13 TeV ones
  and there is no loss of information incurred by not including them.
  Concerning the Tevatron measurements, on the one hand, they are affected by larger uncertainties than those of the LHC, but on the other hand the production of top quarks proceeds mostly via the quark-antiquark process and therefore it provides additional sensitivity to new combinations of four-fermion operators. While in this work we have focused on LHC data, we plan to include the Tevatron constraints in a future iteration of our analysis.

The various experimental datasets used as input in this work are summarised
in Tables~\ref{eq:input_datasets},~\ref{eq:input_datasets3}
and~\ref{eq:input_datasets2}, for inclusive top quark pair production,
$t\bar{t}$ production in association with gauge and Higgs bosons, and single
top production measurements, respectively.
For each dataset, we indicate the type of process, its label, the
centre-of-mass energy $\sqrt{s}$, information on the final state or the
specific production mechanism, the available observables, the number of data
points $N_{\rm dat}$, and the corresponding publication reference.

As we will discuss in more detail below, information on correlations
between systematic uncertainties is available only for a subset of the data,
specifically, for all the 8 TeV distributions in Table~\ref{eq:input_datasets}
(including $W$ helicity fractions)~\cite{Aad:2015mbv,Khachatryan:2015oqa,
Sirunyan:2017azo,Aaboud:2016hsq,Khachatryan:2016fky}
and for the top quark pair production measurement at 13 TeV of
Ref.~\cite{Khachatryan:2016mnb}.
For the rest of the datasets, since this information is missing, we add
statistical and systematic uncertainties in quadrature.
Also, the correlation matrix among various differential distributions
is not usually available.
To avoid double counting, only one distribution per dataset can therefore be
included in the fit.
The ATLAS 8 TeV lepton+jet dataset is an exception,
where such correlations have recently become
available~\cite{ATL-PHYS-PUB-2018-017}.
However, because the effect of correlating all differential distributions
has not been studied yet in the fitting framework used here, we do not
utilise them.

As indicated in Table~\ref{eq:input_datasets}, in the case of differential
distributions we always use absolute rather than normalised cross-sections.
The rationale is that absolute distributions are more sensitive to SMEFT
effects than the normalised ones, except when the corresponding fiducial
cross-sections are included in the fit at the same time.
If the measurements are presented only in terms of normalised
differential distributions, absolute distributions are reconstructed from
the former using the fiducial cross-section.
Uncertainties are added in quadrature.
To avoid double counting, total and/or fiducial cross-sections are excluded
from the fit whenever the corresponding absolute differential distributions
are part of the input dataset.

\begin{figure}[t]
\centering
\includegraphics[width=0.99\linewidth]{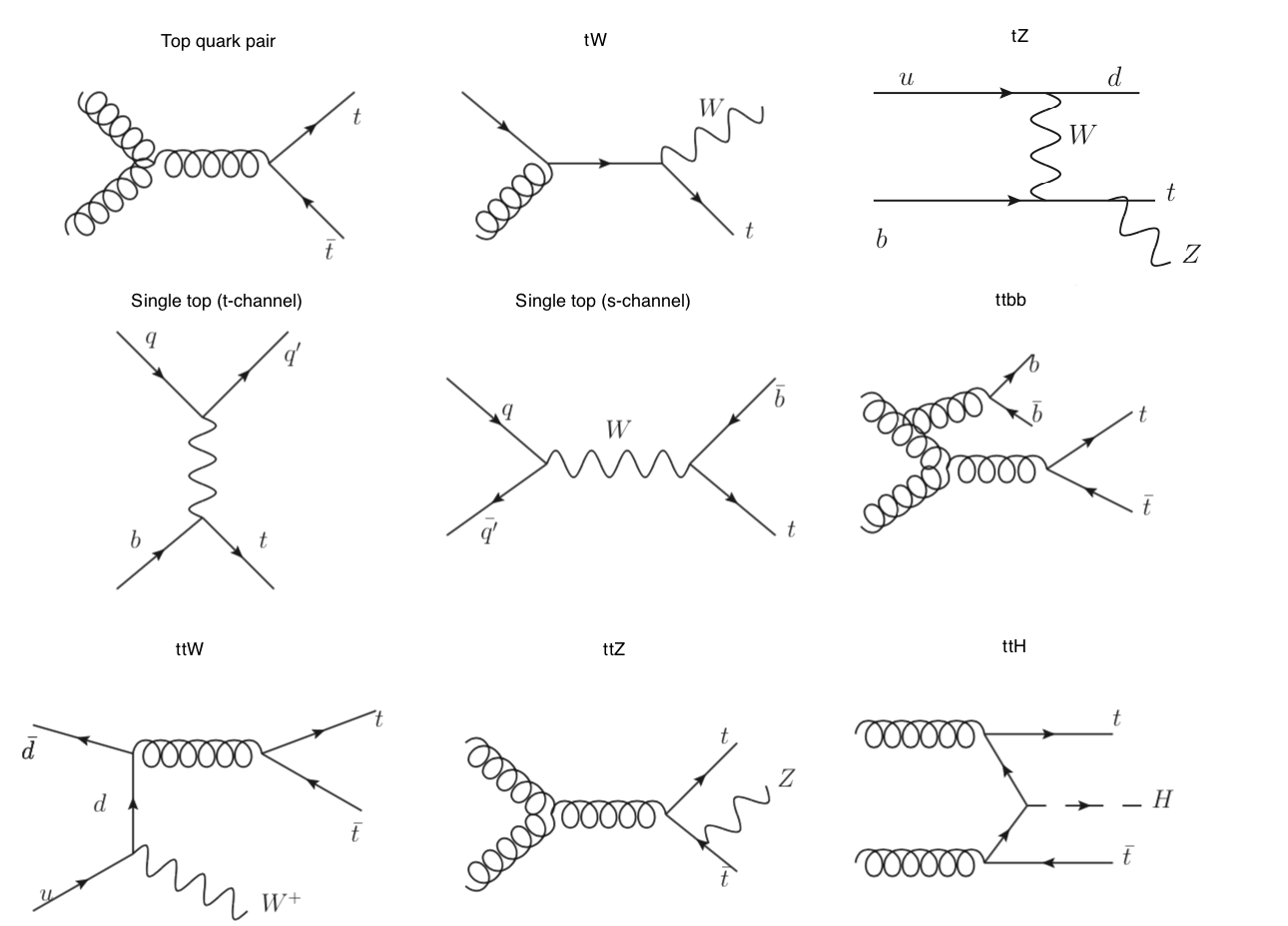}
\caption{\small Representative Feynman diagrams at the Born level for
  the dominant production channels of top quarks at the LHC that are
  considered in the present SMEFT analysis.
  We show top-quark pair production; single-top production in association
  with a $W$ or $Z$ boson and in the $t$- and $s$-channels; $t\bar{t}$
  production in association with $t\bar{t}$ or $b\bar{b}$; and $t\bar{t}$ 
  production in association with a
  $W$ or $Z$ gauge boson or with the Higgs boson $H$.
\label{fig:top-production-LHC} }
\end{figure}

To gain some intuition about the expected sensitivity of the input
dataset to each of the SMEFT operators defined in Sect.~\ref{sec:smefttop},
it is useful to recall what are the dominant production mechanism for each
top-related observable in the SM\@.
In Fig.~\ref{fig:top-production-LHC} we display representative Feynman diagrams
at the Born level for the production of top quarks at the LHC in the channels
that we consider in this analysis.
Specifically, we show top-quark pair production; single-top production
in association with a $W$ or $Z$ boson and in the $t$- and $s$-channels;
$t\bar{t}$ production in association with $t\bar{t}$ or $b\bar{b}$; 
and $t\bar{t}$ production in association with a
$W$ or $Z$ gauge boson or with the Higgs boson $H$.

\begin{table}[t]
  \centering
  \footnotesize
   \renewcommand{\arraystretch}{1.45}
  \begin{tabular}{c|c|c|c|c|c|c}
Process & Dataset   &  $\sqrt{s}$ & Info  &  Observables  & $N_{\rm dat}$ & Ref   \\
    \toprule
 \multirow{2}{*}{$t\bar{t}$}  &   \multirow{2}{*}{ {\tt ATLAS\_tt\_8TeV\_ljets}}      &
    \multirow{2}{*}{{ \bf 8 TeV}}   &  \multirow{2}{*}{lepton+jets}   & $
        d\sigma/d|y_{t}|$, $
        d\sigma/dp^T_{t}$,      & 5, 8,   & \multirow{2}{*}{\cite{Aad:2015mbv}}  \\
  &         &   &   & $
        d\sigma/dm_{t\bar{t}}$, $
        d\sigma/d|y_{t\bar{t}}|$      &  7, 5  &    \\
        \midrule
\multirow{2}{*}{$t\bar{t}$}  &          \multirow{2}{*}{ {\tt CMS\_tt\_8TeV\_ljets}}      &
    \multirow{2}{*}{{\bf 8 TeV}}   &  \multirow{2}{*}{lepton+jets}   & $
        d\sigma/dy_{t}$, $
        d\sigma/dp^T_{t}$,      &  10, 8,  &
        \multirow{2}{*}{\cite{Khachatryan:2015oqa}}  \\
 &          &   &   & $ 
        d\sigma/dm_{t\bar{t}}$, $
        d\sigma/dy_{t\bar{t}}$      & 7, 10 &    \\
        \midrule
\multirow{4}{*}{$t\bar{t}$}  &       \multirow{4}{*}{ {\tt CMS\_tt2D\_8TeV\_dilep}}      &
\multirow{4}{*}{{\bf 8 TeV}}   &  \multirow{4}{*}{dileptons}   &
$d^2\sigma/dy_{t}dp^T_{t}$,
             &  16,   &
        \multirow{4}{*}{\cite{Sirunyan:2017azo}}  \\
        &           &   &   &  $d^2\sigma/dy_{t}dm_{t\bar{t}}$, & 16, &  \\
&           &   &   & $d^2\sigma/dp^T_{t\bar{t}}dm_{t\bar{t}}$, & 16, 
        &    \\
        &           &   &   &
        $d^2\sigma/dy_{t\bar{t}}dm_{t\bar{t}}$  & 16
        &    \\
\midrule
\multirow{2}{*}{$t\bar{t}$}  &      \multirow{2}{*}{ {\tt CMS\_tt\_13TeV\_ljets }}      &
 \multirow{2}{*}{{\bf 13 TeV}}   &  \multirow{2}{*}{lepton+jets}   & $
 d\sigma/d|y_{t}|$, $
  d\sigma/dp^T_{t},$      &  7, 9,  &
   \multirow{2}{*}{\cite{Khachatryan:2016mnb}}  \\
  &         &   &     &    $
 d\sigma/dm_{t\bar{t}}$, $
   d\sigma/d|y_{t\bar{t}}|$      &  8, 6   \\
\midrule
\multirow{2}{*}{$t\bar{t}$}  &      \multirow{2}{*}{ {\tt CMS\_tt\_13TeV\_ljets2 }}      &
 \multirow{2}{*}{{\bf 13 TeV}}   &  \multirow{2}{*}{lepton+jets}   & $
 d\sigma/d|y_{t}|$, $
  d\sigma/dp^T_{t},$      &  11, 12,  &
   \multirow{2}{*}{\cite{Sirunyan:2018wem}}  \\
  &         &   &     &    $
   d\sigma/dm_{t\bar{t}}$,
    $d\sigma/d|y_{t\bar{t}}|$      &  10, 10   \\
  \midrule
\multirow{2}{*}{$t\bar{t}$}  &      \multirow{2}{*}{ {\tt CMS\_tt\_13TeV\_dilep }}      &
 \multirow{2}{*}{{\bf 13 TeV}}   &  \multirow{2}{*}{dileptons}   & $
  d\sigma/dy_{t}$, $
    d\sigma/dp^T_{t},$      &  8, 6,  &
   \multirow{2}{*}{\cite{Sirunyan:2017mzl}}  \\
  &         &   &     &    $
   d\sigma/dm_{t\bar{t}}$, $
   d\sigma/dy_{t\bar{t}}$      &  6, 8   \\
   \midrule
\midrule
$t\bar{t}$   & {\tt ATLAS\_WhelF\_8TeV}  & {\bf 8 TeV}  & $W$ helicity fract &
$F_0, F_L, F_R$  &  3  &  \cite{Aaboud:2016hsq}  \\
\midrule
$t\bar{t}$   & {\tt CMS\_WhelF\_8TeV}  & {\bf 8 TeV}  & $W$ helicity fract &
$F_0, F_L, F_R$  &  3  &  \cite{Khachatryan:2016fky}  \\
\bottomrule
  \end{tabular}
  \caption{\small The experimental measurements of inclusive top quark pair 
    production at the LHC considered in the present analysis to constrain
    the coefficients of the SMEFT dimension-6 operators in the top sector.
    For each dataset, we indicate the type of process, the dataset label, 
    the center of mass energy $\sqrt{s}$, the final state or the specific 
    production mechanism, the available observables, the number of data 
    points $N_{\rm dat}$, and the publication reference.
    Most distributions are statistically correlated among them
    and one needs to be careful to avoid double counting.
\label{eq:input_datasets}
}
\end{table}

\begin{table}[t]
  \centering
  \scriptsize
   \renewcommand{\arraystretch}{1.45}
  \begin{tabular}{c|c|c|c|c|c|c}
Process & Dataset   &  $\sqrt{s}$ & Info  &  Observables  & $N_{\rm dat}$ & Ref   \\
    \toprule
$t\bar{t}b\bar{b}$   & {\tt CMS\_ttbb\_13TeV}  & {\bf 13 TeV}  & total xsec & $\sigma_{\rm tot}(t\bar{t}b\bar{b})$  &  1  &  \cite{Sirunyan:2017snr}  \\
\midrule
$t\bar{t}t\bar{t}$   & {\tt CMS\_tttt\_13TeV}  & {\bf 13 TeV}  & total xsec & $\sigma_{\rm tot}(t\bar{t}t\bar{t})$  &  1  &  \cite{Sirunyan:2017roi}  \\
\midrule
\midrule
 $t\bar{t}Z$   & {\tt CMS\_ttZ\_8\_13TeV}  & {\bf 8+13 TeV}  & total xsec & $\sigma_{\rm tot}(t\bar{t}Z)$  &  2  &  \cite{Khachatryan:2015sha,Sirunyan:2017uzs}  \\
   \midrule
 $t\bar{t}Z$   & {\tt ATLAS\_ttZ\_8\_13TeV}  & {\bf 8+13 TeV}  & total xsec & $\sigma_{\rm tot}(t\bar{t}Z)$  &  2  &  \cite{Aad:2015eua,Aaboud:2016xve}  \\
\midrule
 $t\bar{t}W$   & {\tt CMS\_ttW\_8\_13TeV}  & {\bf 8+13 TeV}  & total xsec & $\sigma_{\rm tot}(t\bar{t}W)$  &  2  &  \cite{Khachatryan:2015sha,Sirunyan:2017uzs}  \\
   \midrule
 $t\bar{t}W$   & {\tt ATLAS\_ttW\_8\_13TeV}  & {\bf 8+13 TeV}  & total xsec & $\sigma_{\rm tot}(t\bar{t}W)$  &  2  &  \cite{Aad:2015eua,Aaboud:2016xve}  \\
\midrule
\midrule
 $t\bar{t}H$   & {\tt CMS\_tth\_13TeV}  & {\bf 13 TeV}  & signal strength & $\mu_{t\bar{t}H}$  &  1  &  \cite{Sirunyan:2018hoz}  \\
\midrule
$t\bar{t}H$   & {\tt ATLAS\_tth\_13TeV}  & {\bf 13 TeV}  & total xsec & $\sigma_{\rm tot}(t\bar{t}H)$  &  1  &  \cite{Aaboud:2018urx}  \\
\bottomrule
  \end{tabular}
  \caption{\small Same as Table~\ref{eq:input_datasets} now for $t\bar{t}$ in association with
    heavy quarks, with weak vector bosons, and with the Higgs boson.
     \label{eq:input_datasets3}
  }
\end{table}

\begin{table}[t]
  \centering
  \footnotesize
   \renewcommand{\arraystretch}{1.45}
  \begin{tabular}{c|c|c|c|c|c|c}
Process & Dataset   &  $\sqrt{s}$ & Info  &  Observables  & $N_{\rm dat}$ & Ref   \\
\toprule
\multirow{1}{*}{Single $t$} &       \multirow{1}{*}{ {\tt CMS\_t\_tch\_8TeV\_inc}}      &
\multirow{1}{*}{{\bf 8 TeV}}   & \multirow{1}{*}{$t$-channel}   &
\multirow{1}{*}{$\sigma_{\rm tot}(t),\sigma_{\rm tot}(\bar{t})~(R_t)$}  & 2~(1)   &
\multirow{1}{*}{\cite{Khachatryan:2014iya}}  \\
\midrule
\multirow{1}{*}{Single $t$} &       \multirow{1}{*}{ {\tt CMS\_t\_sch\_8TeV}}      &
\multirow{1}{*}{{\bf 8 TeV}}   & \multirow{1}{*}{$s$-channel}   &
\multirow{1}{*}{$\sigma_{\rm tot}(t+\bar{t})$} & 1    &
\multirow{1}{*}{\cite{Khachatryan:2016ewo}}  \\
\midrule
\multirow{1}{*}{Single $t$} &       \multirow{1}{*}{ {\tt ATLAS\_t\_sch\_8TeV}}      &
\multirow{1}{*}{{\bf 8 TeV}}   & \multirow{1}{*}{$s$-channel}   &
\multirow{1}{*}{$\sigma_{\rm tot}(t+\bar{t})$}  & 1   &
\multirow{1}{*}{\cite{Aad:2015upn}}  \\
\midrule
  \multirow{2}{*}{Single $t$} &       \multirow{2}{*}{ {\tt ATLAS\_t\_tch\_8TeV}}      &
        \multirow{2}{*}{{\bf 8 TeV}}   & \multirow{2}{*}{$t$-channel}   &
\multirow{1}{*}{$d\sigma(tq)/dp_T^t,d\sigma(\bar{t}q)/dp_T^{\bar{t}}$}
& 5, 4   & \multirow{2}{*}{\cite{Aaboud:2017pdi}}  \\
           &          &   &     &
$d\sigma(tq)/dy_t,d\sigma(\bar{t}q)/dy_t$   &  4, 4  &  \\
     \midrule
   \multirow{1}{*}{Single $t$} &       \multirow{1}{*}{ {\tt ATLAS\_t\_tch\_13TeV}}      &
   \multirow{1}{*}{{\bf 13 TeV}}   & \multirow{1}{*}{$t$-channel}   &
\multirow{1}{*}{$\sigma_{\rm tot}(t),\sigma_{\rm tot}(\bar{t})~(R_t)$}  & 2~(1)   &
\multirow{1}{*}{\cite{Aaboud:2016ymp}}  \\
    \midrule
   \multirow{1}{*}{Single $t$} &       \multirow{1}{*}{ {\tt CMS\_t\_tch\_13TeV\_inc}}      &
   \multirow{1}{*}{{\bf 13 TeV}}   & \multirow{1}{*}{$t$-channel}   &
\multirow{1}{*}{$\sigma_{\rm tot}(t+\bar{t})~(R_t)$}  & 1~(1)   &
\multirow{1}{*}{\cite{Sirunyan:2016cdg}}  \\
  \midrule
 \multirow{2}{*}{Single $t$}  &      \multirow{2}{*}{ {\tt CMS\_t\_tch\_8TeV\_dif}}      &
 \multirow{2}{*}{{\bf 8 TeV}}   &  \multirow{2}{*}{$t$-channel}   & $
     d\sigma/dp_T^{(t+\bar{t})},$  &  6  &
     \multirow{2}{*}{\cite{CMS-PAS-TOP-14-004}}  \\
     &          &   &     & $
     d\sigma/d|y^{(t+\bar{t})}|$   & 6 &   \\
  \midrule
 \multirow{2}{*}{Single $t$}  &      \multirow{2}{*}{ {\tt CMS\_t\_tch\_13TeV\_dif}}      &
 \multirow{2}{*}{{\bf 13 TeV}}   &  \multirow{2}{*}{$t$-channel}   & $
     d\sigma/dp_T^{(t+\bar{t})},$  &  4  &
     \multirow{2}{*}{\cite{CMS:2016xnv}}  \\
     &          &   &     & $
     d\sigma/d|y^{(t+\bar{t})}|$   & 4 &   \\
\midrule
\midrule
 \multirow{1}{*}{$tW$} &       \multirow{1}{*}{ {\tt ATLAS\_tW\_inc\_8TeV}}      &
 \multirow{1}{*}{{\bf 8 TeV}}   & \multirow{1}{*}{inclusive}   &
\multirow{1}{*}{$\sigma_{\rm tot}(tW)$}  &  1  &
\multirow{1}{*}{\cite{Aad:2015eto}}  \\
\midrule
 \multirow{1}{*}{$tW$} &       \multirow{1}{*}{ {\tt CMS\_tW\_inc\_8TeV}}      &
 \multirow{1}{*}{{\bf 8 TeV}}   & \multirow{1}{*}{inclusive}   &
\multirow{1}{*}{$\sigma_{\rm tot}(tW)$}  &  1  &
\multirow{1}{*}{\cite{Chatrchyan:2014tua}}  \\
\midrule
 \multirow{1}{*}{$tW$} &       \multirow{1}{*}{ {\tt ATLAS\_tW\_inc\_13TeV}}      &
 \multirow{1}{*}{{\bf 13 TeV}}   & \multirow{1}{*}{inclusive}   &
\multirow{1}{*}{$\sigma_{\rm tot}(tW)$}  &  1  &
\multirow{1}{*}{\cite{Aaboud:2016lpj}}  \\
\midrule
 \multirow{1}{*}{$tW$} &       \multirow{1}{*}{ {\tt CMS\_tW\_inc\_13TeV}}      &
 \multirow{1}{*}{{\bf 13 TeV}}   & \multirow{1}{*}{inclusive}   &
\multirow{1}{*}{$\sigma_{\rm tot}(tW)$}  &  1  &
\multirow{1}{*}{\cite{Sirunyan:2018lcp}}  \\
\midrule
 \multirow{1}{*}{$tZ$} &       \multirow{1}{*}{ {\tt CMS\_tZ\_inc\_13TeV}}      &
 \multirow{1}{*}{{\bf 13 TeV}}   & \multirow{1}{*}{inclusive}   &
\multirow{1}{*}{$\sigma_{\rm fid}(Wbl^+l^-q)$}  &  1  &
\multirow{1}{*}{\cite{Sirunyan:2017nbr}}  \\
\midrule
 \multirow{1}{*}{$tZ$} &       \multirow{1}{*}{ {\tt ATLAS\_tZ\_inc\_13TeV}}      &
 \multirow{1}{*}{{\bf 13 TeV}}   & \multirow{1}{*}{inclusive}   &
\multirow{1}{*}{$\sigma_{\rm tot}(tZq)$}  &  1  &
\multirow{1}{*}{\cite{Aaboud:2017ylb}}  \\
\bottomrule
  \end{tabular}
  \caption{\small Same as Table~\ref{eq:input_datasets},
    now for inclusive single $t$ production and single $t$ production in association with
    vector bosons.
     \label{eq:input_datasets2}
  }
\end{table}


From these diagrams, one can see that measurements of inclusive top quark pair
production will be particularly sensitive to SMEFT operators that induce or
modify interactions of the form $gt\bar{t}$ and $ggt\bar{t}$, such as the
chromomagnetic operator $c_{tG}$.
In this case, the interference with the most relevant SM production mechanism
will dominate over the small quark-antiquark-initiated contributions.
Likewise, single-top production and associated $tW$ and $tZ$ production will
constrain SMEFT operators that involve both top quarks and electroweak gauge
bosons, such as $c_{tW}$.
As a third example, $t\bar{t}b\bar{b}$ production should provide direct
information on operators involving four heavy quarks, such as
$c^1_{QQ}$ and $c_{QQ}^8$.

Following this overview of our input dataset, we move to describe
in more detail the features of the individual measurements
listed in Tables~\ref{eq:input_datasets},~\ref{eq:input_datasets3}
and~\ref{eq:input_datasets2}.

\subsection{Top quark pair production}

We begin by presenting the LHC datasets of top quark pair production used in
this work.
We consider inclusive production first, and then $t\bar{t}$ production in
association with heavy quarks, with an electroweak gauge boson, and with
the Higgs boson.

\paragraph{Inclusive top-quark pair production.}
At the LHC, the dominant mechanism for the production of top quarks is
through the production of $t\bar{t}$ pairs.
The inclusive $t\bar{t}$ process is dominated
by the gluon-gluon initial state, with a small admixture of the
quark-antiquark partonic luminosity~\cite{Czakon:2013tha}.
In this analysis, we will limit ourselves to parton-level distributions
constructed in terms of the kinematical variables of the top and anti-top
quark, for which NNLO QCD corrections are available 
in the SM~\cite{Czakon:2016dgf}.
See~\cite{Gao:2017goi} for recent progress
in higher order calculations at the particle level for decayed top quarks,
in terms of leptons and $b$-jets.
For all the inclusive $t\bar{t}$ processes computed here,
the SM prediction is computed up to NNLO in the QCD coupling.
Theoretical predictions are obtained at NLO with 
{\tt Sherpa}~\cite{Gleisberg:2008ta}, for 8~TeV measurements, and with
{\tt MCFM}~\cite{Boughezal:2016wmq}, for 13~TeV measurements, and are then 
supplemented with the NNLO QCD $K$-factors computed in 
Ref.~\cite{Czakon:2017dip}.

In the present analysis we include the ATLAS and CMS differential
distributions from $t\bar{t}$ production at $\sqrt{s}=8$ TeV in the
lepton+jets final state~\cite{Aad:2015mbv,Khachatryan:2015oqa}.
These measurements are those used in the study of~\cite{Czakon:2016olj}
to constrain the large-$x$ gluon PDF from the $t\bar{t}$ differential
cross-sections, and are part of the NNPDF3.1 input dataset~\cite{Ball:2017nwa}.
In both cases, the distributions in top quark transverse momentum
and rapidity, $p_T^t$ and $y_t$, as well as in top-quark pair
invariant mass and rapidity, $m_{t\bar{t}}$ and $y_{t\bar{t}}$ are available,
both as absolute cross-sections and normalised to the inclusive results;
only the former are used here.
As discussed in~\cite{Czakon:2016olj}, to avoid double counting only
one distribution per experiment can be added to the fit, as long as
correlations between different distributions are not available or neglected.

Besides these two datasets, we take into account the constraints from
the double-differential distributions from CMS at 8 TeV, which provide
a good handle on the underlying partonic kinematics~\cite{Sirunyan:2017azo}.
Note that this dataset is based on the dilepton final state, therefore
it does not overlap with the dataset used in~\cite{Khachatryan:2015oqa},
which instead is based on the lepton+jets channel.
We also include the CMS differential distributions at $\sqrt{s}=13$ TeV in the
lepton+jets~\cite{Khachatryan:2016mnb} and dilepton~\cite{Sirunyan:2017mzl}
final states based on an integrated luminosity of $\mathcal{L}=2.3$ fb$^{-1}$,
as well as the more recent measurements in the lepton+jet channel
based on $\mathcal{L}=35.8$ fb$^{-1}$~\cite{Sirunyan:2018wem}.
A measurement based on the same dataset but with the dilepton
final state was presented in~\cite{Sirunyan:2018ucr}.
Double-differential distributions from CMS at 13 TeV~\cite{Khachatryan:2016mnb}
are excluded since they overlap with the single-inclusive distributions
from the lepton+jets datasets.

We do not include ATLAS measurements at 13 TeV since the published
differential cross-sections at 13 TeV in the lepton+jets~\cite{Aaboud:2017fha}
and dilepton~\cite{Aaboud:2016syx} channels are provided at the particle level.
In this work, we restrict ourselves to parton-level observables.
Note that in principle ATLAS measurements at 13 TeV are also available
for the fully hadronic final state in the highly boosted
regime~\cite{Aaboud:2018eqg}.
These measurements are not considered here since their analysis requires
jet substructure information alongside the consistent inclusion of
electroweak~\cite{Czakon:2017wor} and threshold
resummation~\cite{Czakon:2018nun} corrections.

\paragraph{Helicity fractions and spin correlations in $t\bar{t}$ production.}
A further window on the underlying dynamics of top quark pair production
is provided by the measurement of observables sensitive to the spin
structure of top quark production and decay.
Among them, polarisation, $W$ helicity fractions, and spin correlations
provide direct constraints on the structure of the $tWb$ vertex.
In this work, we include the helicity fractions $F_{L}$, $F_0$, and $F_R$
of the $W$ bosons from the decay of top quarks measured by
ATLAS~\cite{Aaboud:2016hsq} and CMS~\cite{Khachatryan:2016fky}
at $\sqrt{s}=8$ TeV.
These measurements supersede the previous ATLAS and CMS combined analysis
at $\sqrt{s}=7$ TeV~\cite{ATLAS-CONF-2013-033}.

Related types of available angular observables in $t\bar{t}$ production
include the polarisation asymmetry $A_{P^\pm}$, the spin correlations
variable $A_{\Delta\phi}$, and the $A_{c_1c_2}$ and $A_{\cos\phi}$ asymmetries,
which discriminate between the correlated and uncorrelated $t$ and
$\bar{t}$ spins.
For example, CMS has presented measurements of $t\bar{t}$ spin correlations
and top quark polarisation in the lepton+jet and dilepton final states at
$\sqrt{s}=8$ TeV~\cite{Khachatryan:2016xws}.
Measurements of the $t\bar{t}$ spin correlations at $\sqrt{s}=13$ TeV
in the $e\mu$ final state are also available from ATLAS, specifically
the differential cross-section in the angular separation between the two
leptons.
This measurement deviates from the SM predictions by more than
three sigma~\cite{ATLAS-CONF-2018-027}.
We leave the inclusion of these observables in the SMEFT global fit to
future work.

\paragraph{$t\bar{t}V$ production.}
In this analysis we also include data for the production of a $t\bar{t}$ pair
in association with either a $Z$ or a $W$ boson, which is directly sensitive
to the top quark couplings with the gauge bosons
(see Fig.~\ref{fig:top-production-LHC}).
Specifically, we include the measurements of the total inclusive cross-sections
for $t\bar{t}Z$ and $t\bar{t}W$ production at $\sqrt{s}=8$ TeV
and $\sqrt{s}=13$ TeV from ATLAS~\cite{Aad:2015eua,Aaboud:2016xve}
and CMS~\cite{Khachatryan:2015sha,Sirunyan:2017uzs}.
Note that, for $t\bar{t}W$, the $W$ boson is often emitted from initial-state
light quarks, however, when it is emitted from a final-state leg, it becomes
sensitive to operators involving only one heavy quark, which is a unique
feature of this process.
We do not include the $t\bar{t}\gamma$ production
measurements~\cite{Aaboud:2017era,Aad:2015uwa,Sirunyan:2017iyh},
whose interpretation is hampered by issues related to photon isolation
and fragmentation, as well as to initial- and final-state radiation.
Because of electroweak symmetry, the $t\bar{t}V$ process
is closely related to the $t\bar{t}H$ one, to be discussed next.

\paragraph{Higgs production in association with a $t\bar{t}$ pair.}
The production of a top-antitop pair together with a Higgs boson
allows for a direct probe of the Yukawa coupling of the top quark,
as illustrated by the dominant mechanism indicated in
Fig.~\ref{fig:top-production-LHC}.
Recently, $5\sigma$ evidence for this production mode was presented
by both the ATLAS and CMS collaborations~\cite{Sirunyan:2018hoz,Aaboud:2018urx}.
In the CMS case~\cite{Sirunyan:2018hoz}, we utilise their measurement of the
signal strength $\mu_{t\bar{t}h}$ at $\sqrt{s}=13$ TeV (normalised to the SM
prediction), rather than the cross-section, because the latter is obtained
by combining data at different centre-of-mass energies.
In the ATLAS case~\cite{Aaboud:2018urx}, we utilise their measurement of the
total cross-section for $t\bar{t}h$ production at $\sqrt{s}=13$,
extrapolated to the full phase space.

\paragraph{$t\bar{t}b\bar{b}$ and $t\bar{t}t\bar{t}$ production.}
The production of a top quark pair in association with a bottom-antibottom
pair is a purely QCD process, where a $b\bar{b}$ pair is radiated either
from a gluon emitted from the initial state or from the final state
(see Fig.~\ref{fig:top-production-LHC}).
The production of four top quarks at the LHC, $t\bar{t}t\bar{t}$, obeys
a similar underlying mechanism in the SM, with the cross-section now being
rather smaller due to the heavier top quark mass.
The relevance of this process on the top quark sector of the
SMEFT has been discussed in Ref.~\cite{DHondt:2018cww}.

Concerning $t\bar{t}b\bar{b}$, the total cross-section for $t\bar{t}b\bar{b}$,
production, extrapolated to the full phase space at $\sqrt{13}$ TeV, is
available from CMS~\cite{Sirunyan:2017snr}, together with the corresponding
ratio to $t\bar{t}jj$ production.
This single data point is included in our fit.
Differential cross-sections for $t\bar{t}b\bar{b}$ production at
$\sqrt{s}=8$ TeV have also been presented as a function of
the kinematics of the $b$-jets~\cite{Khachatryan:2015mva}.
These measurements, however, are not included in the fit.
While there is a priori no reason why these measurements
  of $t\bar{t}$ production in association
  with jets cannot be used in a SMEFT fit~\cite{Englert:2018byk},
  this would require to interface the parton level calculation to a
  Monte Carlo shower program and possibly merge samples of different jet multiplicity in order to appropriately model the extra hadronic radiation.
  In this work, we have chosen to focus on inclusive observables, and we have deferred more exclusive observables to future work.

The ATLAS collaboration has presented the results of the measurements
of top quark pair production in association with multiple $b$-jets at
$\sqrt{s}=13$ TeV.
Fiducial cross-sections for $tt$ in association with more than two $b$ quarks, 
both in the lepton+jets
and in the dilepton channels, are provided~\cite{ATLAS-CONF-2018-029},
which supersede a previous measurement at 8 TeV~\cite{Aad:2015yja}.
We do not include these results in the fit as the cuts needed to simulate the
cross-sections are not fully provided, and are therefore not reproducible.

Concerning $t\bar{t}t\bar{t}$, a first measurement
of its cross-section at $\sqrt{s}=13$ TeV has been presented by
CMS~\cite{Sirunyan:2017roi}, albeit with a statistical significance of only
1.6$\sigma$.
This measurement supersedes previous upper bounds at
8 TeV~\cite{Khachatryan:2014sca} and 13 TeV~\cite{Sirunyan:2017tep}.
In the case of ATLAS, upper bounds based on the 2015 dataset
at 13 TeV were presented in~\cite{ATLAS-CONF-2016-020} and
then updated in~\cite{Aaboud:2018jsj} from the 2016 dataset.
In this analysis we utilise the CMS cross-section measurement
of~\cite{Sirunyan:2017roi}.

\subsection{Single top quark production}

We turn to discuss single top quark production, first
inclusively in either the $t-$ or the $s-$channel, and then
in association with an electroweak gauge boson.

\paragraph{Inclusive single top quark production.}
As highlighted in Table~\ref{eq:summaryOperators}, some of the SMEFT $d=6$
operators that contribute to single top production via interference with the
SM amplitudes are different from the corresponding ones in top quark pair
production, whence their relevance in a global fit.
There exist three main modes to produce single top
quarks~\cite{Giammanco:2015bxk}: by means of the exchange of a $W^\pm$ boson,
either in the $t$-channel or in the $s$-channel, and by means of the associated
production with a $W^\pm$ ($Z$) boson that leads to the $q't$, $t\bar{b}$,
and $tW^\pm$ ($tZq$) Born-level final states.
Representative diagrams for these three modes are shown in
Fig.~\ref{fig:top-production-LHC}.

In this work, we include all relevant single top production datasets in the
$t-$ and $s-$channels from ATLAS and CMS at 8 and 13 TeV, see
Table~\ref{eq:input_datasets2}.
We restrict ourselves to parton-level measurements, that is, to un-decayed
top quarks.

From ATLAS, we include the differential cross-sections at
$\sqrt{s}=8$ TeV~\cite{Aaboud:2016ymp}, specifically the $d\sigma(tq)/dp_T^t$
and $d\sigma(tq)/dy_T^{t}$ distributions, as well as
the corresponding measurements for anti-top quarks.
From CMS, we include the inclusive cross-sections for $t$ and for $\bar{t}$
production at $\sqrt{s}=8$ TeV~\cite{Khachatryan:2014iya},
as well as the corresponding differential distributions
in $p_T^{(t+\bar{t})}$ and $|y_{t+\bar{t}}|$~\cite{CMS-PAS-TOP-14-004}.
In the case of the inclusive measurements, the ratio
$R_t=\sigma(tq)/\sigma(\bar{t}q)$ is also provided, the use of which would be
advantageous if the knowledge on correlations were lacking, due to the partial
cancellation of experimental and theoretical systematic uncertainties
between the numerator and the denominator.

We now move to single top $t$-channel based on the Run II dataset at
$\sqrt{s}=13$ TeV.
We include the transverse momentum $p^{t+\bar{t}}_T$ and rapidity $|y_{t+\bar{t}}|$
differential distributions for single top production from
CMS~\cite{CMS:2016xnv}, the ATLAS and CMS measurements of the total
inclusive cross-sections for single $t$ and $\bar{t}$
production~\cite{Aaboud:2016ymp,Sirunyan:2016cdg}.
The ratio $R_t$ is once more provided in both cases.

Concerning single top $s-$channel measurements, we include the CMS total
cross-sections in the $s-$channel at 8 TeV~\cite{Khachatryan:2016ewo}.
We also include the total cross-sections at 8 TeV from
ATLAS~\cite{Aad:2015upn}.
No measurements of $s$-channel single top production at 13 TeV are
available from either experiment.
Neither the ATLAS nor the CMS differential distributions are provided with a
full breakdown of experimental systematic uncertainties.
Therefore, we sum all statistical and systematic uncertainties in quadrature.
To avoid double counting, we do not include total cross-sections if the
corresponding absolute differential distributions are already part of
the input dataset.
For example, using the labelling of Table~\ref{eq:input_datasets2}, if the
{\tt CMS\_t\_tch\_13TeV\_dif} distributions are used, then the associated
{\tt CMS\_t\_tch\_13TeV\_inc} total cross-sections are excluded from the fit.

For both inclusive and differential single-top measurements in the $t$-channel, 
NNLO QCD corrections have been computed~\cite{Berger:2016oht}. 
For all single-top processes for which the measurements
have been published (all measurements except the CMS differential measurements),
we use as a theory input the NNLO calculation.
For unpublished measurements, we use NLO QCD, as the NNLO results 
are not available. 
For the calculation of single top and single top associated channels at NLO, 
we use a $n_f=5$ scheme for consistency for both the SM and the SMEFT calculations.
Given that we are using NLO calculations for all single top and single top 
  associated production processes, we expect the uncertainties related to the choice 
  of number flavour scheme to be under control.

\paragraph{$tV$ associated production.}
The associated production of a top quark and a $W$ boson has a very
distinctive signature that allows one to reconstruct the decay products
from both the top quark and the $W$ decay.
Measurements of $tW$ associated production have been presented by ATLAS and CMS.
Here we include the ATLAS measurements at 8 and 13 TeV of the total
$\sigma(tW)$ cross-section~\cite{Aad:2015eto,Aaboud:2016lpj} extrapolated
to the full phase space.
A measurement of differential distributions at 13 TeV based
on a luminosity of $\mathcal{L}=36$ fb$^{-1}$~\cite{Aaboud:2017qyi}
was also presented by ATLAS.
However, this measurement is at particle level (of leptons and $b$-jets
from the $W$ and top quark decays), therefore we do not include it in the fit.
We also include the CMS measurements of $\sigma_{\rm tot}(tW)$ at 8 and 13
TeV~\cite{Chatrchyan:2014tua,Sirunyan:2018lcp}.
These measurements supersede the previous ones
at 7 and 8 TeV~\cite{Chatrchyan:2014tua,Chatrchyan:2012zca}.

For $tW$ production both the SM and SMEFT part is for 
consistency computed with the DR1 scheme \cite{Frixione:2008yi}. 
We refer the reader to Ref. \cite{Demartin:2016axk} for a detailed 
discussion of the differences between the various diagram subtraction/removal
 schemes. At the inclusive level the difference between the schemes is 
 about 10\% but this is greatly reduced with fiducial cuts. This uncertainty
  is comparable to the scale and PDF uncertainties and we do not
   expect it to significantly alter our findings.

The associated production of a single top quark in association with a
$Z$ boson, shown in Fig.~\ref{fig:top-production-LHC},
is also an interesting probe of the top quark sector of the SMEFT\@.
The $tZ$ production cross-section has been measured by CMS at 13
TeV in the $Wbl^+l^-q$ final state, where the dilepton pair arise
from the decay of the $Z$ boson~\cite{Sirunyan:2017nbr} (see~\cite{CMSstop77} 
for an update based on $\mathcal{L}=77.4$ fb$^{-1}$).
The $tZ$ production cross-section has been measured at 13 TeV
by ATLAS in the tri-lepton final state and extrapolated to the full
phase space~\cite{Aaboud:2017ylb}.
We use these two measurements as data points in the fit, for a total
of four $tV$ input cross-sections.

\subsection{Theory overview and sensitivity to the SMEFT degrees of freedom}
\label{sec:overview}

In Table~\ref{eq:table-processes-theory}, we summarise the details of the
theoretical calculations used for the description of the LHC top quark
production measurements included in the present analysis.
We indicate, for both the SM and the SMEFT contributions to the cross-sections
in Eq.~(\ref{eq:smeftXsecInt}), the perturbative accuracy and the codes used
to produce the corresponding predictions.
In all cases, the same theoretical settings have been used for the calculation
of both the total cross-sections and the differential distributions, where
available.
We emphasise that we have used state-of-the-art theory calculations for
both the SM and the SMEFT pieces, which are instrumental to reduce
the theoretical uncertainties associated to the missing perturbative
higher orders.
We have adopted common input settings for the theory calculations,
in particular all the SM cross-sections are consistently evaluated
with a NNPDF3.1 PDF set at NNLO accuracy that does not include any top
data (henceforth labelled NNPDF3.1NNLO no-top). 
We do not include top quark data in the PDF fit as the datasets
used in NNPDF are also used in the current SMEFT fit. 
Adding $t\bar{t}$ distributions into both fits would imply to 
double-counting the data, otherwise, as MC replicas are also used in NNPDF it 
would require us to keep track of correlations between the two sets of MC
replicas. 
We therefore choose to exclude the top quark data from the PDF fit.

\begin{table}[t]
  \centering
  \footnotesize
  \renewcommand{\arraystretch}{1.30}
  \begin{tabular}{c|c|c|c|c}
   Process 
  & SM   
  & Code  
  & SMEFT 
  & Code 
  \\
  \toprule
    \multirow{2}{*}{$t\bar{t}$}  
  & \multirow{2}{*}{NNLO QCD}   
  & {\tt MCFM/SHERPA} NLO  
  & \multirow{2}{*}{NLO QCD}  
  & \multirow{2}{*}{{\tt MG5\_aMC}} 
  \\
  &    
  & + NNLO $K$-factors  
  &    
  &  
  \\
  \midrule
    \multirow{2}{*}{single-$t$ ($t$-ch)} 
  & \multirow{2}{*}{NNLO QCD} 
  & {\tt MCFM} NLO  
  & \multirow{2}{*}{NLO QCD}  
  & \multirow{2}{*}{{\tt MG5\_aMC}} 
  \\
  &   
  & + NNLO $K$-factors  
  &   
  & 
  \\
  \midrule
    single-$t$ ($s$-ch) 
  & NLO QCD 
  & {\tt MCFM}   
  & NLO QCD  
  & {\tt MG5\_aMC} 
  \\
  \midrule
    $tW$ 
  & NLO QCD 
  & {\tt MG5\_aMC}  
  & NLO QCD  
  & {\tt MG5\_aMC} 
  \\
  \midrule
    \multirow{2}{*}{$tZ$} 
  & \multirow{2}{*}{NLO QCD} 
  & \multirow{2}{*}{{\tt MG5\_aMC}}  
  & LO QCD   
  & \multirow{2}{*}{{\tt MG5\_aMC}} 
  \\
  &  
  &   
  & + NLO SM $K$-factors  
  & 
  \\
  \midrule
    \multirow{2}{*}{ $t\bar{t}W(Z)$} 
  & \multirow{2}{*}{NLO QCD} 
  & \multirow{2}{*}{{\tt MG5\_aMC}}  
  & LO QCD   
  & \multirow{2}{*}{{\tt MG5\_aMC}} 
  \\
  &  
  &   
  & + NLO SM $K$-factors  
  & 
  \\
  \midrule
    \multirow{2}{*}{$t\bar{t}h$ } 
  & \multirow{2}{*}{NLO QCD} 
  & \multirow{2}{*}{{\tt MG5\_aMC}}  
  & LO QCD   
  & \multirow{2}{*}{{\tt MG5\_aMC}} 
  \\
  &  
  &   
  & + NLO SM $K$-factors  
  &
  \\
  \midrule
    \multirow{2}{*}{$t\bar{t}t\bar{t}$} 
  & \multirow{2}{*}{NLO QCD} 
  & \multirow{2}{*}{{\tt MG5\_aMC}}  
  & LO QCD  
  & \multirow{2}{*}{{\tt MG5\_aMC}} 
  \\
  &  
  &   
  & + NLO SM $K$-factors  
  &
  \\
  \midrule
    \multirow{2}{*}{$t\bar{t}b\bar{b}$} 
  & \multirow{2}{*}{NLO QCD} 
  & \multirow{2}{*}{{\tt MG5\_aMC}}  
  & LO QCD   
  & \multirow{2}{*}{{\tt MG5\_aMC}} 
  \\
  &  
  &   
  & + NLO SM $K$-factors  
  &
  \\
  \bottomrule
 \end{tabular}
 \caption{\small Summary of the theoretical calculations used for the 
   description of the LHC top production cross-sections included in the 
   present analysis. We indicate, for both the SM and the SMEFT contributions 
   to the cross-sections, the perturbative accuracy and the codes used to 
   produce the corresponding predictions.}
  \label{eq:table-processes-theory}
\end{table}


It should be clear from the above discussion, as well as from the
considerations presented in Sect.~\ref{sec:smefttop}, that each of the input
LHC processes will have a rather different sensitivity to each of the
$N_{\rm op}=34$ SMEFT degrees of freedom considered in the analysis.
To illustrate this point, in Table~\ref{table:operatorprocess} we indicate the
sensitivity of each of the LHC processes included in the present analysis
along with the degrees of freedom in our fitting basis
(for their definition, see Table~\ref{eq:summaryOperators}).
A check mark outside (inside) brackets indicates that a given
process constrains the corresponding operator
$\mathcal{O}(\Lambda^{-2})$ ($\mathcal{O}( \Lambda^{-4})$). 
A check mark in square brackets indicates that the 
operator enters at $\mathcal{O}(\Lambda^{-2})$ but only at NLO.

\begin{table}[p]
 \centering
 \footnotesize
 \renewcommand{\arraystretch}{1.40}
 \begin{tabular}{c|c|c|c|c|c|c|c|c|c}
   \multirow{1}{*}{Notation} 
 & \multicolumn{9}{c|}{Sensitivity at 
   $\mathcal{O}(\Lambda^{-2})$ ($\mathcal{O}( \Lambda^{-4})$)} 
 \\ 
 \cline{2-10}
 & $t\bar{t}$ 
 & single-top 
 & $tW$ 
 & $tZ$ 
 & $t\bar{t}W$ 
 & $t\bar{t}Z$ 
 & $t\bar{t}H$ 
 & $t\bar{t}t\bar{t}$ 
 & $t\bar{t}b\bar{b}$ 
 \\
 \toprule
   {\tt OQQ1} 
 & 		
 &	 
 &	 
 &
 &		 
 & 
 & 
 & \checkmark 
 & \checkmark
 \\
   {\tt OQQ8} 
 & 		
 &	 
 &	 
 &
 &		 
 & 
 & 
 & \checkmark
 & \checkmark
 \\
   {\tt OQt1} 
 & 		
 &	 
 &	 
 &
 &		 
 & 
 & 
 & \checkmark 
 & \checkmark
 \\
   {\tt OQt8} 
 & 		
 &	 
 &	 
 &
 &		 
 & 
 & 
 & \checkmark 
 & \checkmark
 \\
   {\tt OQb1} 
 & 		
 &	 
 &	 
 &
 &		 
 & 
 & 
 &  
 & \checkmark
 \\
   {\tt OQb8} 
 & 		
 &	 
 &	 
 &
 &		 
 & 
 & 
 & 
 & \checkmark
 \\
   {\tt Ott1} 
 & 		
 &	 
 &	 
 &
 &		 
 & 
 & 
 & \checkmark
 &  
 \\
   {\tt Otb1} 
 & 		
 &	 
 &	 
 &
 &		 
 & 
 & 
 &  
 & \checkmark
 \\
   {\tt Otb8} 
 & 		
 &	 
 &	 
 &
 &		 
 & 
 & 
 &  
 & \checkmark
 \\
 {\tt OQtQb1}
  & 		
 &	 
 &	 
 &
 &		 
 & 
 & 
 &
  & (\checkmark)
 \\
 {\tt OQtQb8} 
  & 		
 &	 
 &	 
 &
 &		 
 & 
 & 
 &
  &(\checkmark)
 \\ 
 \midrule
   {\tt  O81qq } 
 & \checkmark 
 & 	
 & 
 & 
 & \checkmark 
 & \checkmark 
 & \checkmark
 & \checkmark
 & \checkmark 
 \\
   {\tt O11qq }  
 & [\checkmark]
 &     
 & 
 & 
 & [\checkmark] 
 & [\checkmark]
 & [\checkmark] 
 & \checkmark 
 & \checkmark 
 \\
   {\tt O83qq } 
 & \checkmark
 & [\checkmark]
 & 
 & [\checkmark] 
 & \checkmark 
 & \checkmark 
 & \checkmark 
 & \checkmark
 & \checkmark 
 \\
   {\tt  O13qq } 
 & [\checkmark]
 & \checkmark	 
 & 
 & \checkmark
 & [\checkmark] 
 & [\checkmark]
 & [\checkmark]
 & \checkmark 
 & \checkmark
 \\
   {\tt O8qt } 
 & \checkmark
 & 		
 & 
 & 
 & \checkmark
 & \checkmark
 & \checkmark
 & \checkmark 
 & \checkmark
 \\
   {\tt O1qt } 
 & [\checkmark]
 & 		
 & 
 & 
 & [\checkmark]
 & [\checkmark]
 & [\checkmark]
 & \checkmark 
 & \checkmark
 \\
   {\tt O8ut }  
 & \checkmark 
 &		 
 & 
 & 
 & 
 & \checkmark
 & \checkmark 
 & \checkmark 
 & \checkmark
 \\
   {\tt O1ut } 
 & [\checkmark]
 & 		
 & 
 & 
 & 
 & [\checkmark]
 & [\checkmark] 
 & \checkmark 
 & \checkmark
 \\
   {\tt O8qu } 
 & \checkmark 
 & 		
 &
 & 
 & 
 & \checkmark
 & \checkmark
 & \checkmark
 & \checkmark 
 \\
   {\tt O1qu } 
 & [\checkmark]
 & 		
 & 
 & 
 & 
 & [\checkmark]
 & [\checkmark] 
 & \checkmark&\checkmark
 \\
   {\tt O8dt }
 & \checkmark
 & 		
 & 
 & 
 & 
 & \checkmark
 & \checkmark
 & \checkmark
 & \checkmark
 \\
   {\tt O1dt } 
 & [\checkmark]
 & 		
 & 
 & 
 & 
 & [\checkmark] 
 & [\checkmark]
 & \checkmark 
 & \checkmark
 \\
   {\tt O8qd }
 & \checkmark
 & 		
 & 
 & 
 & 
 & \checkmark
 & \checkmark
 & \checkmark 
 & \checkmark
 \\
   {\tt O1qd} 
 & [\checkmark]
 & 		
 &
 &
 &
 & [\checkmark]
 & [\checkmark] 
 & \checkmark&\checkmark\\
 \midrule
   {\tt OtG }    
 & \checkmark 
 & 	
 & \checkmark
 & 
 & \checkmark
 & \checkmark
 & \checkmark
 & \checkmark 
 & \checkmark
 \\
   {\tt  OtW} 
 & 
 & \checkmark 
 & \checkmark
 & \checkmark
 & 
 & 
 & 
 & 
 &
 \\
   {\tt  ObW} 
 & 		
 & (\checkmark)  
 & (\checkmark)  
 & (\checkmark)
 & 
 & 
 &  
 & 
 &
 \\
   {\tt OtZ} 
 & 		
 & 
 & 
 & \checkmark
 & 
 & \checkmark
 & 
 & 
 & 
 \\
   {\tt  Off} 
 & 		
 & (\checkmark) 
 & (\checkmark) 
 & (\checkmark)  
 & 
 &
 & 
 & 
 &
 \\
   {\tt  Ofq3} 
 & 
 & \checkmark
 & \checkmark
 & \checkmark
 & 
 &  
 &  
 & 
 &
 \\
   {\tt OpQM} 
 & 		
 & 
 & 
 & \checkmark 
 & 
 & \checkmark
 & 
 & 
 & 
 \\ 
   {\tt Opt} 
 & 		
 &	 
 &	 
 & \checkmark
 &		 
 & \checkmark 
 & 
 & 
 &
 \\
   {\tt Otp}     
 & 
 & 
 & 
 & 
 &  
 & 
 & \checkmark 
 & 
 &
 \\
 \bottomrule
 \end{tabular}
 \caption{\small The sensitivity of each of the LHC processes included
    in the present analysis with each of the $N_{\rm obs}=34$ dimension-six
    SMEFT operators that constitute our fitting
    basis (see Table~\ref{eq:summaryOperators} for their definition).
    A check mark outside (inside) brackets indicates that a given
    process constrain the corresponding operator $\mathcal{O}( \Lambda^{-2})$ 
    ($\mathcal{O}( \Lambda^{-4})$), and operators in square brackets indicate
    that an operator contributes at $\mathcal{O}( \Lambda^{-2})$ but only at NLO, see text for details.
 }
  \label{table:operatorprocess}	
\end{table}


The comparison in Table~\ref{table:operatorprocess} illustrates the importance
of a global approach to the SMEFT analysis of top quark production.
On the one hand, several operators are constrained by many different
processes, and this allows independent and complementary constraints.
For instance, the chromomagnetic operator $c_{tG}$ is relevant for the
description of all the input processes with the exception of single-top
production.
On the other hand, other operators are constrained by one
or two processes at most, so that information on them can be obtained
only by including a wide range of different input observables.
For instance, $\ccc{}{t\varphi}{}$, constrained only by $t\bar{t}H$
production; $\ccc{}{bW}{}$, sensitive only to single-top production
at $\mathcal{O}\lp \Lambda^{-4}\rp$; and the four-heavy-quark operators
for which the only available information is from $t\bar{t}b\bar{b}$ and
$t\bar{t}t\bar{t}$.

From Table~\ref{table:operatorprocess} we also observe that
adding the formally subleading $\mathcal{O}( \Lambda^{-4})$
contributions from the dimension-six operators increases
the sensitivity of many different processes.
For example, the {\tt ObW} and {\tt Off} operators can only be constrained
once $\mathcal{O}( \Lambda^{-4})$ terms are included in the fit.
This is also true for several of the four-fermion operators, for which 
additional constraints can be obtained from the $t\bar{t}$, $t\bar{t}V$ and
$t\bar{t}H$ production processes once $\mathcal{O}(\Lambda^{-4})$
corrections are taken into account.

Needless to say, it is in principle inconsistent to account only for
the $\mathcal{O}( \Lambda^{-4})$ effects arising from the dimension-six
operators and not from the dimension-eight operators.
However, there are good reasons why it could be worth including
them in the analysis.
First, including or not the $\mathcal{O}( \Lambda^{-4})$ corrections provides
an estimate of whether the SMEFT fit results are stable upon higher orders in the effective
field theory parameter expansion.
Second, one might consider scenarios where the dimension-eight operators do not
interfere with the SM amplitudes.
In this case the only physically relevant $\mathcal{O}(\Lambda^{-4})$ effects
are those arising from the dimension-six operators, which we include here.

\section{The SMEFiT fitting methodology}
\label{sec:fitsettings}

In this section, we describe the SMEFiT fitting approach that we adopt here to 
constrain the SMEFT operators summarised in Table~\ref{eq:summaryOperators}.
First, we explain how the MC replica method can be used 
to construct the probability density in the space of the fitted SMEFT 
coefficients $\{c_i\}$. 
Then, we motivate the choice of the figure of merit 
used for the minimisation, discuss the propagation of experimental
and theoretical uncertainties, and explain how we determine the best-fit
parameters in a way that avoids over-fitting.
Finally, we describe how the fitting methodology can be
validated by means of closure tests,
analogously to the PDF case, and apply this strategy to study the robustness of
the results and their dependence with a number of fit settings.

\subsection{The Monte Carlo replica method}
\label{sec:mcmethod}

In this work, we adopt the MC replica method to propagate the 
experimental uncertainties from the input experimental cross-sections
to the fitted SMEFT coefficients $\{c_i\}$.
The idea underlying this method is to construct a sampling of the probability
distribution in the space of the experimental data, which then translates 
into a sampling of the probability distribution in the space of the SMEFT 
coefficients by means of the fitting procedure.
This strategy can be implemented by generating a large number ($N_{\rm rep}$) 
of artificial replicas of the original data.
The replica generation is based on the available information on the 
experimental central values, uncertainties, and correlations associated to 
each of the input data points.
It can then be shown that averages, variances, and correlations
computed over the sample of $N_{\rm rep}$ MC replicas reproduce 
the corresponding experimental values.

In practice, the MC replica method works as follows.
Given an experimental measurement of a hard-scattering
cross-section, denoted by $\mathcal{O}_i^{\rm (exp)}$, with
total uncorrelated uncertainty $\sigma_{i}^{\rm (stat)}$, $N_{\rm sys}$ 
correlated systematic uncertainties $\sigma^{\rm (sys)}_{i,\alpha}$, and
$N_{\rm norm}$ normalisation uncertainties $\sigma^{\rm (norm)}_{i,n}$ (such as 
those associated to the luminosity), the artificial replicas are generated as
\be
\label{eq:replicas}
\mathcal{O}_{i}^{(\art)(k)}
=
S_{i,N}^{(k)} 
\mathcal{O}_{i}^{\rm (\mrexp)}\lp 1
+
r_{i}^{(k)}\sigma_{i}^{\rm (stat)}
+
\sum_{\alpha=1}^{N_{\rm sys}}r_{i,\alpha}^{(k)}\sigma^{\rm (sys)}_{i,\alpha}\rp
\ , \quad k=1,\ldots,N_{\rep} \ , 
\ee
where the index $i$ runs from 1 to $N_{\rm dat}$, the total
number of points in a specific dataset, and
where the normalisation prefactor is given by
\be
\label{eq:totalnorm}
S_{i,N}^{(k)}\equiv \prod_{n=1}^{N_{\rm norm}}\lp
1+r_{i,n}^{(k)}\sigma^{\rm (norm)}_{i,n}\rp \, .
\ee 
In Eqns.~(\ref{eq:replicas}) and~(\ref{eq:totalnorm}), 
$r_{i}^{(k)}$, $r_{i,\alpha}^{(k)}$, and $r_{i,n}^{(k)}$
are univariate Gaussian random numbers.
Correlations between data points induced by systematic uncertainties 
are accounted for by ensuring that $r^{(k)}_{i,\alpha}=r^{(k)}_{i',\alpha}$.
A similar condition is applied for
multiplicative normalisation uncertainties if the $n$-th
normalisation uncertainty is common to the entire dataset, 
i.e.~$r^{(k)}_{i,n}=r^{(k)}_{i',n}$.

The MC approach is conceptually different from the commonly adopted Hessian 
method, based on the expansion of the $\chi^2$ around its 
best-fit minimum assuming a quadratic behaviour.
Nevertheless, under specific conditions, the two methods
can be shown to reproduce equivalent results for the determination 
of the uncertainties in fitted parameters, see {\it e.g.} 
Ref.~\cite{Gao:2017yyd} for studies in the PDF context.
The main advantage of the MC method is that it does not require any 
assumption about the underlying probability distribution of the parameters, 
and in particular it is not restricted to Gaussian distributions.
Moreover, it is suited to problems where the parameter space is large
and complicated, with a large number of quasi-degenerate minima and flat
directions.
For these reasons, adopting the MC approach rather than the Hessian method is 
rather advantageous in the case of SMEFT fits.

\begin{figure}[t]
  \begin{center}
\includegraphics[width=0.65\linewidth]{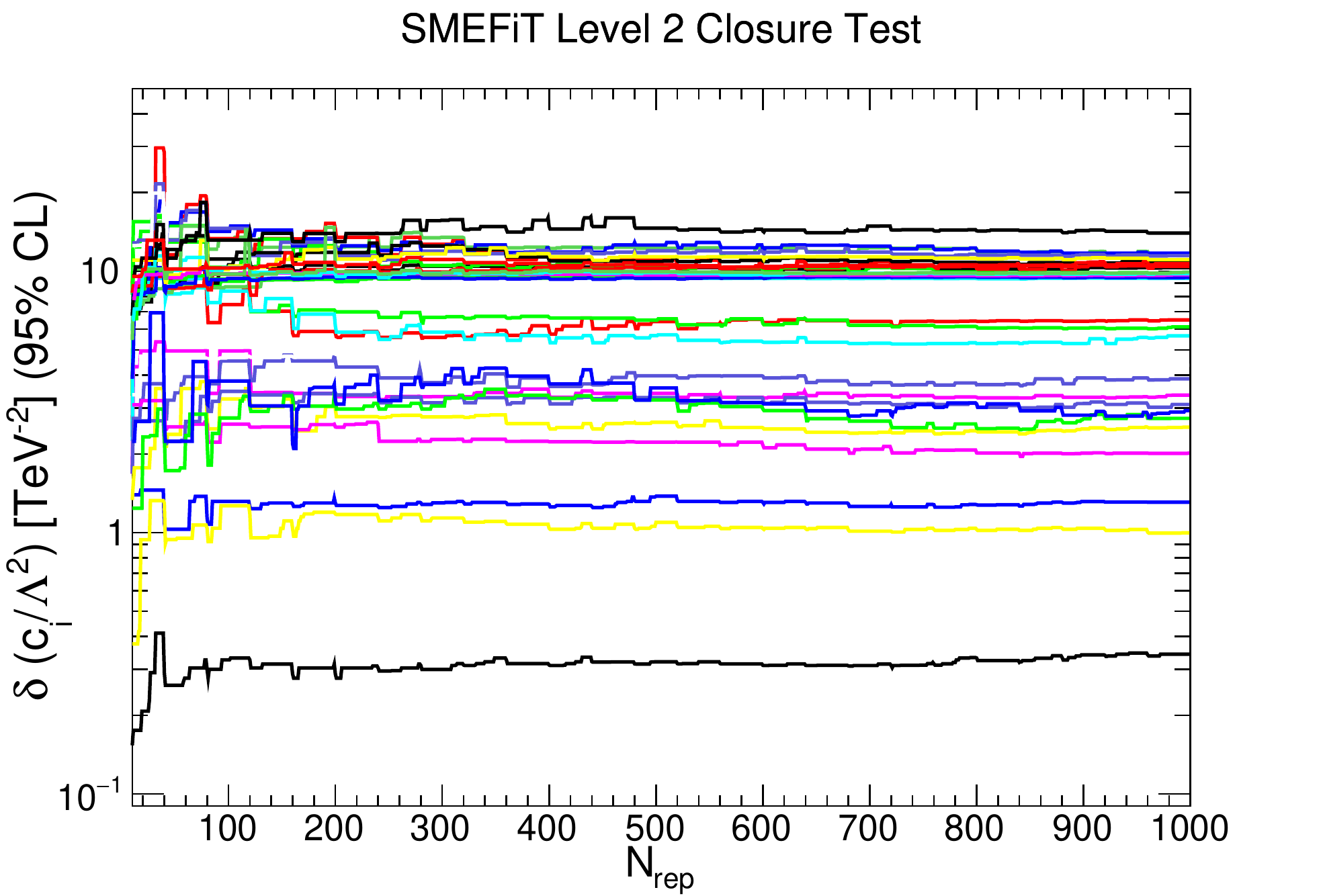}
\caption{\small The dependence of the 95\% CL bounds
  on the fit parameters $\delta c_i/\Lambda^2$
   with the
  number of MC replicas $N_{\rm rep}$ used in the fit,
  as determined in a level 2 closure test.
  Each of the lines shown in the plot correspond to one of the
  $N_{\rm rep}=34$ degrees of freedom defined in
  Table~\ref{eq:summaryOperators}. 
     \label{fig:dependence_nrep_68cl} }
  \end{center}
\end{figure}

An important aspect to address in the MC method
is how many replicas $N_{\rm rep}$ need to be generated for each
specific application.
In order to determine this, we assess the robustness of our results with 
respect to the number of MC replicas used in the fit.
To do so, in Fig.~\ref{fig:dependence_nrep_68cl} we show the dependence of the 
bounds $\delta c_i/\Lambda^2$, determined at the 95\% confidence
level, on the value of $N_{\rm rep}$ from a level 2 closure test, 
discussed in detail in Sect.~\ref{sec:closuretest}.
Each line corresponds to one of the $N_{\rm op}=34$ degrees of freedom 
defined in Table~\ref{eq:summaryOperators}.
From Fig.~\ref{fig:dependence_nrep_68cl}, we find that for 
$N_{\rm rep}\lsim 100$ the fit estimate for the bounds is affected by large 
fluctuations. These fluctuations are dampened as the number of replicas 
increases, and for $N_{\rm rep}\gsim 500$ the results become independent
from $N_{\rm rep}$.
In order to ensure that no residual MC fluctuations remain, we will use 
$N_{\rm rep}= 1000$ as our baseline.
We note however that the validity of this conclusion, in general, 
will depend on the input dataset, 
and should therefore be reconsidered if this is modified, in particular
if the dataset is significantly extended.

\subsection{$\chi^2$ definition}

For each of the MC replicas generated with  Eq.~\eqref{eq:replicas}, we 
perform a fit to the $N_{\rm op}=34$ degrees of freedom $c_l/\Lambda^2$,
$l=1,\dots , N_{\rm op}$, associated to the dimension-6 SMEFT operators 
defined in Sect.~\ref{sec:smefttop}.
This process results in a set of $\{ c_l^{(k)}\}$ best-fit values
for each replica, from which estimators such as expectation values, variances
and correlations can be readily evaluated.

For each MC replica, the corresponding best-fit values are determined from
the minimisation of a figure of merit, the error function, defined as
\begin{equation}
  E(\{c_l^{(k)}\})\equiv \frac{1}{N_{\rm dat}}\sum_{i,j=1}^{N_{\rm dat}}\lp 
  \mathcal{O}^{(\rm th)}_i\lp \{c_l^{(k)}\} \rp-\mathcal{O}^{{(\rm art)}(k)}_i\rp ({\rm cov}^{-1})_{ij}
  \lp \mathcal{O}^{(\rm th)}_j\lp \{c_l^{(k)}\} \rp-\mathcal{O}^{{(\rm art)}(k)}_j\rp
  \label{eq:chi2definition}
    \; ,
\end{equation}
where $N_{\rm dat}$ is the number of data points used in the fit,
and $\mathcal{O}^{(\rm th)}_i$ is the theoretical
prediction for the $i-$th cross-section evaluated using the $\{ c_l^{(k)}\}$  
values for the SMEFT degrees of freedom.
Note that in Eq.~(\ref{eq:chi2definition}) the theory predictions
are compared to the MC replicas, rather than to the original
experimental central values.
Once the best-fit parameters have been determined for all the $N_{\rm rep}$ 
replicas, the overall fit 
quality can be quantified by means of the $\chi^2$
\begin{equation}
  \chi^2 \equiv \frac{1}{N_{\rm dat}}\sum_{i,j=1}^{N_{\rm dat}}\lp 
  \mathcal{O}^{(\rm th)}_i\lp \{ \la c_l\ra \} \rp
  -\mathcal{O}^{(\rm exp)}_i\rp ({\rm cov}^{-1})_{ij}
\lp 
  \mathcal{O}^{(\rm th)}_j\lp \{ \la c_l\ra \} \rp
  -\mathcal{O}^{(\rm exp)}_j\rp
 \label{eq:chi2definition2}
    \; ,
\end{equation}
where now the theoretical predictions, computed using the expectation value 
(the mean) for the degree of freedom $c_l$, are compared to the central 
experimental data.
This is evaluated as the average over the resulting MC best-fit sample 
$\{c_l^{(k)}\}$ 
\be
\label{eq:meancoefficient}
\la c_l\ra \equiv \frac{1}{N_{\rm rep}}\sum_{k=1}^{N_{\rm rep}} c_l^{(k)}\, .
\ee

Both the error function, Eq.~(\ref{eq:chi2definition}), and the $\chi^2$, 
Eq.~(\ref{eq:chi2definition2}), are expressed in terms of the total covariance 
matrix, ${\rm cov}_{ij}$, which should contain all the relevant sources of
experimental and theoretical uncertainties.
Assuming that theoretical uncertainties follow an underlying Gaussian
distribution, and that they are uncorrelated
to the experimental uncertainties, it can be shown~\cite{Ball:2018odr} that 
the total covariance matrix can be expressed as
\be
\label{eq:covmatsplitting}
{\rm cov}_{ij} = {\rm cov}^{(\rm exp)}_{ij} + {\rm cov}^{(\rm th)}_{ij} \, ,
\ee
that is, as the sum of the experimental and theoretical covariance matrices.

Concerning the experimental covariance matrix,
we use the so-called `$t_0$' definition~\cite{Ball:2009qv}
\begin{eqnarray}
  ({\rm cov_{t_0}})^{(\rm exp)}_{ij}\equiv
  \lp \sigma^{\rm (stat)}_i\rp^2\delta_{ij}
  &+&
  \Bigg(\sum_{\alpha=1}^{N_{\rm sys}}\sigma_{i,\alpha}^{\rm (sys)}
  \sigma_{j,\alpha}^{\rm (sys)}\mathcal{O}_{i}^{(\rm exp)}\mathcal{O}_{j}^{(\rm exp)} \nonumber\\
  &+&\sum_{\beta=1}^{N_{\rm norm}}\sigma_{i,\beta}^{\rm (norm)}
  \sigma_{j,\beta}^{\rm (norm)}\mathcal{O}_{i}^{(\rm th,0)}\mathcal{O}_{j}^{(\rm th,0)}\Bigg)\; ,
  \label{eq:t0covmat}
\end{eqnarray}
where one treats the additive (`sys') relative experimental systematic errors 
separately from the multiplicative (`norm') ones.
In the additive case, one uses the central value of the experimental
measurement, $\mathcal{O}_i^{({\rm exp})}$.
In the multiplicative case, one uses instead a fixed set of theoretical 
predictions, $\{\mathcal{O}_{i}^{(\rm th,0)}\}$.
These theoretical predictions are typically obtained from a previous fit; the 
fit is then iterated until consistency is reached.
The use of the $t_0$ covariance matrix defined in Eq.~(\ref{eq:t0covmat})
avoids the bias associated to multiplicative uncertainties, which would lead 
to a systematic undershooting of the best-fit values as compared to their true 
values~\cite{dagos}.

As mentioned in Sect.~\ref{sec:settings_expdata}, we construct
the experimental covariance matrix, Eq.~(\ref{eq:t0covmat}), from all 
available sources of statistical and systematic uncertainties
for a given dataset.
Information on the bin-by-bin correlations of systematic uncertainties is 
available only for a subset of the data listed in 
Tables~\ref{eq:input_datasets}-\ref{eq:input_datasets3}, specifically, for 
all the 8 TeV top-quark pair differential distributions in 
Table~\ref{eq:input_datasets} and for the corresponding CMS distributions 
at 13 TeV from Ref.~\cite{Khachatryan:2016mnb}.
For all the other measurements, we add all uncertainties in quadrature; 
our analysis can be easily updated should more correlations become available.

In addition to the experimental uncertainties, there are at least two main 
classes of theoretical uncertainties that are in principle relevant
for the present fits: (i) uncertainties
associated to missing higher orders (MHOs) in the perturbative calculation,
and (ii) PDF uncertainties.
The impact of the former is not expected to affect this analysis significantly,
because we perform the SM calculation at the highest available perturbative 
order.
In particular, we take into account NNLO QCD corrections for the two families 
of processes that are more precisely known experimentally, namely the absolute 
differential distributions in inclusive $t\bar{t}$ and single top ($t$-channel) 
production.
Furthermore, as discussed in Sect.~\ref{sec:nlosmeft},
for most of the SMEFT contributions, the NLO QCD calculation is used.

The inclusion of PDF uncertainties, instead, is more important.
In this work, we use as input to all our theory calculations the NNPDF3.1 NNLO
no-top PDF set~\cite{Ball:2017nwa}, which differs from the NNPDF3.1 baseline 
set only for the 
exclusion of the top-quark pair production data from the dataset.
As explained in Sect.~\ref{sec:settings_expdata}, this is necessary to avoid
double-counting in the fit.
However, this implies that the SM calculation of top quark pair production 
could be affected by sizeable PDF uncertainties, especially in the tails of the
differential distributions, which are not constrained by alternative 
gluon-sensitive processes in the fit such as transverse momentum 
$Z$-boson~\cite{Boughezal:2017nla}, jet~\cite{Rojo:2014kta}, and 
direct photon production~\cite{Campbell:2018wfu}.
Therefore, not accounting for PDF uncertainties
may bias the results of the fit.

With this motivation, we construct the theoretical
covariance matrix from the contributions of 
the PDF uncertainty as
\be
\label{eq:PDFcovmat}
{\rm cov}^{(\rm th)}_{ij} = \la 
\mathcal{O}^{(\rm th)(r)}_i \mathcal{O}^{(\rm th)(r)}_j
\ra_{\rm rep} -
\la 
\mathcal{O}^{(\rm th)(r)}_i \ra_{\rm rep} \la\mathcal{O}^{(\rm th)(r)}_j
\ra_{\rm rep} \, ,
\ee
where the theoretical predictions $\mathcal{O}^{(\rm th)(r)}_i$
are computed using the SM theory and the $r$-th replica
from the NNPDF3.1NNLO no-top PDF set, and averages 
$\langle \cdot\rangle_{\rm rep}$ are performed over the 
$\widetilde{N}_{\rm rep}=100$ replicas of this PDF set.
Note that replicas in the PDF set are not directly related to replicas
in the SMEFT set, since the two sets represent different 
probability distributions.

In general, the theoretical covariance matrix, Eq.~(\ref{eq:PDFcovmat}),
induces correlations between all the datasets included in the fit.
However we account for them only within a given dataset, in the same way as 
for experimental measurements.
If the PDF-induced correlations between data points $i$
and $j$ are neglected, Eq.~(\ref{eq:PDFcovmat}) reduces to
\be
\label{eq:PDFcovmat2}
{\rm cov}^{(\rm th)}_{ii} = \la \lp
\mathcal{O}^{(\rm th)(k)}_i \rp^2
\ra_{\rm rep} -
\la 
\mathcal{O}^{(\rm th)(k)}_i \ra_{\rm rep}^2  \, ,
\ee
and vanishes for $i\ne j$.
This corresponds to adding the PDF error in quadrature to the experimental
uncertainties.

For consistency, PDF uncertainties should be included in the fit not only via 
the covariance matrix in 
Eqs.~(\ref{eq:chi2definition})-(\ref{eq:chi2definition2}), 
but also in the MC replica generation.
That is, the generation of the data replicas according to 
Eq.~(\ref{eq:replicas}) includes an additional source of fluctuation determined 
from the theoretical covariance matrix, Eq.~(\ref{eq:PDFcovmat}).
Note that, for the $k$-th data replica, the theory predictions 
$\mathcal{O}^{(\rm th)}_i( \{c_l^{(k)}\})$ are evaluated using a different 
PDF replica from the NNPDF3.1NNLO no-top set.
Since in general the number of data replicas, $N_{\rm rep}=1000$, is much
larger than the number of PDF replicas, $\widetilde{N}_{\rm rep}=100$,
the latter are selected at random with repetition for each data replica.

\subsection{Minimisation and stopping}
\label{sec:stopping}

In the case of current SMEFT fits, the minimisation of the error function, 
$E$, Eq.~\eqref{eq:chi2definition}, may be achieved 
by exploiting gradient descent methods, which rely on variations of $E$. 
This is because the relationship between the theory cross-sections and the
fitted parameters is at most quadratic, 
see Eq.~(\ref{eq:smeftXsecInt}).
Taking this into account, the optimiser that we use here to 
determine the best-fit values of the degrees of freedom $\{c_i\}$ is the 
sequential least squares programming algorithm
{\tt SLSQP}~\cite{Kraft:1988aa} available in the {\tt SciPy} package.
It belongs to the family of sequential quadratic programming methods, which
are based on solving a sequence of optimisation subproblems, where each of them
optimises a quadratic model.
An advantage of using {\tt SLSQP} is that it allows one to provide the 
optimiser with any combination of constraints on the 
coefficients, including existing bounds,
a feature that might become useful for future studies.

Since the dimensionality of this parameter space is not that different from
the total number of input cross-sections ($N_{\rm dat}= 103$ points),
one needs to avoid over-fitting, i.e.~fitting the statistical fluctuations 
of the experimental data rather than the underlying physical law.
Such an effect is particularly dangerous in a situation like the current
one, where there are a large number of flat directions with several
parameters strongly (anti)-correlated.

To prevent the minimiser from over-fitting the data, we use
(MC) cross-validation.
For each replica, the data is randomly split with equal probability into two 
disjoint sets, known as the training and validation sets.
Only the data points in the training set are then used to compute the figure of 
merit being minimised, Eq.~(\ref{eq:chi2definition}), while the data points in 
the validation set are monitored alongside the fit.
The random assignment of the data points to the training or validation sets
is different for each MC replica.

\begin{figure}[!t]
\begin{center}
\includegraphics[scale=0.75]{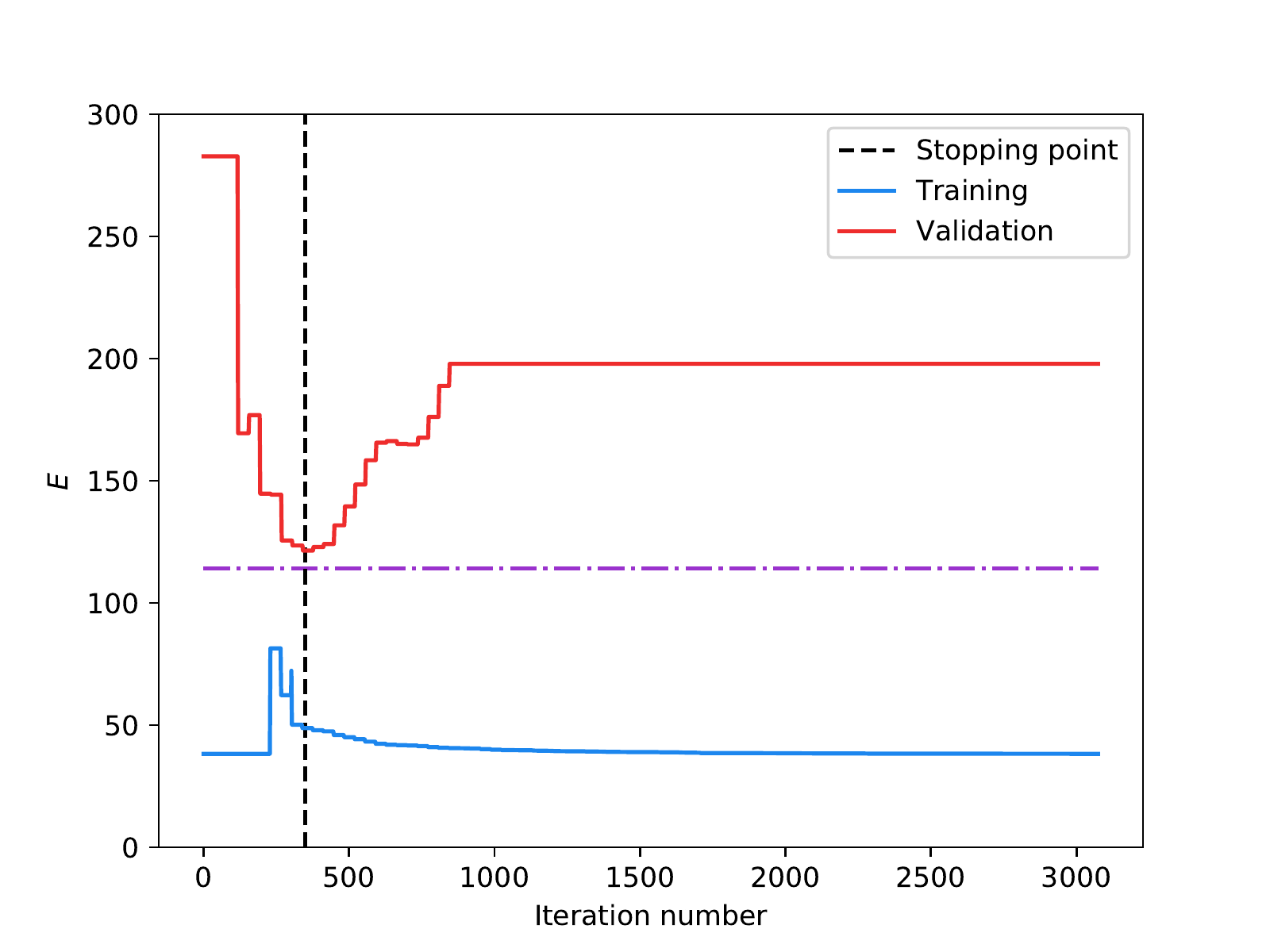}
\caption{\small The values of the training and validation error functions, 
   $E_{\rm tr}$ and $E_{\rm val}$ respectively, as a function of the number of 
   iterations of the minimisation algorithm for a representative data replica.
   The optimal stopping point, as determined by the cross-validation algorithm, 
   is indicated by the dashed vertical line.
   The horizontal dot-dashed line indicates the value 
   $E/N_{\rm dat}=2$, see text.
 \label{fig:trainvalsplit} }
\end{center}
\end{figure}

The optimal stopping point of 
the fit is reached when the figure of merit evaluated on the validation set, 
$E^{(k)}_{\rm val}$, starts to increase. This is not equivalent to the absolute 
minimum of the error function evaluated on the training set, $E_{\rm tr}^{(k)}$.
To illustrate how cross-validation works,
in Fig.~\ref{fig:trainvalsplit} we show the values of the training 
and validation error functions, $E_{\rm tr}$ and $E_{\rm val}$
respectively, as a function of the number of iterations of the minimisation 
algorithm for a representative data replica.
The optimal stopping point is 
indicated by the dashed vertical line.
For reference, the horizontal dot-dashed line indicates the value 
$E/N_{\rm dat}=2$, corresponding to the expectation value of the error function 
close to the optimal fit.
Note that $E \sim 1$ would indicate over-fitting as well, since we have two 
independent sources of fluctuations in the artificial data: the original data 
fluctuations, and the fluctuations induced by the MC replicas.

From this comparison, we can observe the expected behaviour for both the 
training and validation sets; namely that the validation $\chi^2$, once it 
reaches its lowest value, increases rapidly,
while the training $\chi^2$ continues to decrease.
We find $E_{\rm tr}\ll E_{\rm val}$ as the number of iterations increase, which is an indication that the optimisation algorithm 
is over-fitting.
It is therefore clear that, without adopting cross-validation, the 
absolute minimum found by the optimisation algorithm would not correspond to 
the true underlying law, but rather to fitting statistical noise.
We will quantify the importance of cross-validation in the SMEFT fit 
results in Sect.~\ref{sec:methtests}, where we will show that without it one 
obtains unreliable results, including spurious deviations from 
the SM predictions.

A peculiar behaviour in Fig.~\ref{fig:trainvalsplit} is
  that the cost function $E$ remains flat for the initial
  part of the minimisation.
This feature arises due to the fact that for some replicas the (randomly chosen) initial starting point of the fit may either turn out to be close enough to the Standard Model values, or to induce
    compensations between operators that effectively lead to a cost function $E$ 
    which is similar to the SM one.
    For those replicas for which the initial value of  $E$ is
    higher than the SM result, we observe the expected behaviour
    where the training cost function decreases rapidly at low iteration numbers.

\subsection{Closure test validation}
\label{sec:closuretest}

A reliable fitting framework should be able to fit a wide range of different 
datasets without tuning the methodology and without
biasing the results.
Validating a new methodology can be complicated by issues
such as potential inconsistencies (internal or external) in the experimental
data, or by limitations in the theoretical calculations.
To validate the fitting methodology used in this SMEFT analysis, we carry out a series of closure tests, 
based on pseudo-data generated with a known underlying physical law, 
see~\cite{Ball:2014uwa} for more details.

The basic idea underlying a closure test is to test the SMEFT fitting 
procedure by performing fits where the ``correct'' result is known, 
i.e.~by fitting pseudo-data generated from a fixed reference set 
of values for the SMEFT degrees of freedom, $\{c_i^{(\rm ref)}\}$.
Closure tests allow one to check that the fitting methodology 
can reproduce the underlying law, which is known by 
construction.
The SM and SMEFT theory calculations can be assumed to be exact, since we use 
the same theory settings to generate and fit the pseudo-data.
As a consequence, the theoretical uncertainties associated to MHOs and PDFs
do not enter closure tests, where only methodological and experimental
uncertainties are checked.
In the case of SMEFT fits, we can perform a closure test 
assuming that the underlying truth is the SM, 
i.e.~$\{c_i^{(\rm ref)}=0\}$, or any BSM scenario, i.e.~$\{c_i^{\rm (ref)}\ne 0\}$.
This allows for the validation of potential BSM anomalies 
identified in the SMEFT fit to the actual experimental data.

In the following, we consider three levels of closure tests according to 
the type of pseudo-data that is used as input to the fit.
\begin{itemize}

\item In a level zero (L0) closure test the pseudo-data
  coincides with the true underlying law, without any additional fluctuations.
  Then $N_\text{rep}$ fits are performed to exactly the same pseudo-data,
  with the only difference being the random initial conditions in each case.
  For instance, if the pseudo-data is generated with the SM hypothesis
  $\{c_i^{(\rm ref)}=0\}$, then the same values should be reproduced
  at the fit level within uncertainties.
  For a L0 closure test, the training/validation partition is not necessary, 
  since the information contained in both sets would be identical.

  In a L0 closure test, one expects the error function $E$ to tend to zero for 
  a large enough number of iterations.
  Therefore, direct evidence that a L0 test is successful is to show
  how the error function decreases with the number of iterations.
  A L0 closure test therefore allows one to check that the minimiser is 
  efficient enough to properly explore the entire parameter space.

\item In a level one (L1) closure test, one adds noise on top of the 
  pseudo-data. 
  Two types of noise may be added; in a L1a closure test we generate 
  MC replicas of the pseudo-data generated in a L0 closure test in the same 
  way as in a real fit to data.
  Alternatively, in a L1b closure test one adds stochastic noise directly to 
  the pseudo-data, in order to replicate experimental uncertainties included in 
  the fits to data. In this work we adopt a L1a-type closure test, 
  in contrast to NNPDF, where L1b-type closure tests are used as a default. 
  We note that adopting a L1a-type test over L1b simply means that a different 
  type of uncertainty is being probed at L1 - we discuss the various types of 
  uncertainty in Sec.~\ref{char_fit_unc} where we characterise the types of 
  fit uncertainties.

  In comparison to a L0 closure test, a L1 closure test propagates the 
  experimental uncertainties into the fitted coefficients, and can therefore 
  be used to demonstrate that the quoted uncertainties in the fit parameters 
  admit a robust statistical interpretation.
  One expects $E\sim 1$ for a successful closure test.

\item In a level two (L2) closure test, one adds the aforementioned stochastic 
  noise on top of the MC replicas included in the L1 closure test.
  This statistical noise is generated according to the experimental 
  covariance matrix of the real data.
  A L2 closure test is therefore equivalent to a fit to the real data, 
  the only difference being that data and theory are perfectly
  consistent by construction.
\end{itemize}

\begin{figure}[!t]
  \begin{center}
\includegraphics[width=0.7\linewidth]{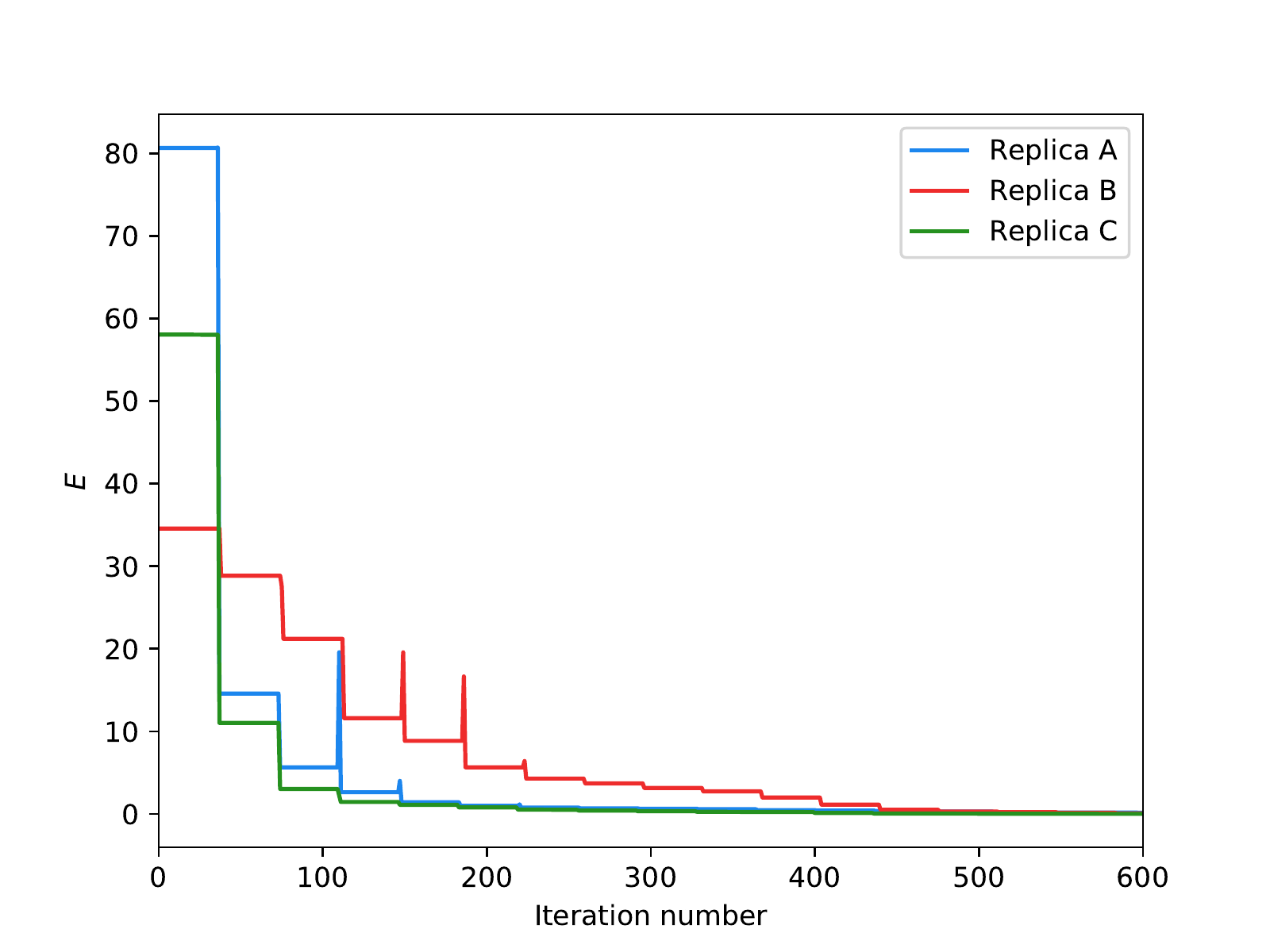}
\caption{\small The error function $E$ for L0 closure tests based on the SM 
  scenario as a function of the number of iterations.
  We show the results for three fits to the same pseudo-data but with
  different initial boundary conditions.
  \label{fig:chi2_LO} }
  \end{center}
\end{figure}

\paragraph{L0 closure tests.}
First of all, we want to demonstrate that the optimiser
is efficient enough to explore the full 34-dimensional parameter space.
With this motivation, pseudo-data corresponding to the SM has been
generated for all the cross-sections described in
Tables~\ref{eq:input_datasets}-\ref{eq:input_datasets2} and fitted
without introducing any additional noise.
As mentioned above, here all data is fitted since the training/validation 
separation is not required.
In Fig.~\ref{fig:chi2_LO} we show the error function $E$ for L0 closure tests based on the SM scenario as a function of 
the number of iterations in the minimiser for three replicas with
different initial boundary conditions.
We see how the error function decreases with the number of 
iterations, approaching the limit $E\to 0$ which corresponds to the case 
where the fit results reproduce the reference values $\{ c_i=0\}$.

\begin{figure}[t]
\begin{center}
\includegraphics[width=0.49\linewidth]{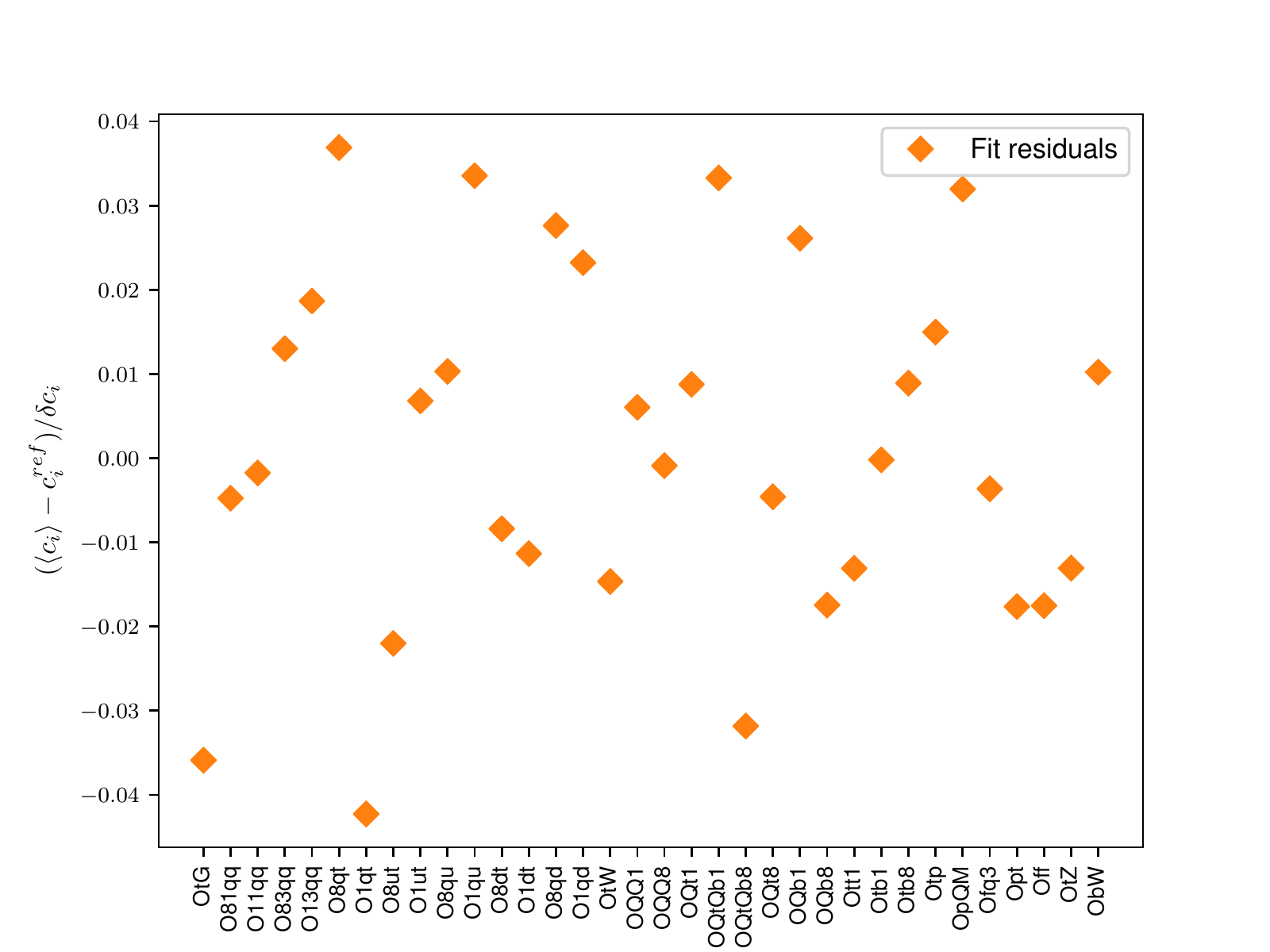}
\includegraphics[width=0.49\linewidth]{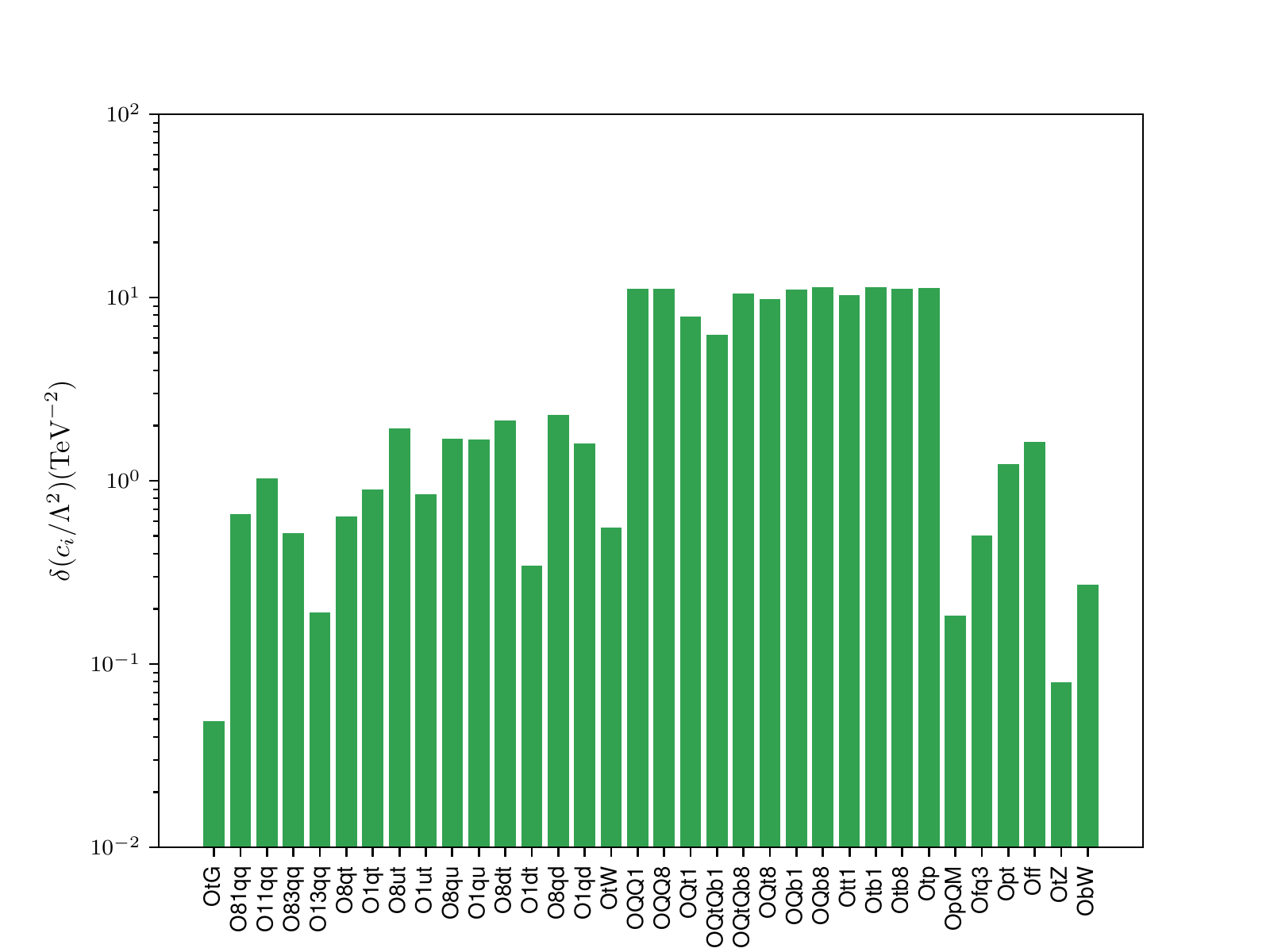}
\caption{\small Left: the fit residuals, defined as in 
  Eq.~(\ref{eq:fitresiduals}) for the 34 SMEFT degrees of freedom included 
  in the fit for the L0 closure test.
  Right: the corresponding values of the 95\% CL for the fit parameters 
  $\delta c_i$. Note that since in general the bounds on each degree of freedom 
  will be rather different, this comparison is performed on a log scale.
  \label{fig:closuretestsL0} }
\end{center}
\end{figure}

In Fig.~\ref{fig:closuretestsL0} we show the results of the L0 closure tests; 
in the left plot we show the fit residuals for the $N_{\rm op}=34$
degrees of freedom included in the fit.
They are defined as
\be
\label{eq:fitresiduals}
r_i \equiv \frac{\lp \la c_i\ra - c_i^{\rm (ref)}\rp}{\delta c_i} \, ,
\ee
where $\la c_i\ra $ and $\delta c_i$ indicate the expectation value,  
Eq.~(\ref{eq:meancoefficient}), and the 95\% CL.

As before, $c_i^{\rm (ref)}$ represent the reference values of
the SMEFT degrees of freedom used to generate the pseudo-data, which here are
set to zero.
We find that the residuals are all very close to zero, i.e.~the
optimiser has managed to identify with good accuracy the true
underlying values of the fit parameters.

In the same figure, we also show 
the corresponding values of the 95\% CL on the fit parameters $\delta c_i$.
The units of the $\delta c_i$ are TeV$^{-2}$, and as in the rest of this work
for reference we are assuming that $\Lambda=1$ TeV.
While the bounds $\delta c_i$ span
  up to two orders of magnitude, in all cases they will
  be smaller or comparable to those obtained from the
  corresponding Level 2 closure tests, to be discussed
  below.
  Indeed, as will be shown in Fig.~\ref{fig:CTlevel012compare},
  the L0 bounds can be an order of magnitude smaller (or even more)
  for specific degrees of freedom.

\begin{figure}[t]
  \begin{center}
  \includegraphics[width=0.49\linewidth]{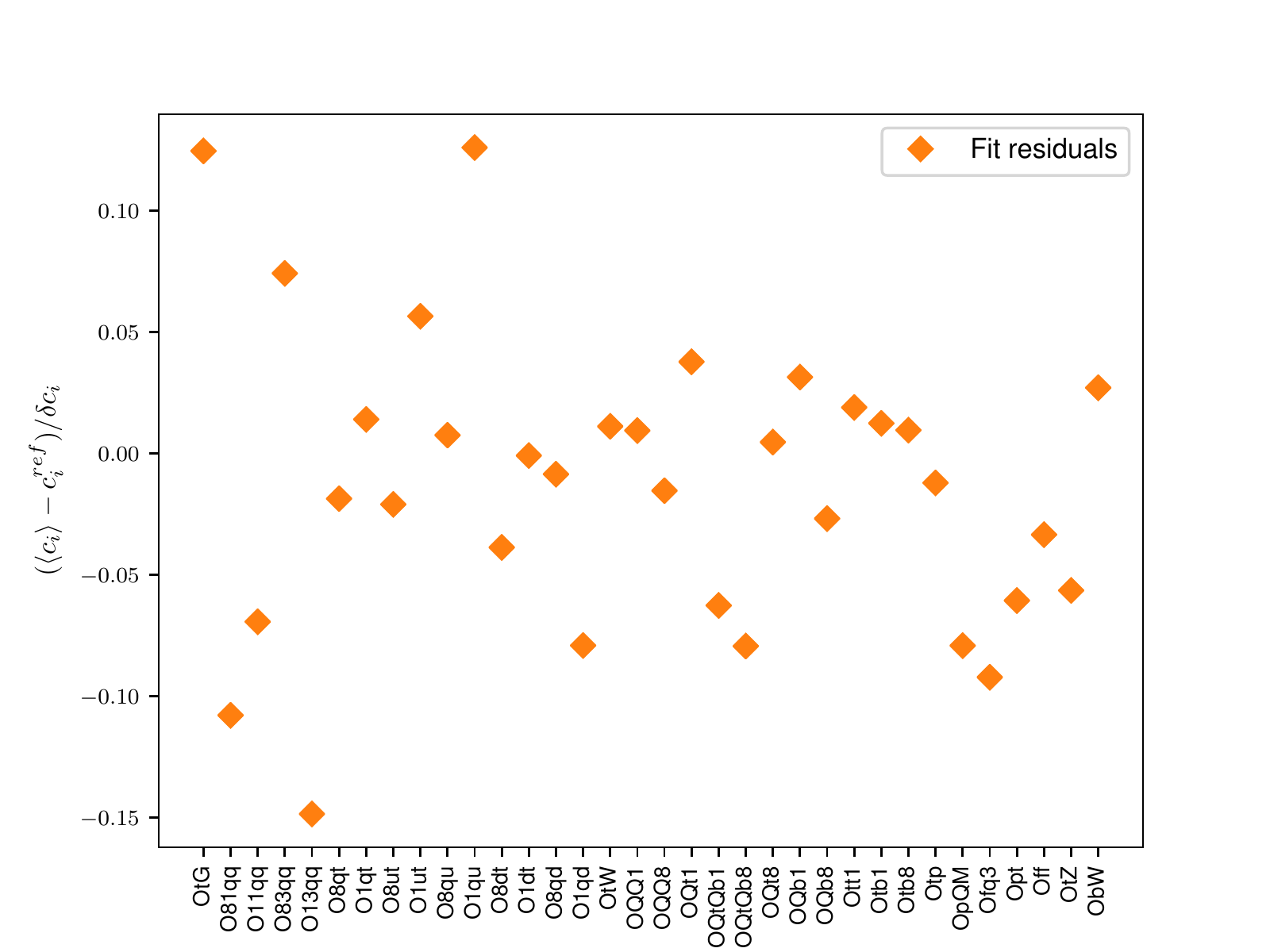}
\includegraphics[width=0.49\linewidth]{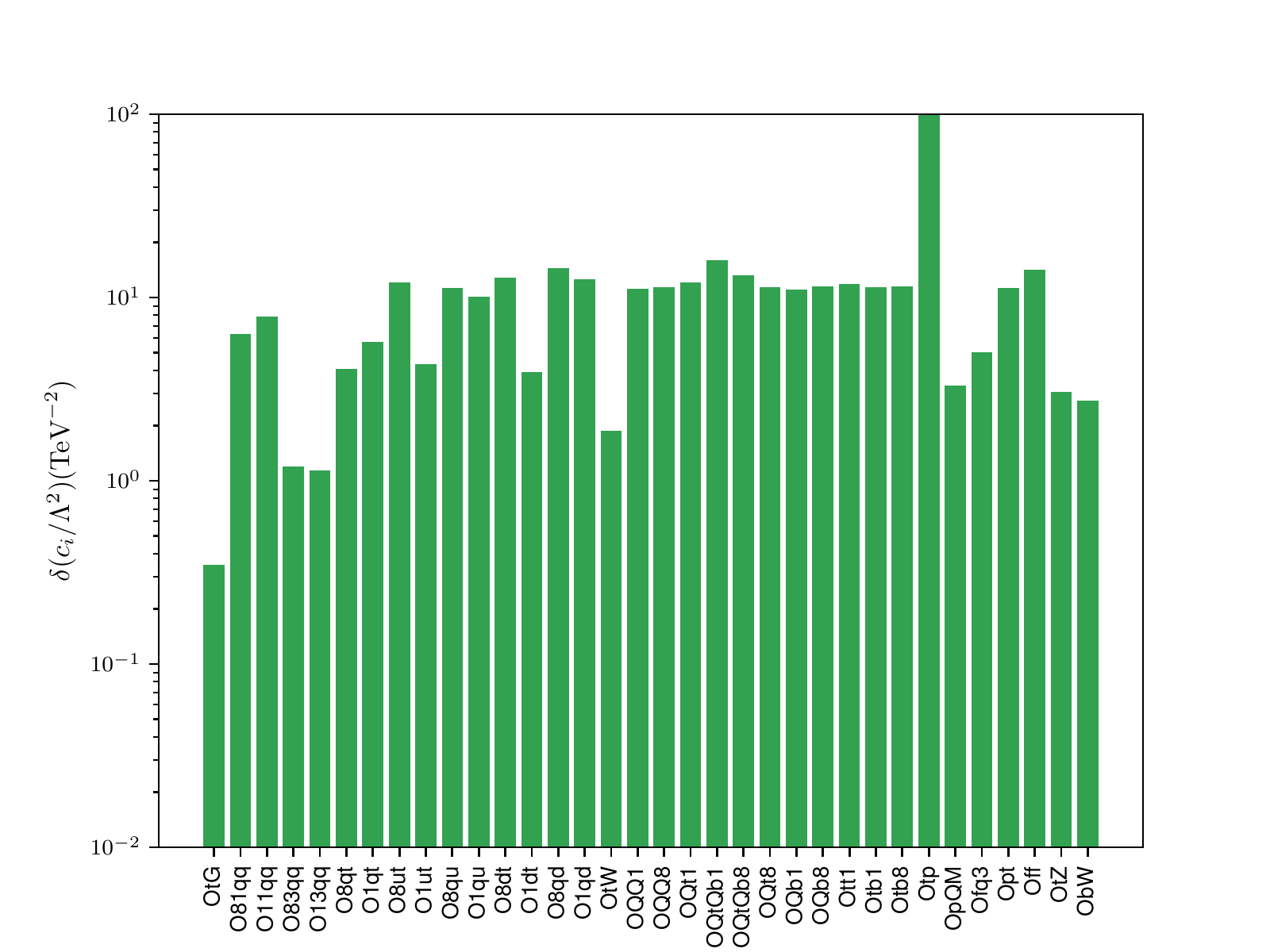}
\caption{\small Same as in Fig.~\ref{fig:closuretestsL0}, but now
  for the L2 closure tests.
       \label{fig:closuretestsL2} }
  \end{center}
\end{figure}

\paragraph{L2 closure tests.}
In this case we compare the results based on two different hypotheses 
for the underlying theory; one in which all degrees of freedom are set to their 
SM values; and another in which all degrees of freedom are set to their 
SM values, except $c^8 _{tu} /\Lambda^2=20$ TeV$^{-2}$.
The goal of this exercise is to verify if the fit can successfully identify 
a BSM deviation once it is built into the pseudo-data.
The results of the L2 closure test to the SM is shown in 
Fig.~\ref{fig:closuretestsL2};
we find that the fit manages to reproduce the underlying law,
since the residuals $r_i$ are very close to zero indicating that the
fit results $\la c_i\ra$ agree with the corresponding references values
$c_i^{(\rm ref)}$.

Another important output of the L2 closure tests is the size of the bounds
on the SMEFT degrees of freedom reported in Fig.~(\ref{fig:closuretestsL2}) 
(right).
These bounds reflect the constraints on the fit parameters that can be expected
based on the input experimental dataset in the case of perfect consistency 
between data and theory.
They provide a baseline to later compare the corresponding results at the
level of the fits to the real data.
We see from this comparison that some degrees of freedom will be constrained
rather better than others: for instance one expects the bounds on
{\tt OtG} to be in the range of $\delta c_i \simeq 0.1$ TeV$^{-2}$,
while the bounds on {\tt Otp} to be in the range 
$\delta c_i \simeq 100$ TeV$^{-2}$.

\begin{figure}[t]
  \begin{center}
    \includegraphics[width=0.49\linewidth]{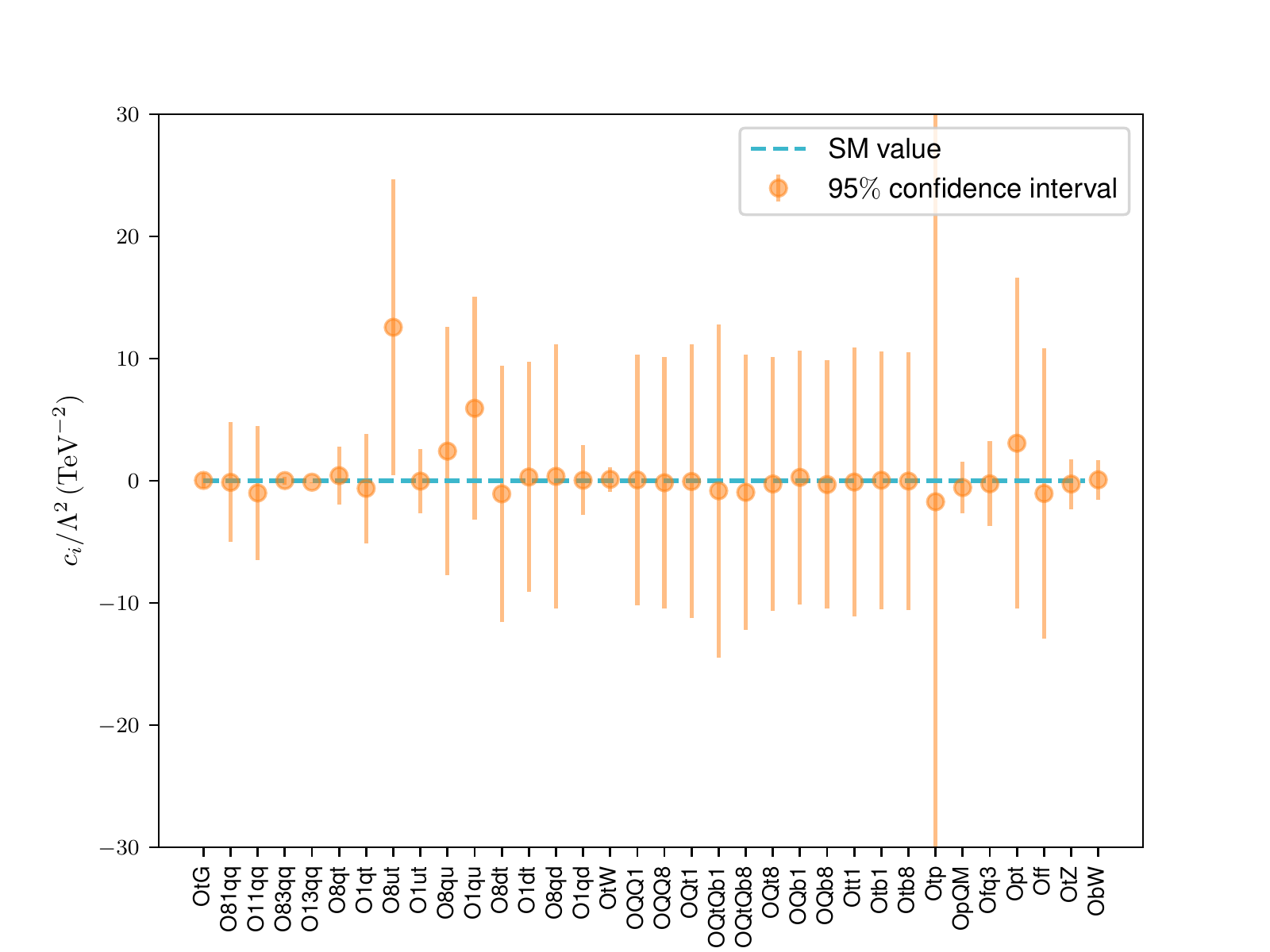}
        \includegraphics[width=0.49\linewidth]{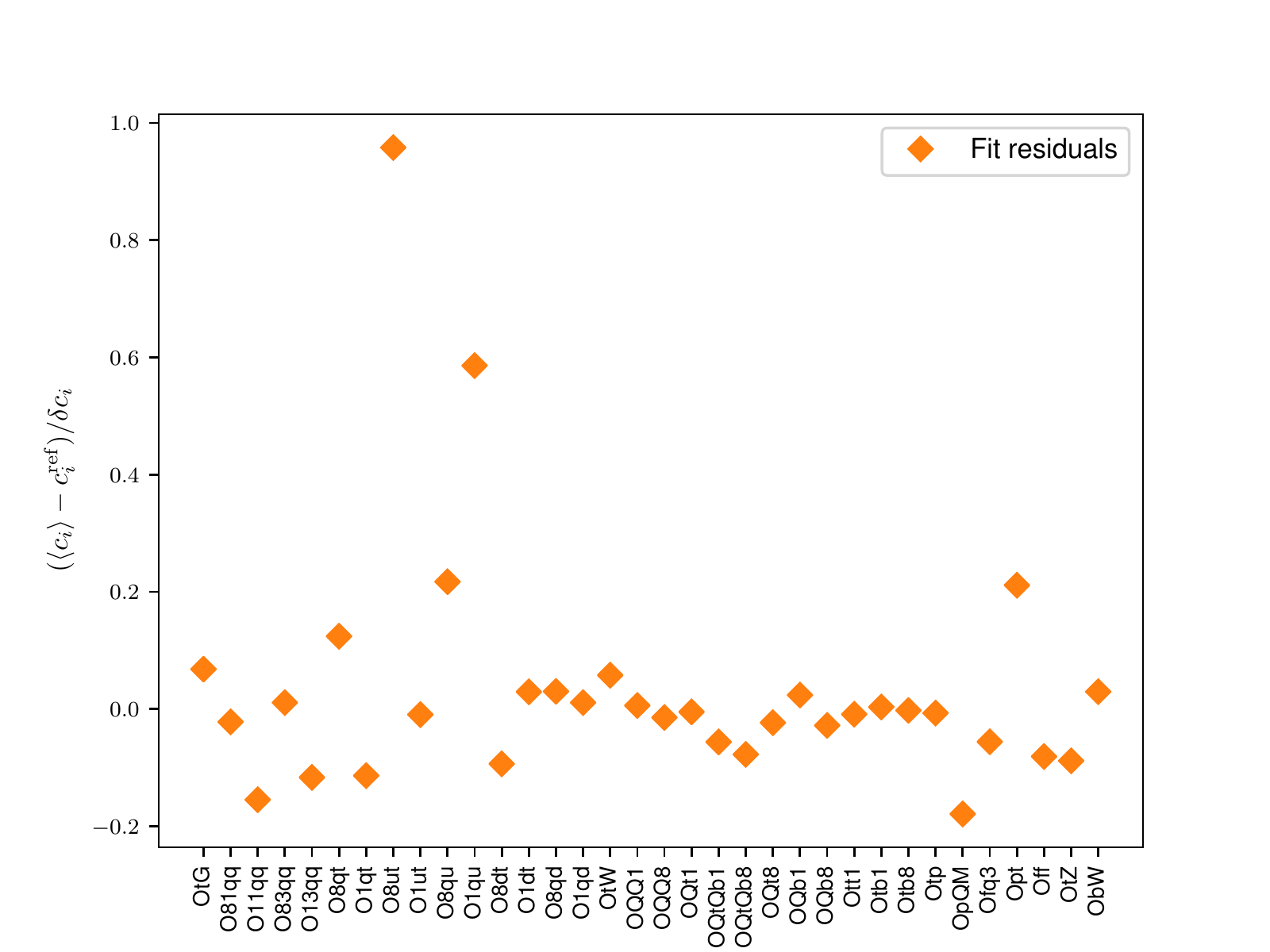}

    \caption{\small Left: the values of the SMEFT degrees of freedom and their
      uncertainties, $\la c_i\ra \pm \delta c_i$, for the L2 closure test
      in the BSM scenario where one has set
      $c^8 _{tu} /\Lambda^2=20$ TeV$^{-2}$. Right: the corresponding fit residuals.
       \label{fig:ctBSM} }
  \end{center}
\end{figure}

In  Fig.~\ref{fig:ctBSM} we show
the values of the SMEFT degrees of freedom and their
uncertainties, $\la c_i\ra \pm \delta c_i$, for the L2 closure test
in the BSM scenario where one has set
$c^8 _{tu}/\Lambda^{2}=20$ TeV$^{-2}$, which is approximately twice as large as the 95\% CL found by the 
L2 closure test to the SM. This allows us to ensure the starting point of the fit is BSM for this operator, but is not so far away from the SM
as to make the closure test redundant. 
We firstly observe that the closure test does indeed find a best-fit value for $c^8 _{tu}/\Lambda^{2} \approx 13$ TeV$^{-2}$, which is outside the error for this operator reported in the SM closure test. 
If one computes the fit residuals, Eq.~\eqref{eq:fitresiduals}
we find that the central value lies outside the 95\% CL, which roughly corresponds to 2$\sigma$.
Therefore at the fit level we would expect to find at least a 2$\sigma$ deviation from the SM, and larger if the value of the coefficient is much larger than the size of the error associated to it.


\begin{figure}[t]
  \begin{center}
    \includegraphics[width=0.7\linewidth]{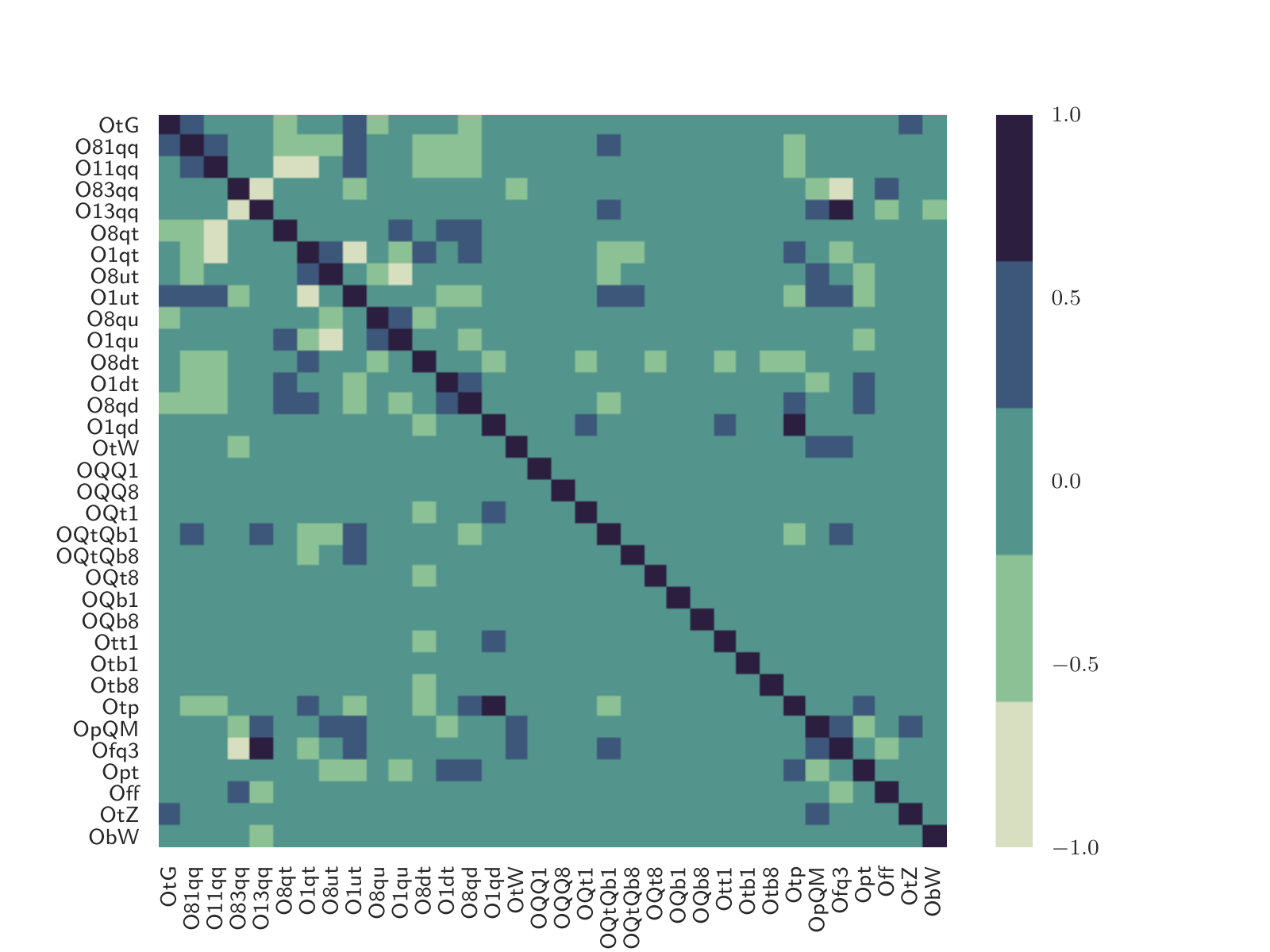}
     \caption{\small The values of the correlation coefficient between
      the different degrees of freedom $c_i$, Eq.~(\ref{eq:correlationL2CT}),
      for the L2 closure test
      with SM reference values.
      \label{fig:CTheatmap}}
  \end{center}
\end{figure}

It is however important to emphasise that the bounds
reported in Figs.~(\ref{fig:closuretestsL2},\ref{fig:ctBSM}) need to be taken with a grain of salt,
since some degrees of freedom are highly (anti-)correlated.
To quantify this, in Fig.~\ref{fig:CTheatmap} we show the values of the correlation coefficient between
the different degrees of freedom $c_i$ for the L2 closure test
with SM reference values.
The correlation coefficient between two of the degrees of freedom in the fit
$c_i$ and $c_j$ is computed using the standard MC expression, namely
\be
\label{eq:correlationL2CT}
\rho\lp c_i,c_j\rp=\frac{\frac{1}{N_{\rm rep}}\sum_{k=1}^{N_{\rm rep}}
c_i^{(k)} c_j^{(k)} -\la c_i\ra \la c_j\ra
}{\delta c_i \delta c_j} \, .
\ee
From this comparison we see  that some degrees of freedom
are very correlated, for example the chromomagnetic operator {\tt OtG}
is highly correlated with the two-heavy-two-light operators
{\tt O81qq} and {\tt O11qq}.
   
\begin{figure}[t]
  \begin{center}
    \includegraphics[width=0.44\linewidth,angle=-90]{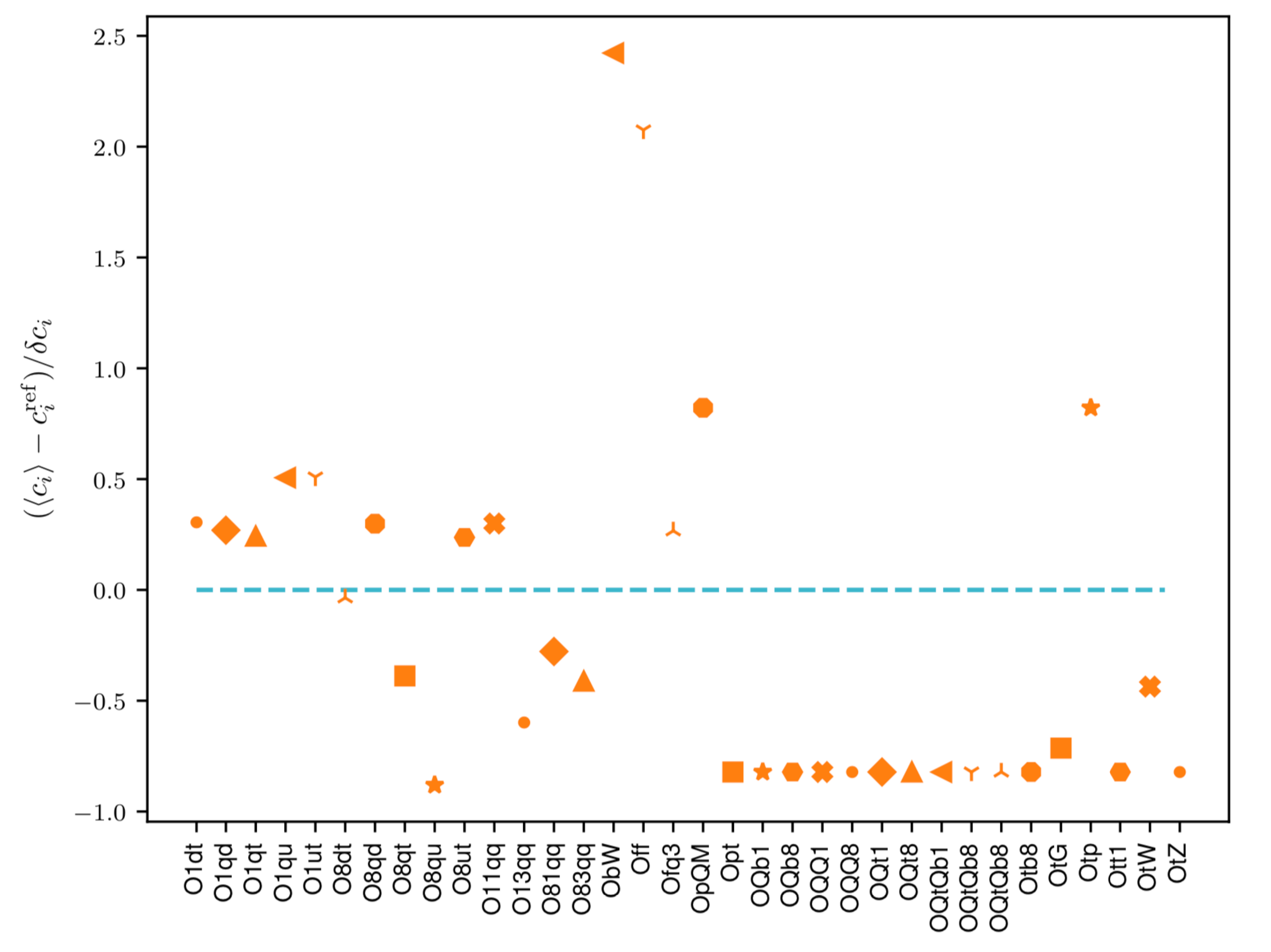}
    \caption{\small  Same as Fig.~\ref{fig:closuretestsL2}
        (left) for the closure tests in which each of the $N_{\rm op}=34$
        operators has been constrained separately from
        all the others.
        The dashed blue line indicates the Standard Model value
        $\{c_i^{{\rm ref}}\}=0$,
        which is the true underlying law in these closure tests.
         \label{fig:CTL2_residuals_individualfits}
      }
  \end{center}
\end{figure}

One might argue that the residuals reported in Fig.~\ref{fig:closuretestsL2}
(left) for the Level 2 closure test appear
to be unnaturally small, even taking into account that the bounds
$\delta c_i$ are computed as 95\% CL ranges.
Indeed, if the $N_{\rm op}=34$ degrees of freedom that we are considering here
were truly independent random variables, then one would expect to find
at least one or two operators with residuals $|r_i|\gsim 1$.
It is possible to  demonstrate that the limited spread of the residuals
in Fig.~\ref{fig:closuretestsL2} is a direct consequence
  of both the correlations between the operators and of the redundancies in the SMEFT
  parameter space, given the input dataset used in the analysis.
  To show this, in Fig.~\ref{fig:CTL2_residuals_individualfits} we report
  the same residuals as in Fig.~\ref{fig:closuretestsL2}  but now
  with the results of  $N_{\rm op}=34$ independent closure tests where
  each operator has been constrained separately, setting the contribution
  from all others to zero.
  As one can see, fitting one operator
  at a time results in a  greater spread of the residuals, with several
  instances in which $|r_i|\gsim 1$.
  In other words, the reduced spread of the residuals in the global
  fit is a genuine effect, arising
  from the correlations and degeneracies in the explored
  SMEFT parameter space, rather than an artifact of the fit.

  Note that in Fig.~\ref{fig:CTL2_residuals_individualfits} the residuals
  for the four-heavy-quark operators all take a similar value.
  The reason is that all these operators are just constrained by the same
  two datapoints, so in individual fit they all result in the same residual.
  One can check that both $\la c_i\ra$ and $\delta_i$ are different
  operator by operator, as expected since in each case the size
  of the SMEFT corrections is different, but that at the residual level
  one ends up with the same result.

\subsection{Methodological variations}
\label{sec:methtests}

We now turn to study the robustness of the 
baseline results with respect to a number of variations in the fitting 
methodology.
In particular: (i) the impact of cross-validation; (ii) the effects
of experimental uncertainties in determining the bounds on the SMEFT degrees
of freedom; and (iii) the role of $\mathcal{O}\lp \Lambda^{-2}\rp$ corrections
on these same bounds.
We will always assume the SM; as we have shown
above, closure tests will likewise work in the case of BSM scenarios.

\paragraph{Cross-validation.}
As discussed in Sect.~\ref{sec:stopping}, it is important to ensure that
over-fitting is avoided, and, to do so, we adopt cross-validation.
To quantify the role that cross-validation plays on 
fit results, we perform two L2 closure tests, with
the only difference that cross-validation is absent in one of them.

\begin{figure}[t]
\begin{center}
\includegraphics[width=0.49\linewidth]{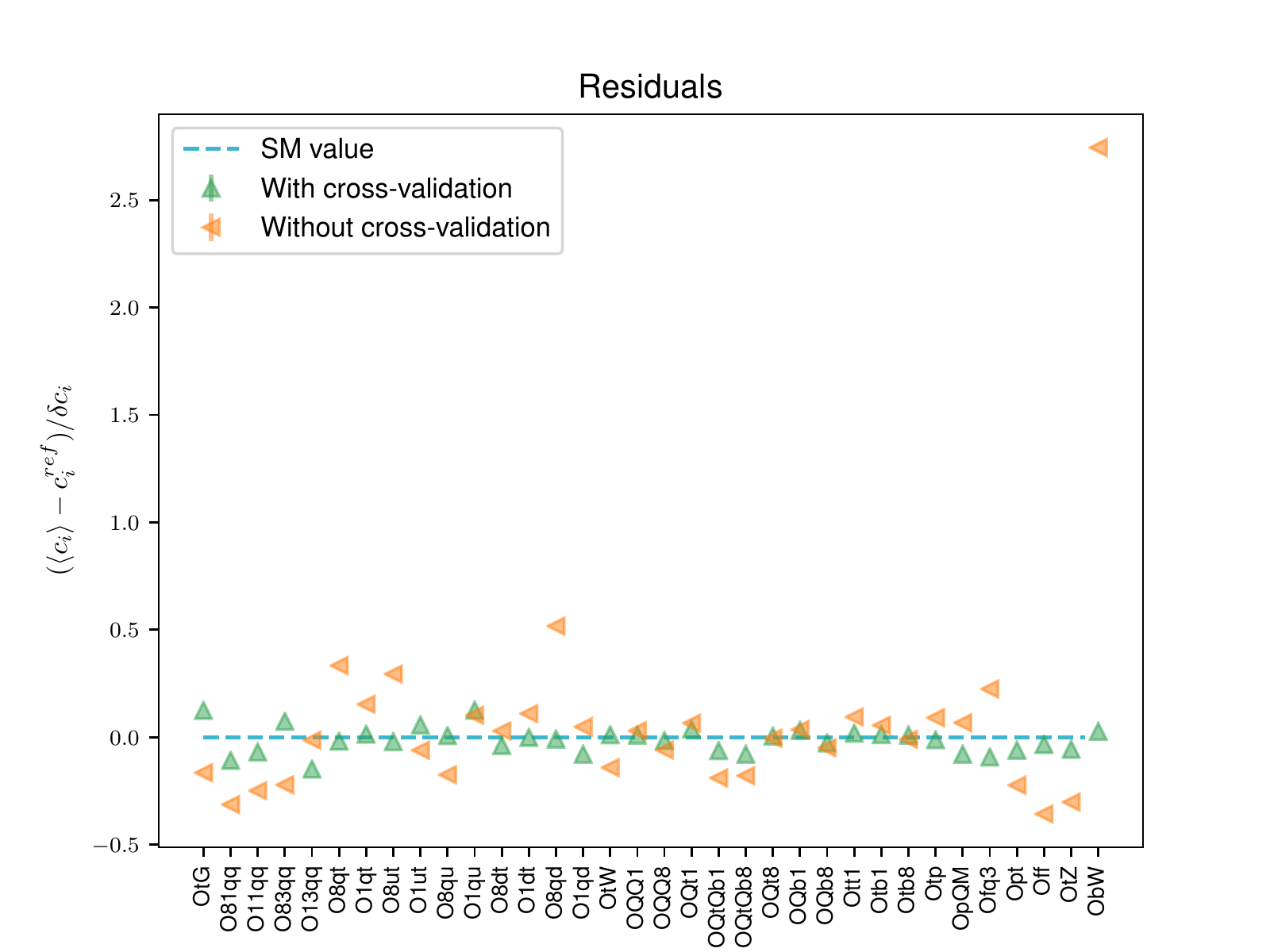}
\includegraphics[width=0.49\linewidth]{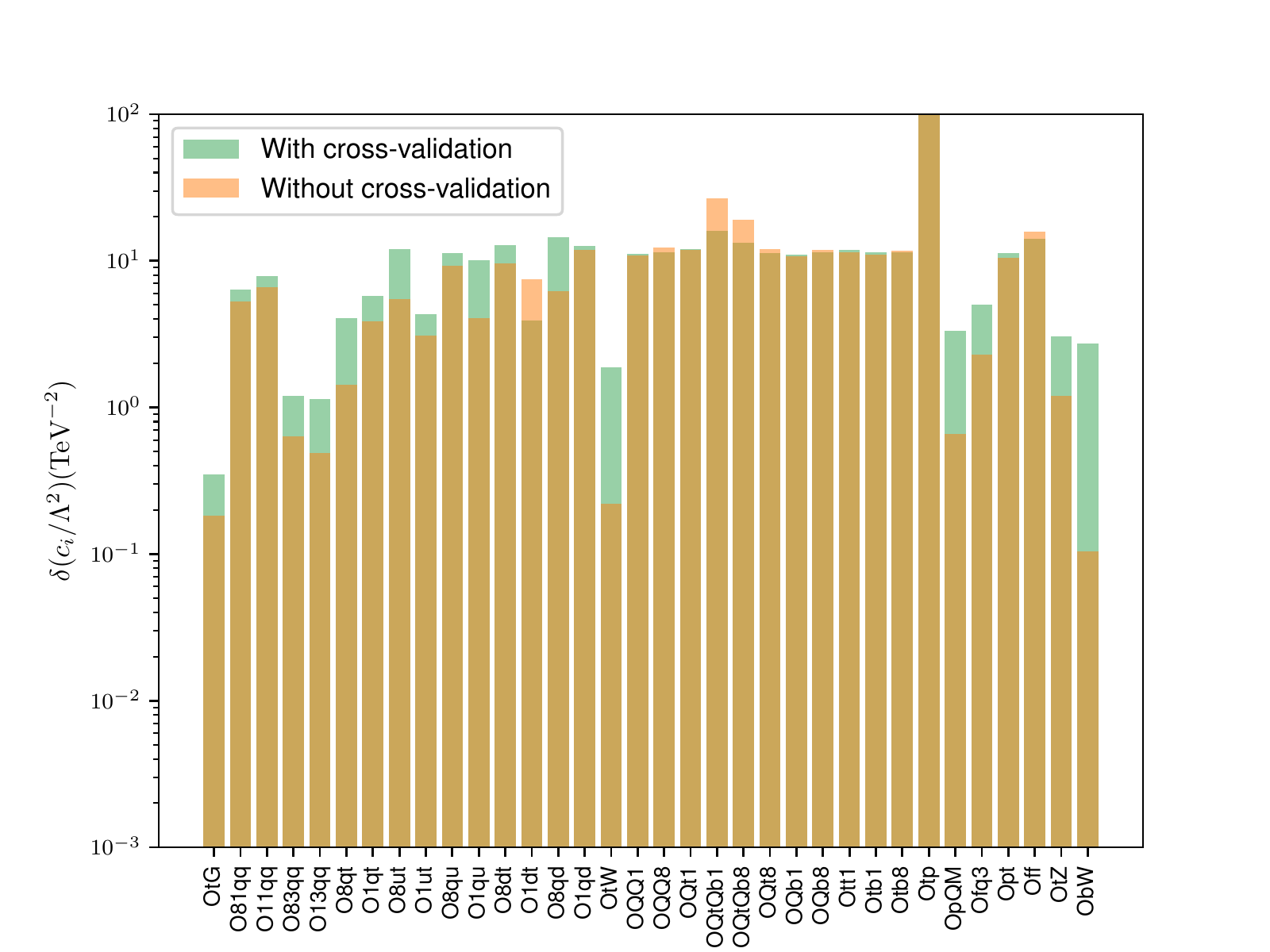}
\caption{\small Same as Fig.~\ref{fig:closuretestsL2}
  comparing the L2 closure tests with and without cross-validation.
     \label{fig:crossvalcompare} }
  \end{center}
\end{figure}

In Fig.~\ref{fig:crossvalcompare} we compare the fit residuals and the 
95\% CL of the fit parameters obtained from the two closure
tests.
When cross-validation is absent, the central values of the fitted degrees of 
freedom $c_i$ fluctuate around the true result (the SM in this case) rather
more than when cross-validation is used.
This is a consequence of the fact that the fit without cross-validation
has overfitted the experimental data, and therefore the fluctuations around 
the true result have been enhanced.
For example, $r_{\rm bW}\simeq 2.5$ without cross-validation, while
it should be $r_{\rm bW}\simeq 0$ as we can see from the left panel.
Moreover, from the right panel of Fig.~\ref{fig:crossvalcompare} we see 
that the bounds obtained by the fit are usually stronger
when cross-validation is switched off.
However, in this case, they are methodologically biased, and one would 
incorrectly claim to have derived more stringent limits than the truth.
These comparisons highlight that reliable results in a global SMEFT analysis
can be obtained only if overfitting is avoided.
Otherwise, deviations between experimental data and theory calculations, 
and/or stringent bounds on the fitted degrees of freedom can be misinterpreted
as a sign of new physics, while they are instead a sign of methodological bias.

\paragraph{Characterising fit uncertainties.}
\label{char_fit_unc}
As explained in Sect.~\ref{sec:closuretest}, L2 closure tests differ from L1 
closure tests for the introduction of an additional set of fluctuations.
Comparing closure tests at different levels allows one to identify the 
different components that build up the total uncertainty on the fit parameters 
$\delta c_i$, for a more-in depth discussion applied to PDFs, see 
Ref.~\cite{Ball:2014uwa}.
To begin with, L0 closure test results might have {\it interpolation} and
{\it extrapolation} uncertainties: even if the fit to the data points is 
perfect, there will be non-zero uncertainties in-between and outside the data 
region.
In the SMEFT case, however these uncertainties vanish in L0 closure tests, 
since the associated parameter space is discretised over the $N_{\rm op}=34$ 
independent degrees of freedom, and additional directions are never explored.

The comparison between the values of $\delta c_i$ in L1 and L2 
closure tests is more subtle.
In the L1 case, the data uncertainty is propagated into the fit,
see Eq.~(\ref{eq:replicas}).
Therefore, the component of $\delta c_i$, that L1 closure tests identify, is 
associated to the finite precision of the input experimental measurements,
and hence we call this the {\it experimental} component of the uncertainty.
At L2, we additionally account for the fact that there are infinite different 
sets of $\{c_i\}$ that optimise the error function equally well.
The spread among these solutions represent the irreducible {\it redundant}
component of the uncertainty.

\begin{figure}[t]
  \begin{center}
\includegraphics[width=0.99\linewidth]{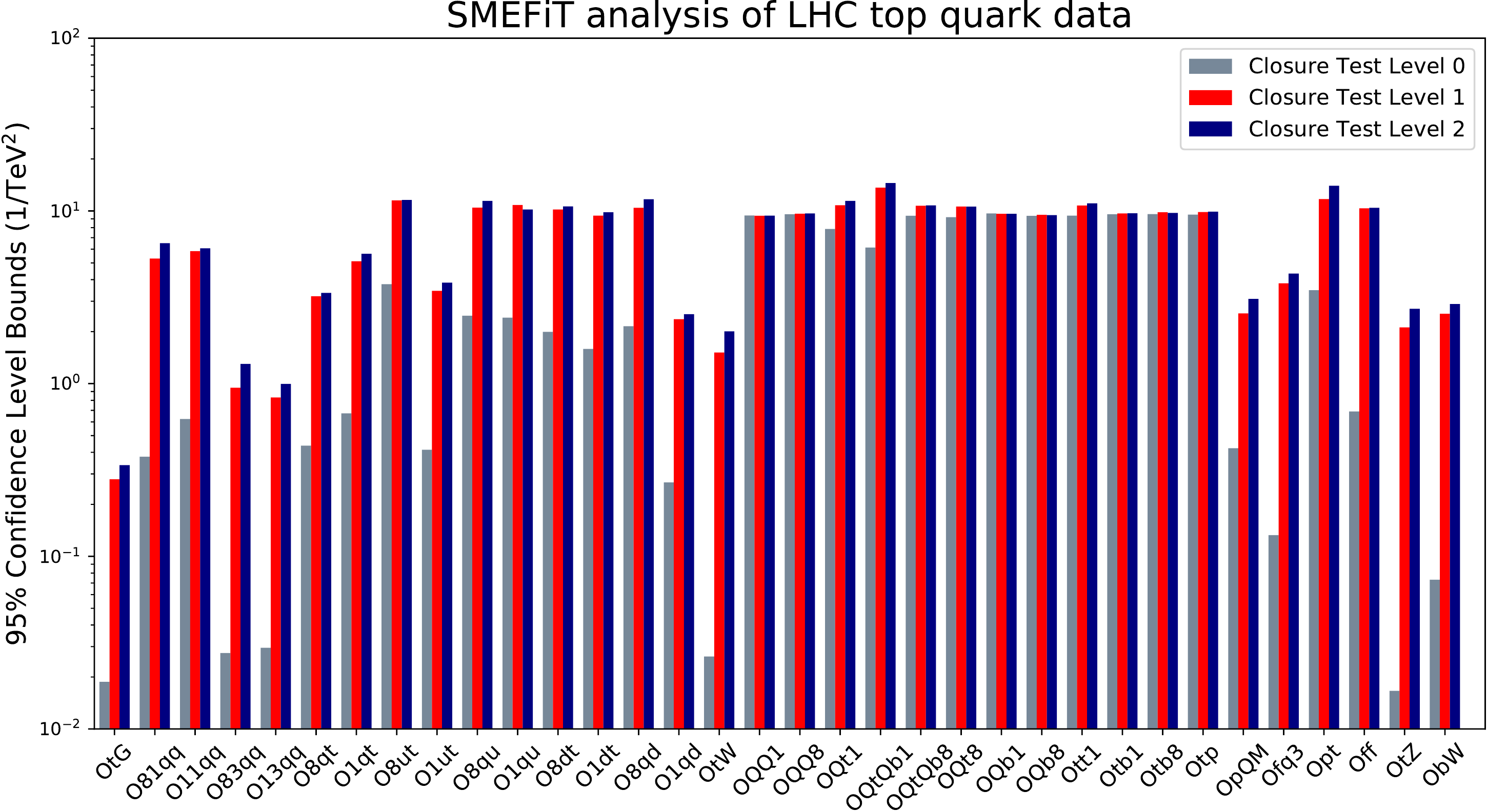}
\caption{\small The symmetrised
  95\% CL bounds for the fit parameters, $\delta c_i$,
  obtained in the closure tests at levels 0, 1a and 2.
     \label{fig:CTlevel012compare} }
  \end{center}
\end{figure}

To illustrate the relative weight of these two components on the overall size
of $\delta c_i$, in Fig.~\ref{fig:CTlevel012compare} we
show the bounds that are obtained in L0, L1 and L2 closure tests, 
leaving everything else unchanged.
We find that there is a significant increase in the size of $\delta c_i$
when going from L0 to L1, but then there is only a very slight increase
when going from L1 to L2.

\paragraph{The role of $\mathcal{O}\lp \Lambda^{-4}\rp$ corrections.}
Closure tests can also be used to assess the dependence of the fit results 
upon variations of the details of the theory calculations.
Specifically, we are interested in the role played by 
$\mathcal{O}\lp \Lambda^{-4}\rp$ corrections in the determination of 
the bounds on the fitted degrees of freedom.
As highlighted in Table~\ref{table:operatorprocess}, including 
$\mathcal{O}\lp \Lambda^{-4}\rp$ terms in the theoretical model
modifies rather significantly the parameter space, by opening up new directions 
and by enhancing the sensitivity to those directions already covered by
$\mathcal{O}\lp \Lambda^{-2}\rp$ terms.
Therefore, despite the fact that pseudo-data are generated according to a
given theory in a closure test, including or not 
$\mathcal{O}\lp \Lambda^{-4}\rp$ corrections implies that the corresponding 
results should in general be different.

\begin{figure}[t]
  \begin{center}
\includegraphics[width=0.49\linewidth]{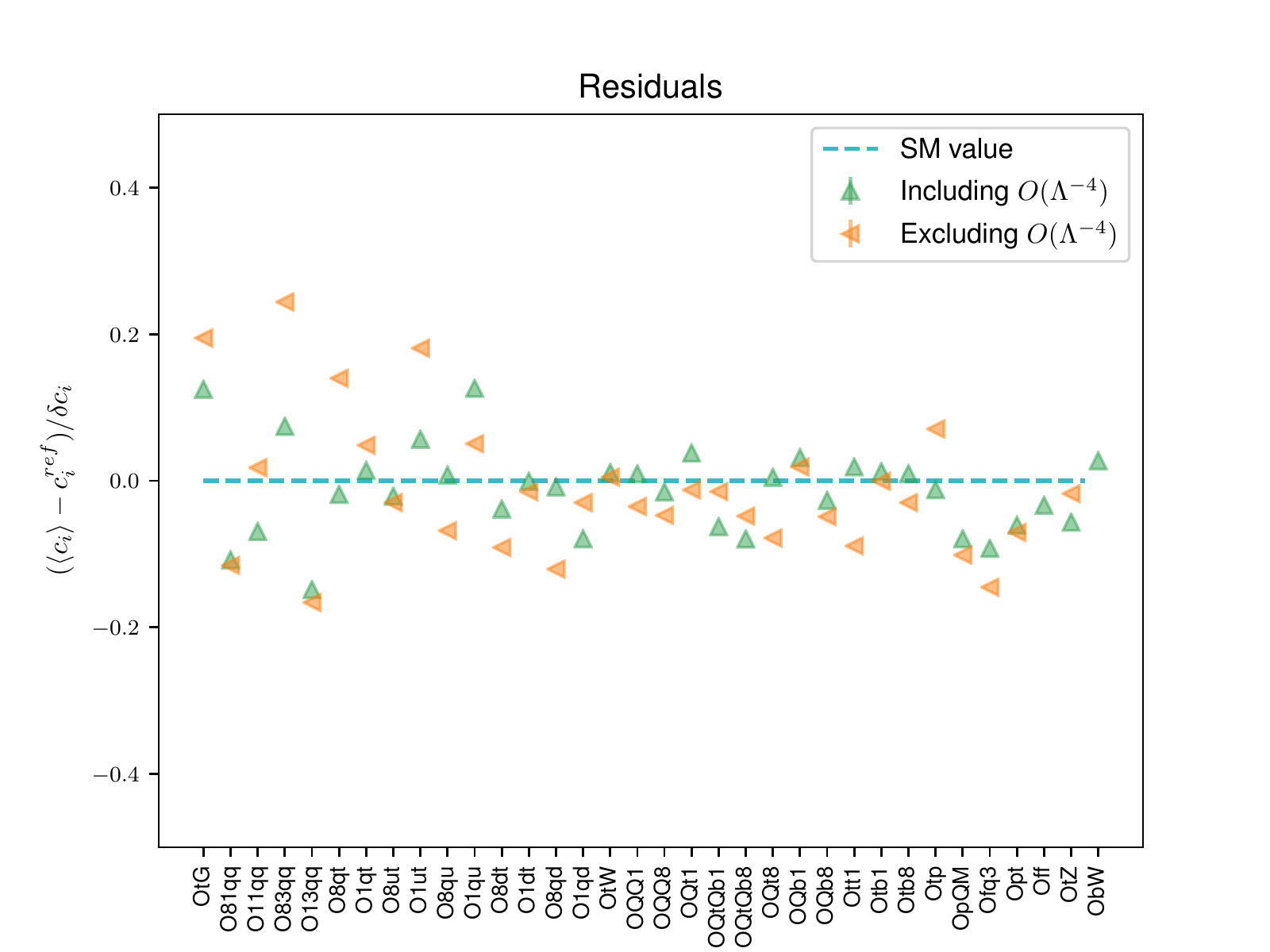}
\includegraphics[width=0.49\linewidth]{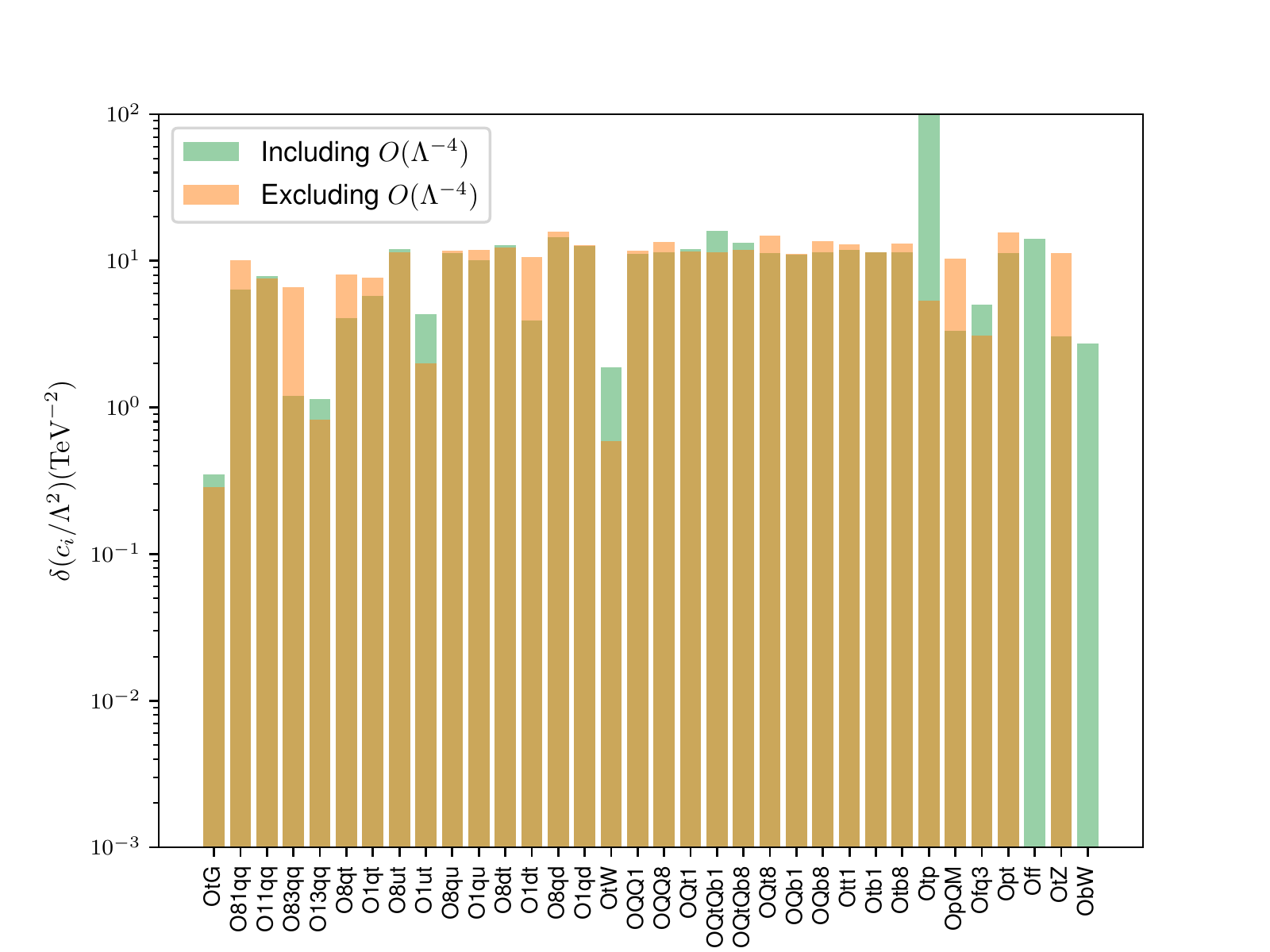}
\caption{\small Same as Fig.~\ref{fig:closuretestsL2},
  now comparing the results of two L2 closure test fits with and without 
  $\mathcal{O}\lp \Lambda^{-4}\rp$ SMEFT corrections. Our baseline results 
  include these corrections.
  \label{fig:SMEFit-L2CT-Lambda2vs4.pdf} }
  \end{center}
\end{figure}

In Fig.~\ref{fig:SMEFit-L2CT-Lambda2vs4.pdf} we show
the comparison of the residuals $r_i$ (left panel)
and of the bounds $\delta c_i$ (right panel) for L2 closure tests
between two fits that differ only for the inclusion (or not) of the 
$\mathcal{O}\lp \Lambda^{-4}\rp$ terms.
Two degrees of freedom, {\tt Off} and {\tt ObW}, are
not constrained in the fit without $\mathcal{O}\lp \Lambda^{-4}\rp$ terms, 
and are therefore set to zero.
From this comparison, we see that the bounds on the coefficients $\delta c_i$ 
generally improve when $\mathcal{O}\lp \Lambda^{-4}\rp$ corrections
are included in the theoretical calculation.
For example, the bound on {\tt OtZ} decreases from $\delta c_{tZ}\simeq 6$ 
TeV$^{-2}$ to  $\delta c_{tZ}\simeq 2$ TeV$^{-2}$.
The slight worsening observed for the bounds on some few operators
when only linear terms are included is consistent with 
statistical fluctuations, and is therefore not significant.
In any case, the fit results are qualitatively similar irrespective of the 
inclusion of $\mathcal{O}\lp \Lambda^{-4}\rp$ corrections.
Note that some of the degrees of freedom are highly correlated,
therefore the interpretation of the results at the individual bound level 
should be taken with care.

\clearpage

\section{The top quark sector of the SMEFT at NLO}
\label{sec:results}

In this section, we present the main results of this work, namely we derive
the constraints on the $N_{\rm op}=34$ SMEFT dimension-6
degrees of freedom relevant for the interpretation of top quark production
measurements at the LHC.
We first discuss the fit quality and the agreement between experimental
data and theoretical predictions for individual processes.
We then present the best-fit values, the 95\% confidence level intervals
and the correlations for these degrees of freedom, and
we compare our results with other related analyses in the literature.
We also study the impact that both NLO QCD perturbative corrections
and quadratic $\mathcal{O}\lp \Lambda^{-4}\rp$ terms have on the results.
Finally, we assess the dependence of the fit results
on the choice of input dataset, and
quantify the dependence of the derived bounds on the high-energy
limit of the cross-sections included in the fit.

\subsection{Fit quality and comparison with data}
\label{sec:datacomparison}
\label{sec:fitquality}

We will first assess the quality of the fit at the level of both
the total dataset and of individual measurements, and then
compare the fit results with the input experimental cross-sections.
In the following, as discussed in Sect.~\ref{sec:smefttop},
our baseline fit is based on $N_{\rm rep}=1000$ MC
replicas and includes both NLO QCD corrections for the SMEFT contributions
and the quadratic $\mathcal{O}\lp \Lambda^{-4}\rp$ higher order terms.

In Table~\ref{tab:chi2table} we indicate
the values of the $\chi^2$ per datapoint for each
of the datasets included in the fit.
In each case, we indicate the values of $\chi^2/n_{\rm dat}$ first
when the theory calculations include only the SM contributions
(second column)
and then once they account for the SMEFT corrections
after the fit (third column).
In the last column, we indicate the number of data
points $n_{\rm dat}$
for each dataset.
The datasets are classified into three groups following the structure of
Tables~\ref{eq:input_datasets}--\ref{eq:input_datasets2}: inclusive
$t\bar{t}$, $t\bar{t}$ in association with $V$, $H$, or heavy quarks, and
single top production.
In the case of datasets consisting of multiple differential distributions,
we indicate the one that has been included in this analysis.

\begin{table}[p]
  \centering
  \footnotesize
   \renewcommand{\arraystretch}{1.34}
   \begin{tabular}{l|c|c|c}
      \toprule
Dataset   &  $\chi^2/n_{\rm dat}$ (prior) & $\chi^2/n_{\rm dat}$ (fit)  &  $n_{\rm dat}$  \\
\midrule
{\tt ATLAS\_tt\_8TeV\_ljets} [ $m_{t\bar{t}}$ ]  &   1.51      & 1.25     &    7     \\
{\tt CMS\_tt\_8TeV\_ljets} [ $y_{t\bar{t}}$ ]   &    1.17    &   1.17     &    10    \\
{\tt CMS\_tt2D\_8TeV\_dilep } [ $\lp m_{t\bar{t}},y_t\rp$ ] &   1.38    &   1.38     &    16   \\
{\tt CMS\_tt\_13TeV\_ljets2 } [ $m_{t\bar{t}}$ ]    &    1.09    &  1.28      &    8    \\
{\tt CMS\_tt\_13TeV\_dilep } [ $m_{t\bar{t}}$ ]  &    1.34   &  1.42      &    6    \\
{\tt CMS\_tt\_13TeV\_ljets\_2016} [ $m_{t\bar{t}}$ ]	    &    1.87    &   1.87    &    10   \\
{\tt ATLAS\_WhelF\_8TeV}    &     1.98   &  0.27      &    3    \\
{\tt CMS\_WhelF\_8TeV}     &    0.31    &  1.18     &    3   \\
\midrule
%
{\tt CMS\_ttbb\_13TeV}    &    5.00    &    1.29     &    1    \\
{\tt CMS\_tttt\_13TeV}    &    0.05     &  0.02     &    1    \\
{\tt ATLAS\_tth\_13TeV}      &   1.61     &  0.55      &    1    \\
{\tt CMS\_tth\_13TeV}    &    0.34    &   0.01     &    1    \\
{\tt ATLAS\_ttZ\_8TeV}    &   1.32      &   5.29    &    1    \\
{\tt ATLAS\_ttZ\_13TeV}    &   0.01     &  1.06     &    1    \\
{\tt CMS\_ttZ\_8TeV}    &  0.04      &   0.06      &    1     \\
{\tt CMS\_ttZ\_13TeV}   &    0.90     &   0.67     &    1    \\
{\tt ATLAS\_ttW\_8TeV}    &   1.34    &    0.27     &    1    \\
{\tt ATLAS\_ttW\_13TeV}    &    0.82     &   0.65      &    1    \\
{\tt CMS\_ttW\_8TeV}   &    1.54     &    0.54    &    1    \\
{\tt CMS\_ttW\_13TeV}&      0.03   &  0.09     &    1    \\
\midrule
{\tt CMS\_t\_tch\_8TeV\_dif}    &   0.11     &  0.32      &    6   \\
{\tt ATLAS\_t\_tch\_8TeV} [ $y_t$ ]    & 0.91       &  0.43     &    4    \\
{\tt ATLAS\_t\_tch\_8TeV} [ $y_{\bar{t}}$ ]   &  0.39      &   0.45     &    4   \\
{\tt ATLAS\_t\_sch\_8TeV}    &   0.08     &   1.92    &    1  \\
{\tt ATLAS\_t\_tch\_13TeV} &     0.02         & 0.09     & 2   \\
{\tt CMS\_t\_tch\_13TeV\_dif} [ $y_t$ ]    &  0.46      & 0.49       &    4    \\
{\tt CMS\_t\_sch\_8TeV}    &  1.26      & 0.76      &    1   \\
{\tt ATLAS\_tW\_inc\_8TeV}    &   0.02     &  0.06     &    1   \\
{\tt CMS\_tW\_inc\_8TeV}    &     0.00   &  0.07     &    1   \\
{\tt ATLAS\_tW\_inc\_13TeV}    &   0.52     &   0.82     &    1    \\
{\tt CMS\_tW\_inc\_13TeV}    &   4.29     &  1.68     &    1    \\
{\tt ATLAS\_tZ\_inc\_13TeV}    &   0.00      &  0.00      &    1    \\
{\tt CMS\_tZ\_inc\_13TeV}    &  0.66      &    0.34    &    1    \\
\midrule
Total &	      {\bf 1.11}    &	  {\bf 1.06}       &  {\bf 103}     \\
\bottomrule
  \end{tabular}
  \caption{\small The values of the $\chi^2$ per data point for each
    of the datasets included in the fit.
    In each case, we indicate the values of $\chi^2/n_{\rm dat}$ first
    when the theory calculations include only the SM contributions (second 
    column) and then once they account for the SMEFT corrections, after the 
    fit (third column).
    In the last column we indicate the number of data points $n_{\rm dat}$.
    Datasets are classified in three groups following the structure of
    Tables~\ref{eq:input_datasets}--\ref{eq:input_datasets2}:
    inclusive top quark pair production; $t\bar{t}$ production in association
    with heavy quarks, vector bosons, and Higgs bosons; and inclusive
    and associate production of single top quarks.
    In the case of datasets made of multiple differential distributions,
    we indicate the one that has been used in the analysis.
     \label{tab:chi2table}
  }
\end{table}

From the values in Table~\ref{tab:chi2table}
we find that the overall fit quality to the $n_{\rm dat}=103$ data
points included in the fit is satisfactory, with
of $\chi^2/n_{\rm dat}=1.06~(1.11)$ after (before) the fit.
We find therefore a slight improvement in the overall fit quality
once the dimension-6 SMEFT corrections are taken into account.
Note however that this improvement is not inconsistent
with statistical fluctuations, since for 103 points
one expects $\Delta\lp \chi^2/n_{\rm dat}\rp\simeq 0.1$.

For most of the individual datasets,
the SM description of the input measurements is already good
to begin with.
In several cases, the $\chi^2$ decreases once the SMEFT corrections
are accounted for.
For instance, the ATLAS $m_{t\bar{t}}$ distribution at 8 TeV improves from
$\chi^2/n_{\rm dat}=$1.51 to 1.25,
and the CMS $t\bar{t}b\bar{b}$ cross-section improves from 5.0 to 1.29.
As expected in a global fit, given that the figure of merit being
optimised is the total $\chi^2$, Eq.~(\ref{eq:chi2definition2}), for some
datasets the overall fit quality is unchanged or slightly worsened
as compared to the SM prediction.

\begin{figure}[t]
  \begin{center}
      \includegraphics[width=0.49\linewidth]{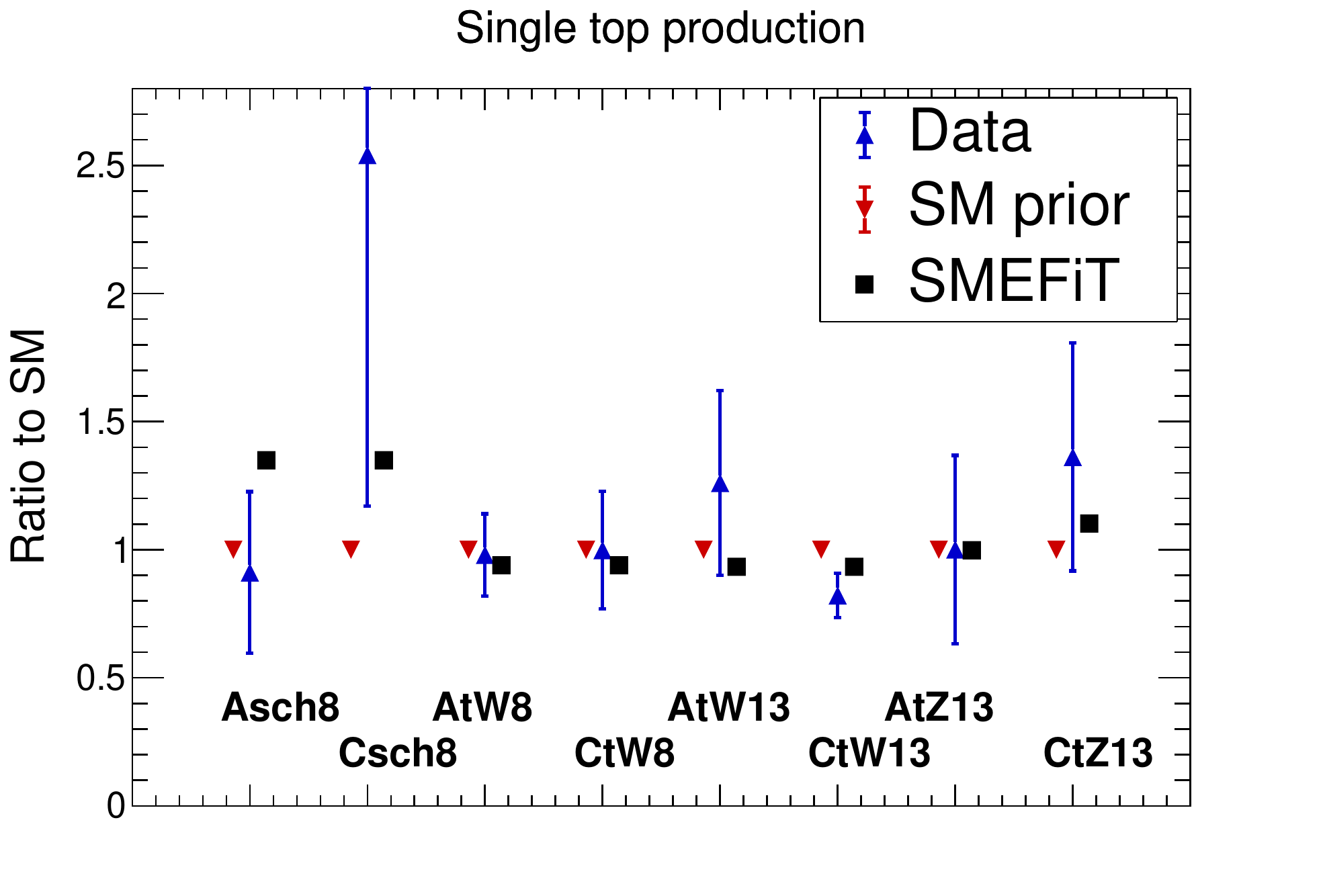}
    \includegraphics[width=0.49\linewidth]{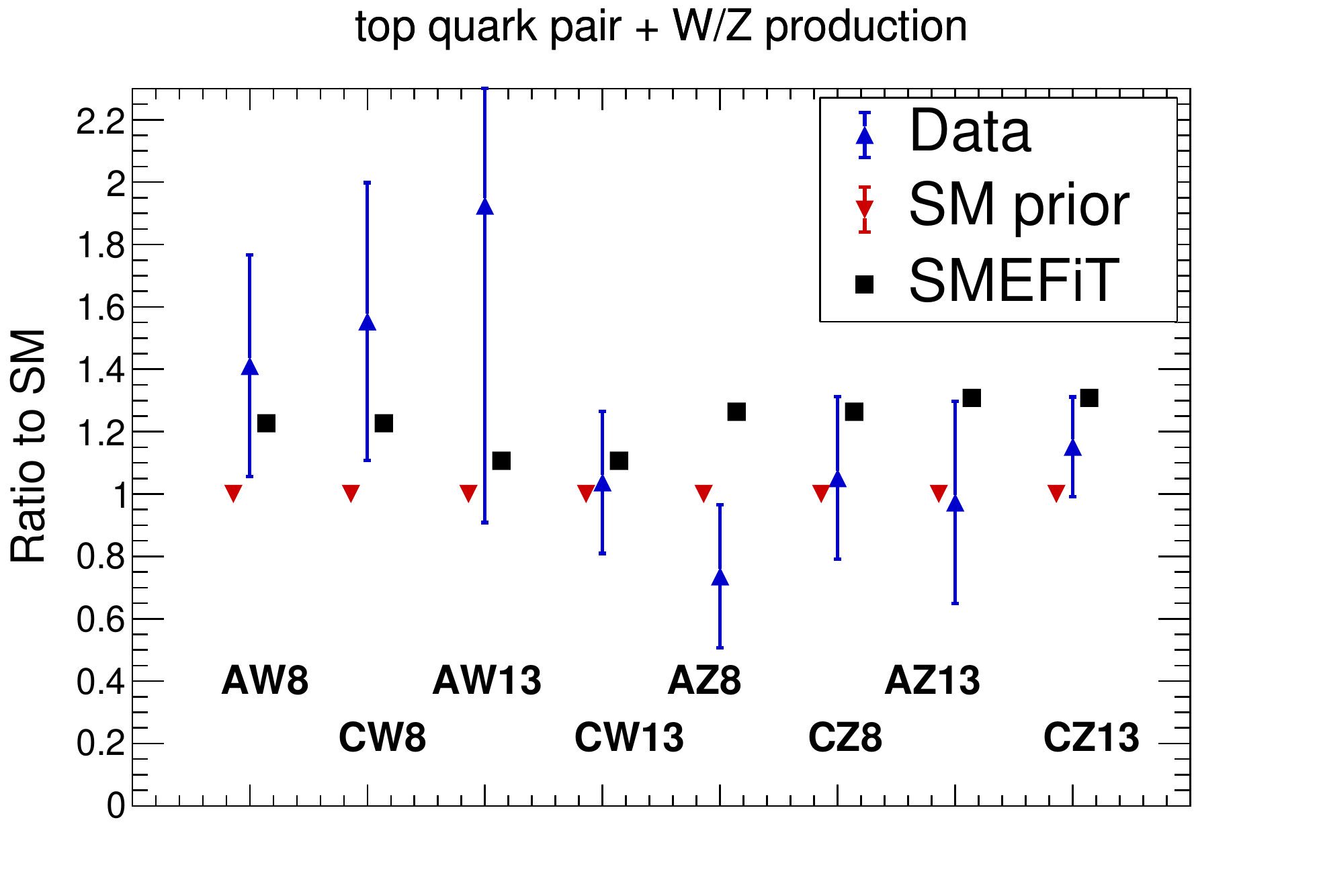}
    \caption{\small Left: the ATLAS (A) and CMS (C) measurements for single top 
      production in the $s$-channel and in the $tW$ and $tZ$ associated 
      production channel.
      Right: the corresponding comparison of the cross-section
      of $t\bar{t}$ associated
      production with $W$ and $Z$ vector bosons.
      We show the results for the most updated measurements both at 8 TeV
      and at 13 TeV.
     \label{fig:datath_3} }
  \end{center}
\end{figure}

From Table~\ref{tab:chi2table}, we
notice that the only experiment for which the $\chi^2/n_{\rm dat}$
worsens significantly after the fit is the ATLAS  $t\bar{t}Z$ cross-section
measurement at 8 TeV, whose SM value of $\chi^2/n_{\rm dat}=1.32$
increases to $5.29$ after the fit.
The origin of this poor $\chi^2$ value
can be traced back to some tension between
the ATLAS and CMS measurements of the same observable.
Indeed, as shown in Fig.~\ref{fig:datath_3}, the ATLAS $t\bar{t}Z$
cross-section at 8 TeV lies somewhat below other measurements of
the same quantity, in particular of the precise CMS measurement at 13 TeV.
This exception aside, we find overall a 
good agreement between the theory calculations and the data used in the fit.

We now turn to present the comparisons
between the results of the present SMEFT fit and the ATLAS
and CMS input experimental data.
We will also show comparisons for observables
that are not included
in the fit to avoid double counting,
but which are anyway interesting to visualise in order to understand the
main features of our results.

\begin{figure}[t]
  \begin{center}
    \includegraphics[width=0.49\linewidth]{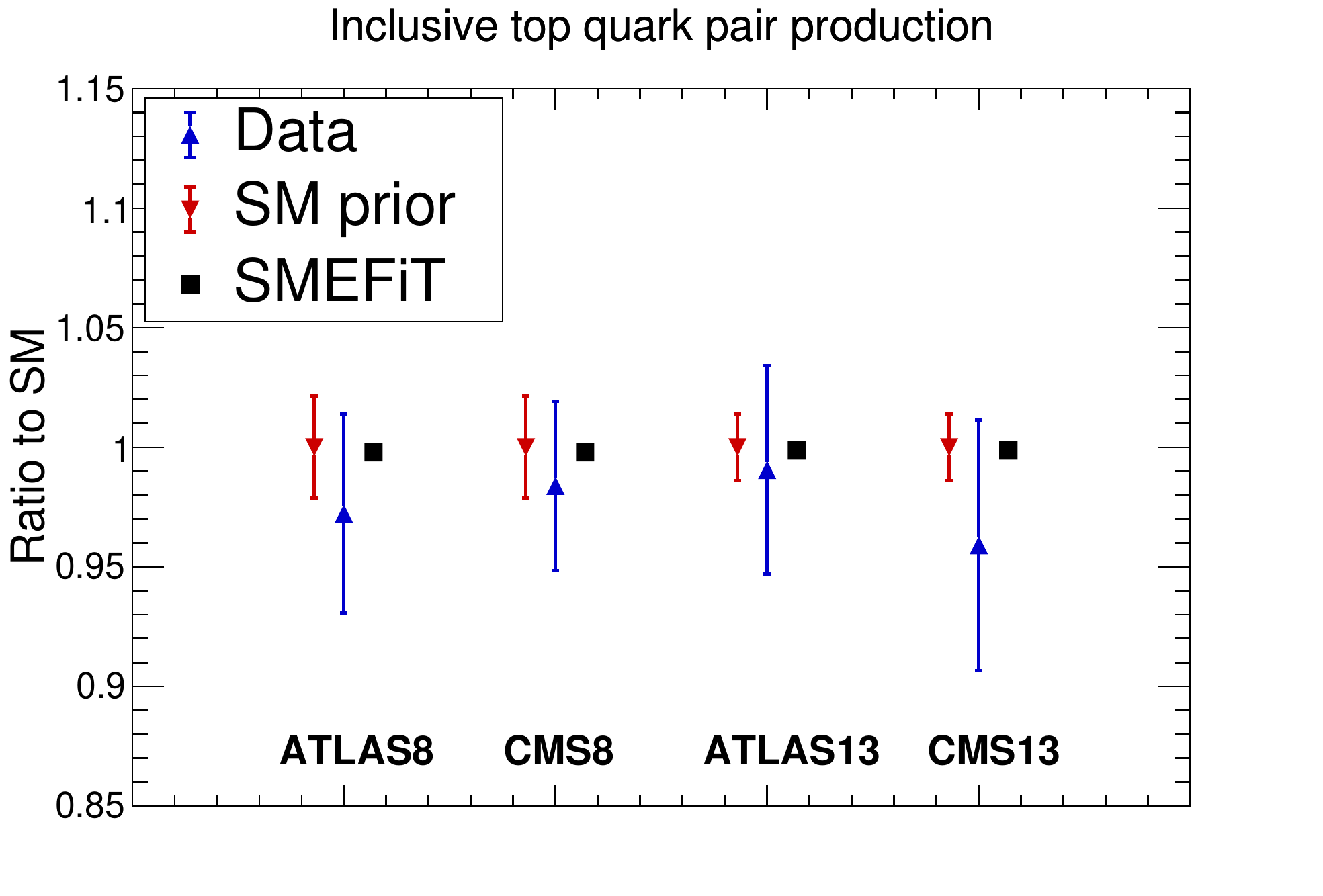}
    \includegraphics[width=0.49\linewidth]{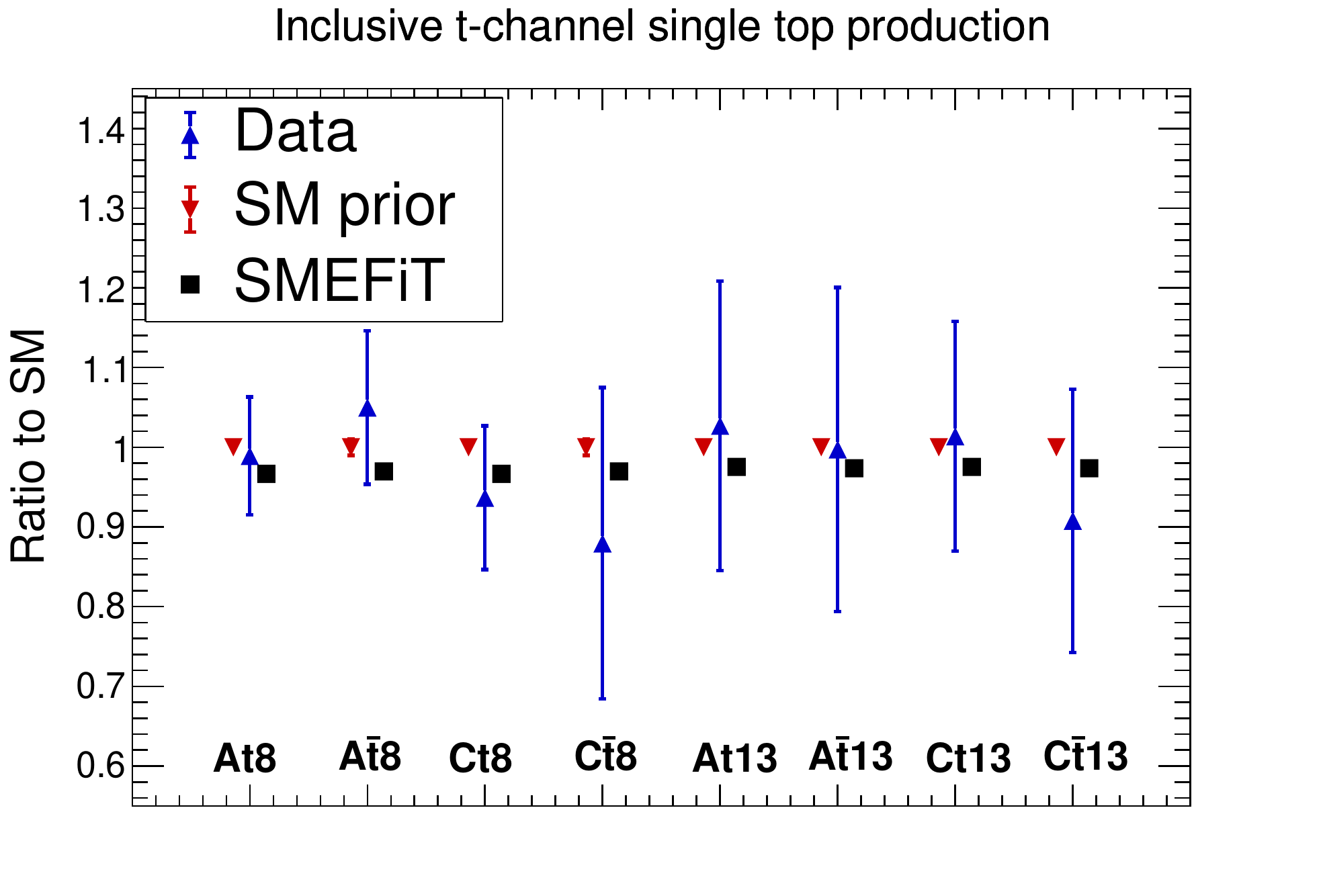}
    \caption{\small Comparison between ATLAS and CMS experimental data
      on the total inclusive $t\bar{t}$ (left) and single top
      $t$-channel (right) production cross-sections at 8 TeV and 13 TeV
      with the corresponding
      SM calculations and with the results of the SMEFT analysis.
      In the case of the SM calculations, we also
      shown the associated PDF uncertainties.
      Results are shown normalised to the central value
      of the SM prediction.
      Note that these inclusive cross-sections are not used as
      input to the fit (to avoid double counting with the
      corresponding differential distributions).
     \label{fig:datath_1} }
  \end{center}
\end{figure}

To begin with, in Fig.~\ref{fig:datath_1}
we show a comparison between the ATLAS and CMS experimental data
on the total inclusive $t\bar{t}$ and single top
$t$-channel production cross-sections at 8 TeV and 13 TeV
with the corresponding
SM calculations and with the results of the SMEFT fit.
For the single top case, we show separately the top and the anti-top
cross-sections.
In the case of the SM calculations, we also
show the associated PDF uncertainties.
Note that none of these total cross-sections
(apart from the ATLAS 13 TeV single-top cross-section) are included 
in the SMEFT fit, since we already include
the corresponding differential distributions.
See Sect.~\ref{sec:settings_expdata} for more details
about measurements shown in this comparison.
In Fig.~\ref{fig:datath_1}, and in all subsequent
comparisons, results are shown normalised to the central SM prediction.

From the comparisons in Fig.~\ref{fig:datath_1}, we find 
good agreement between the data and the SM predictions.
The SMEFT fit result typically moves towards the direction
of the central experimental data point by an amount which
corresponds to at most $|\delta_{\rm th}| \simeq 1\%$ and $\simeq 3\%$ of 
the SM prediction
for inclusive $t\bar{t}$ and single-top production respectively,
well below the experimental uncertainties.
This SMEFT-induced shift in the theory
predictions at the fit level is defined as
\be
\delta_{\rm th} \equiv \frac{\lp \sigma_{\rm SMEFT}-\sigma_{\rm SM}\rp }{\sigma_{\rm SM}}
= \lp \sum_i^{N_{d6}}
\kappa_i \frac{\la c_i\ra}{\Lambda^2} +
\sum_{i,j}^{N_{d6}}  \widetilde{\kappa}_{ij} \frac{ \la c_ic_j\ra}{\Lambda^4} \rp\Bigg/ \sigma_{\rm SM}  \, ,
\ee
with $\la c_i \ra$ and $\la c_ic_j\ra$  represent the averages
of the fitted SMEFT coefficients computed over the MC replica sample.
While this shift is small for these precisely
measured inclusive processes,
 this is not necessarily the case
for differential distributions and for rarer top
production processes, such as for single and top-pair production in association with
vector bosons, as we will show below.

\begin{figure}[t]
  \begin{center}
    \includegraphics[width=0.49\linewidth]{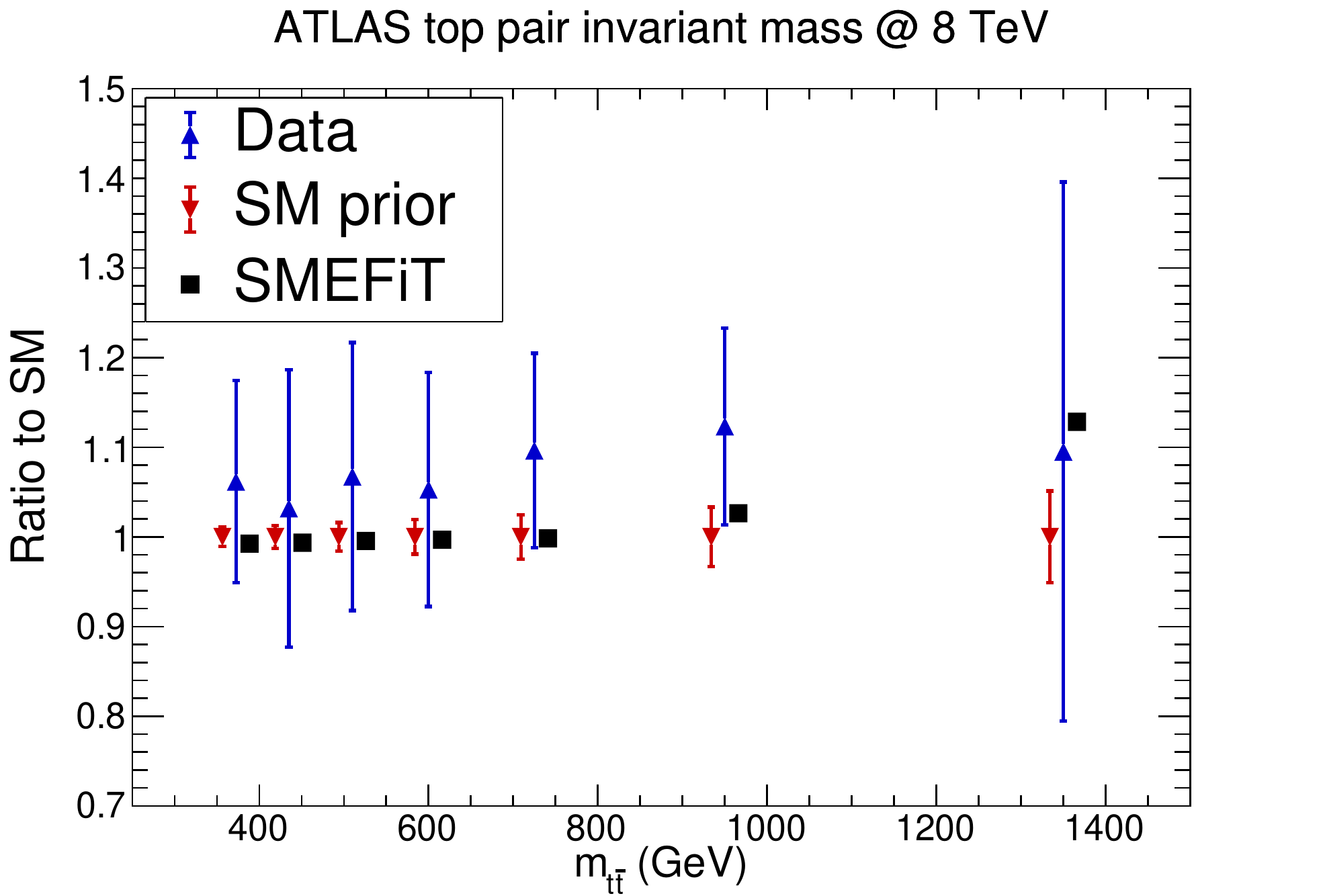}
    \includegraphics[width=0.49\linewidth]{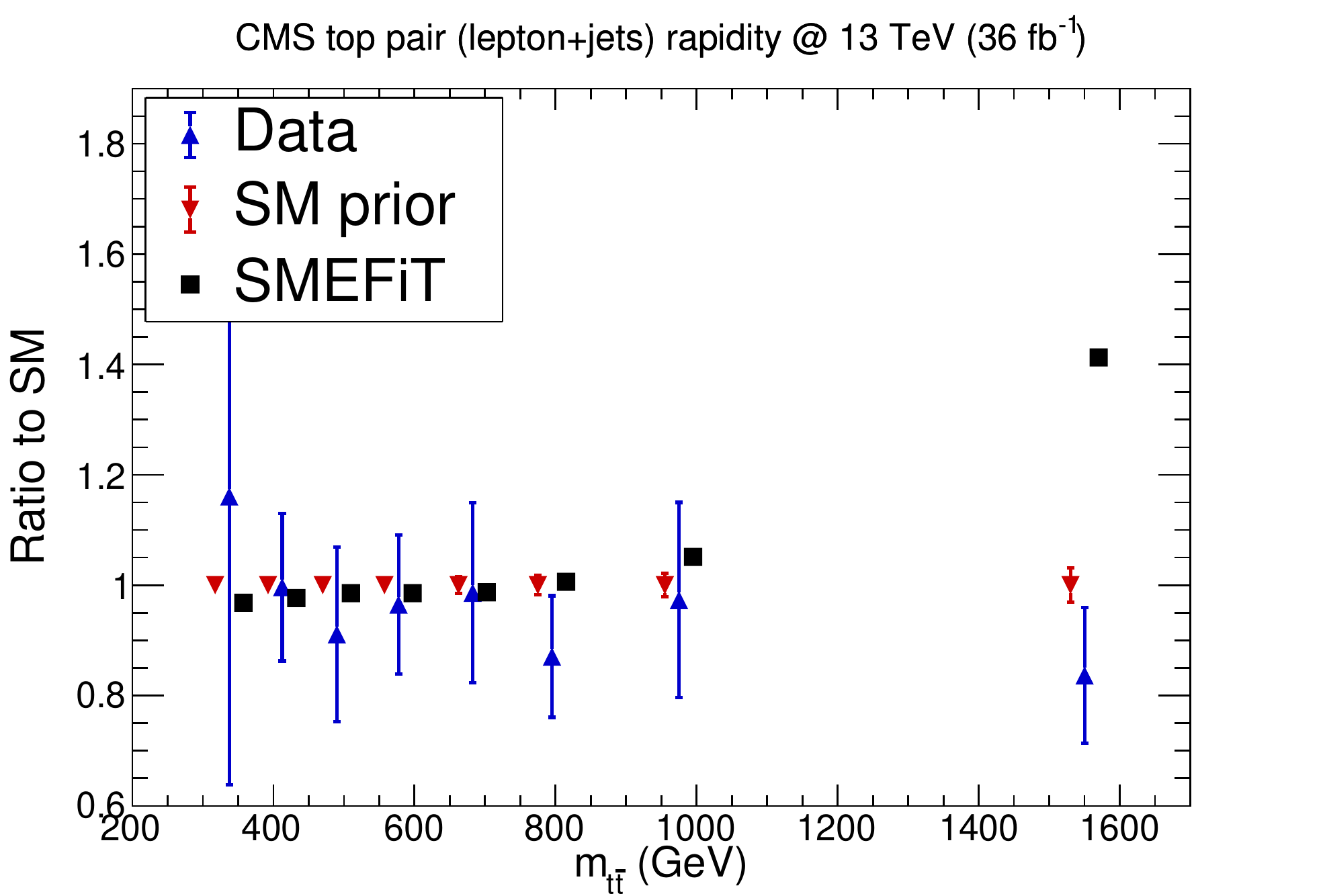}
    \includegraphics[width=0.49\linewidth]{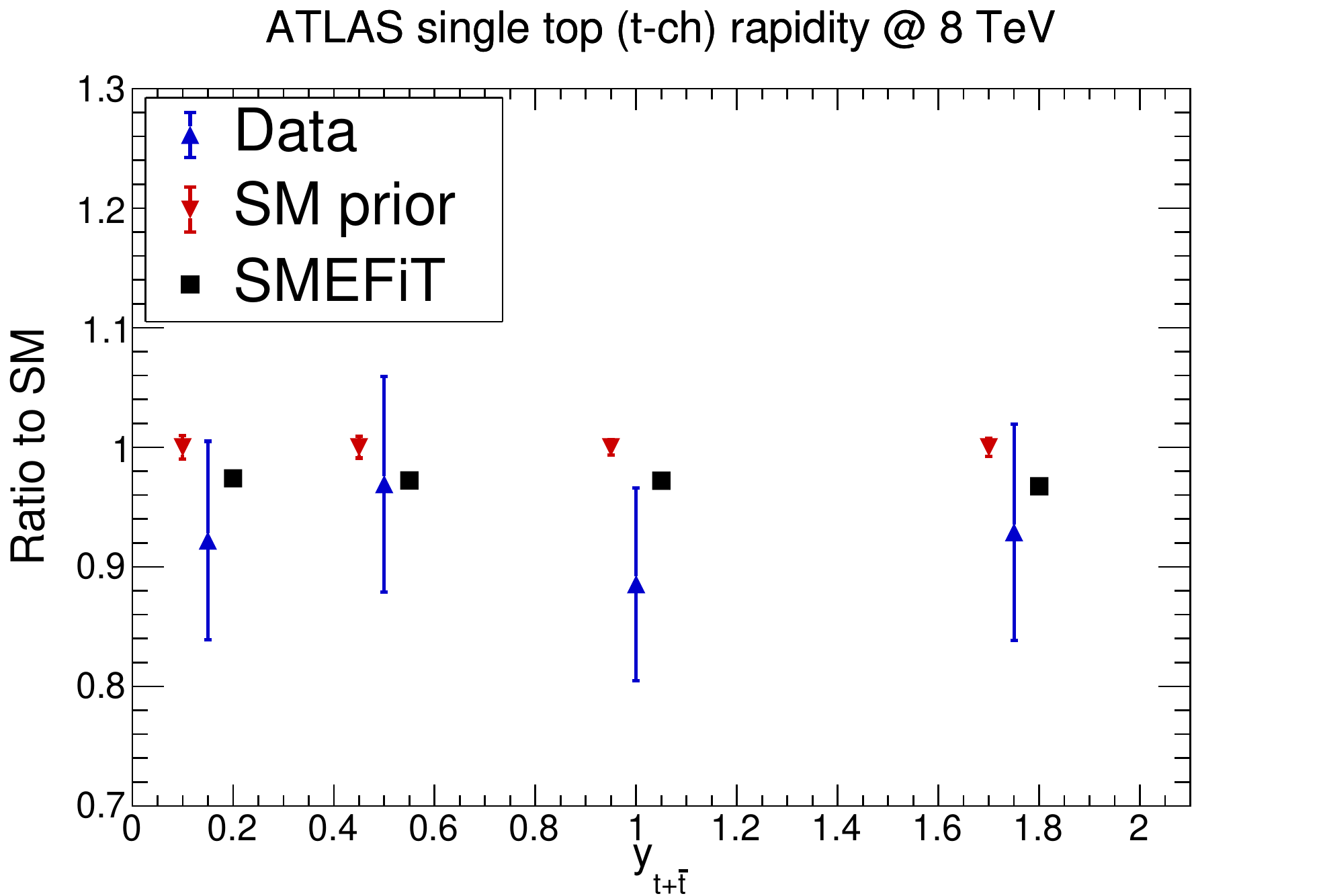}
    \includegraphics[width=0.49\linewidth]{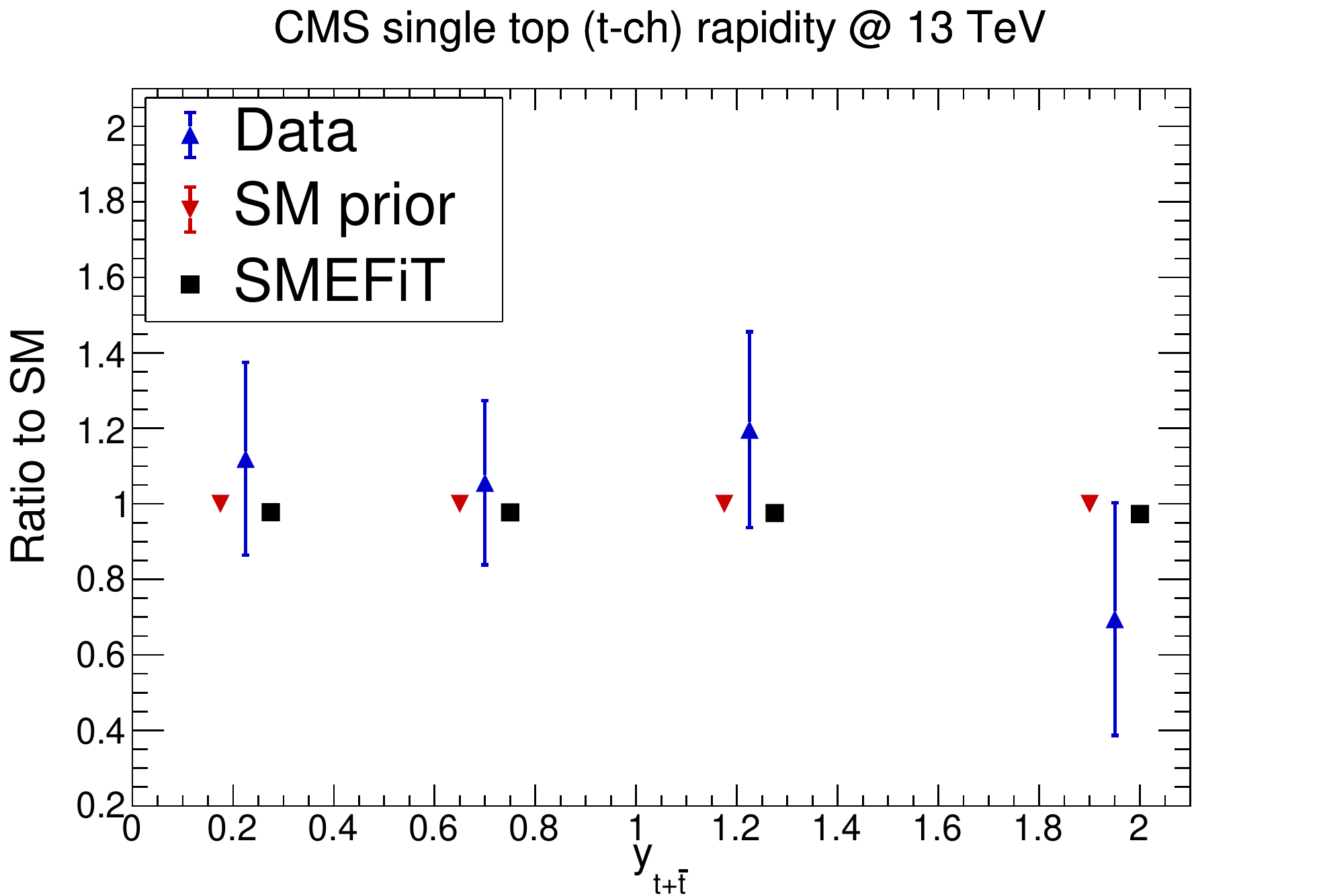}
    \caption{\small Same as Fig.~\ref{fig:datath_1},
    now for differential
    distributions in inclusive top quark pair and single top $t$-channel
    production.
    From left to right and from top to bottom we show the invariant mass
    distribution in $t\bar{t}$ production in ATLAS at 8 TeV, the
    rapidity distribution in $t\bar{t}$ production in CMS at 13 TeV
    (from the 2016 dataset based on $\mathcal{L}=36$ fb$^{-1}$),
    and the rapidity distributions for
    the $t+\bar{t}$ sum in single top quark production
    in the $t$-channel from ATLAS at 8 TeV and from CMS at 13 TeV.
    \label{fig:datath_2}
  }
  \end{center}
\end{figure}

Next, in Fig.~\ref{fig:datath_2} we show a similar comparison as in 
Fig.~\ref{fig:datath_1} now for differential distributions in inclusive 
top quark pair and single top $t$-channel production.
Specifically, we show the invariant mass distribution in $t\bar{t}$ production 
from ATLAS at 8 TeV and CMS at 13 TeV (2016 dataset), and
the rapidity distributions in single top quark production
in the $t$-channel from ATLAS at 8 TeV and from CMS at 13 TeV.
In the latter case the top and anti-top quarks are combined into
a single distribution.

From these comparisons, we find a similar level of agreement
for the differential distributions as for the inclusive
cross-sections.
In the case of the $m_{t\bar{t}}$ distributions from ATLAS and CMS,
the most marked effect comes from the rightmost bin of the
distributions, where energy-growing effects are more important.
We find that the SMEFT-induced shift is $\delta_{\rm th}=+13\%$~($+40\%$)
at $m_{t\bar{t}}\simeq 1.4$~TeV~(1.6~TeV) for the ATLAS 8 TeV (CMS 13 TeV)
measurements.
In Sect.~\ref{sec:highenergy} we will show that results do not change if
the $m_{t\bar{t}}$ distributions are replaced by the corresponding $y_{t\bar{t}}$ 
ones where the energy-growing effects are absent.
In the case of the ATLAS $y_{t+\bar{t}}$ distribution
in single top $t$-channel production, we observe how
the data pulls the fit results.
For this process, the SMEFT-induced
shifts are around $\delta_{\rm th}\simeq -2.5\%$ for all the data bins
for the rapidity distributions in $t$-channel single-top production,
both at 8 TeV and at 13 TeV.

In Fig.~\ref{fig:datath_3} we show the corresponding
comparison between experimental data and theory predictions
for the ATLAS (labelled as `A') and CMS (labelled as `C')
measurements of the cross-sections for single top production
in the $s$-channel and in the $tW$ and $tZ$ associated production channels.
We include in this comparison the results for the most updated
measurements both at 8 TeV
and at 13 TeV.
In general, there is good agreement between the
theory calculations and experimental data.
The biggest SMEFT-induced shift is found for the $s-$channel
cross-sections at 8 TeV, where $\delta_{\rm th}\simeq +35\%$.
For single top production in association with a $W$ boson,
there is a negative shift of $\delta_{\rm th}\simeq -6\%$,
similar for the two centre-of-mass energies.
From the comparison of Fig.~\ref{fig:datath_3} we can also observe
how in some cases the SMEFT fit 
interpolates between the ATLAS and CMS measurements, for
instance for the $t+W$ cross-sections at 13 TeV
and the $s-$channel cross-sections at 8 TeV.
     
Considering now the $t\bar{t}V$ processes, in Fig.~\ref{fig:datath_3}
we show the corresponding plot for the measurements of the production
cross-section of a top quark pair associated
with a $W$ or $Z$ vector boson.
We may observe here the origin of the poor agreement of the ATLAS $t\bar{t}Z$
measurement at 8 TeV with the theory prediction after the fit
reported in Table~\ref{tab:chi2table}.
Indeed, we find that for this process
the ATLAS 8 TeV measurement (normalised to the SM prediction) barely
agrees within uncertainties with the corresponding CMS 13 TeV cross-section,
which exhibits the smallest uncertainties and thus dominates in the fit.
For these $t\bar{t}+V$ processes, the SMEFT-induced shifts
are $\delta_{\rm th}\simeq +23\%~(+11\%)$ for $t\bar{t}+W$ at 8 TeV (13 TeV)
and $\delta_{\rm th}\simeq +26\%~(+31\%)$ for $t\bar{t}+Z$ at 8 TeV (13 TeV).
These shifts are rather larger than for the corresponding
inclusive cross-sections shown in Fig.~\ref{fig:datath_1}, as allowed
by the larger experimental uncertainties.

\begin{figure}[t]
  \begin{center}
    \includegraphics[width=0.49\linewidth]{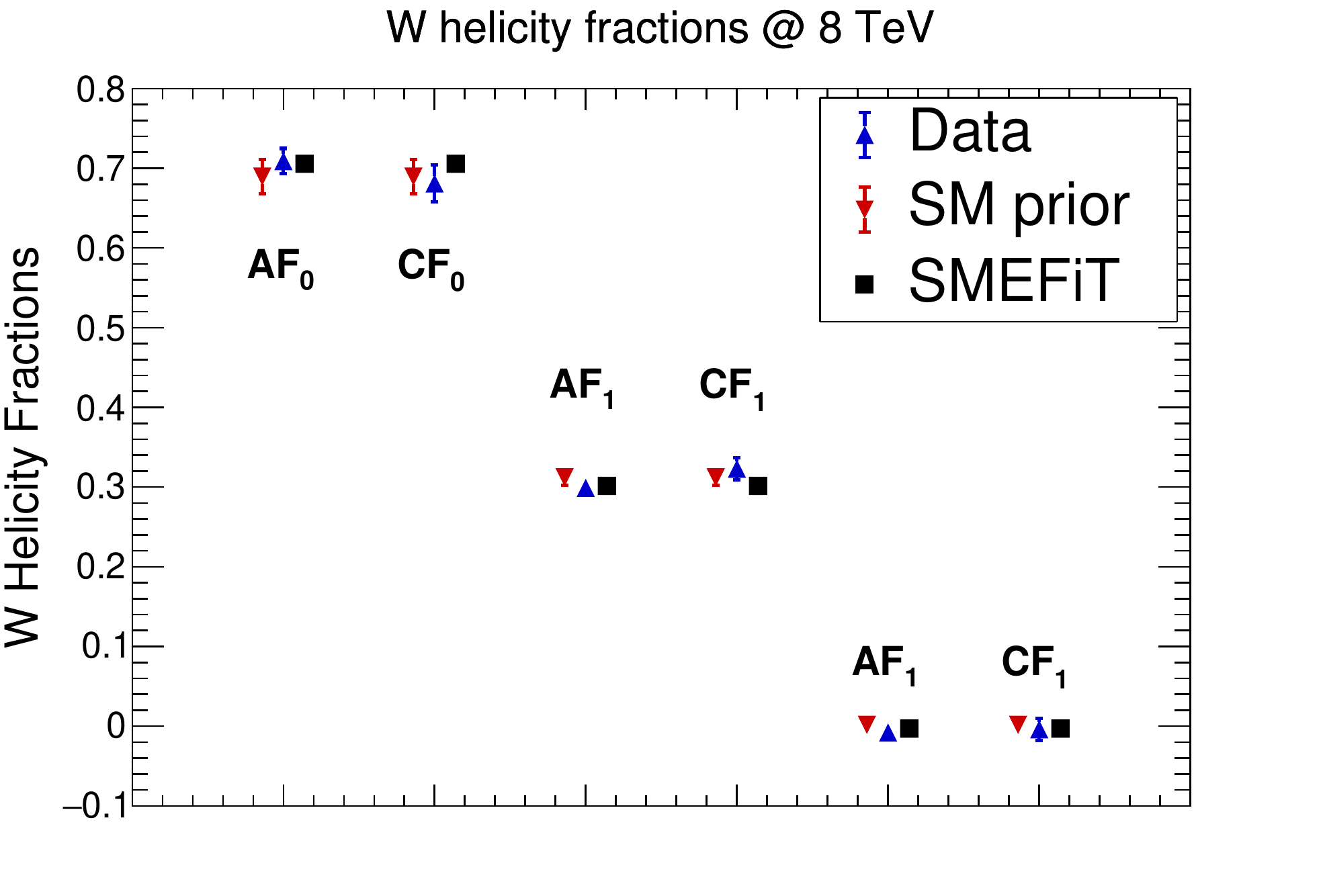}
    \includegraphics[width=0.49\linewidth]{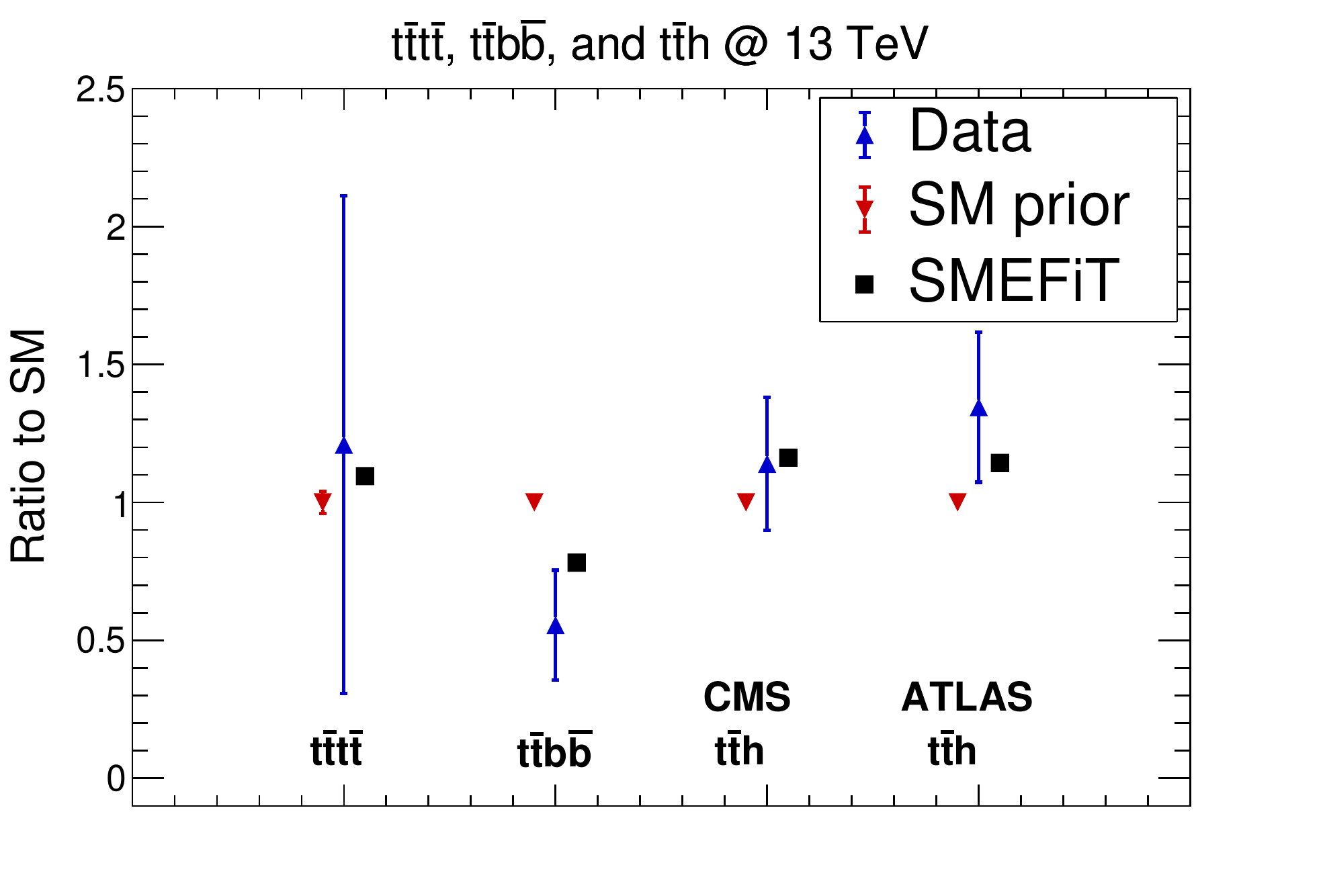} 
      \caption{\small Left: same as Fig.~\ref{fig:datath_1}
        for the $W$ helicity fractions $F_0$, $F_1$, and $F_2$
        from ATLAS and CMS.
        Right: the corresponding comparison between data and theory predictions
        for the CMS measurements of $t\bar{t}b\bar{b}$ and $t\bar{t}t\bar{t}$
        at 13 TeV, as well as for the $t\bar{t}h$ cross-section measurements
        from ATLAS and CMS at 13 TeV.
     \label{fig:datath_4} }
  \end{center}
\end{figure}

Finally, to complete this set of comparisons between the input experimental
data and the corresponding theory calculations before and after the fit,
we show in Fig.~\ref{fig:datath_4}  the $W$ helicity fractions $F_0$, $F_1$, 
and $F_2$ from ATLAS and CMS.
There is good agreement between data and theory, and the $\delta_{\rm th}$ 
shifts are quite small.
In the same figure, we also show the corresponding comparisons
between data and theory predictions
for the CMS measurements of $t\bar{t}b\bar{b}$ and $t\bar{t}t\bar{t}$
at 13 TeV, as well as for the $t\bar{t}H$ cross-section measurements
from ATLAS and CMS at 13 TeV.
Here the SMEFT-induced shifts are larger than for
other processes, and we find $\delta_{\rm th}\simeq +10\%$ for 
$t\bar{t}t\bar{t}$ production,
$\delta_{\rm th}\simeq -21\%$ for $t\bar{t}b\bar{b}$ production,
and $\delta_{\rm th}\simeq +15\%$ for Higgs boson production
in association with a $t\bar{t}$ pair.

As expected from the good agreement between the experimental
data and the theory calculations already at the SM level
reported in Table~\ref{tab:chi2table},
the overall pattern that is observed in these data/theory comparisons
is that the SMEFT-induced shifts are (in relative terms) larger
for observables with larger experimental uncertainties, and smaller
for more precisely measured cross-sections such as in inclusive $t\bar{t}$
production.
In all cases, these shifts $\delta_{\rm th}$ are smaller
or at most comparable
to the corresponding uncertainties of the experimental data.

\subsection{The top quark degrees of freedom of the SMEFT}
\label{sec:resultsbounds}

We now discuss the main results of this work.
In the following, we present the fit results
for the central values $\la c_i\ra$, Eq.~\eqref{eq:meancoefficient},
and the corresponding 95\% CL uncertainties, $\delta c_i$,
for the $N_{\rm op}=34$ dimension-6 SMEFT degrees of freedom relevant
for the interpretation of top quark production measurements at the LHC.
We also study the cross-correlations between these degrees of freedom.
They provide an important piece of information
since we know from the closure tests of Sect.~\ref{sec:closuretest}
that these correlations
might be large because of flat directions in the parameter space.

\begin{figure}[t]
  \begin{center}
\includegraphics[width=0.49\linewidth]{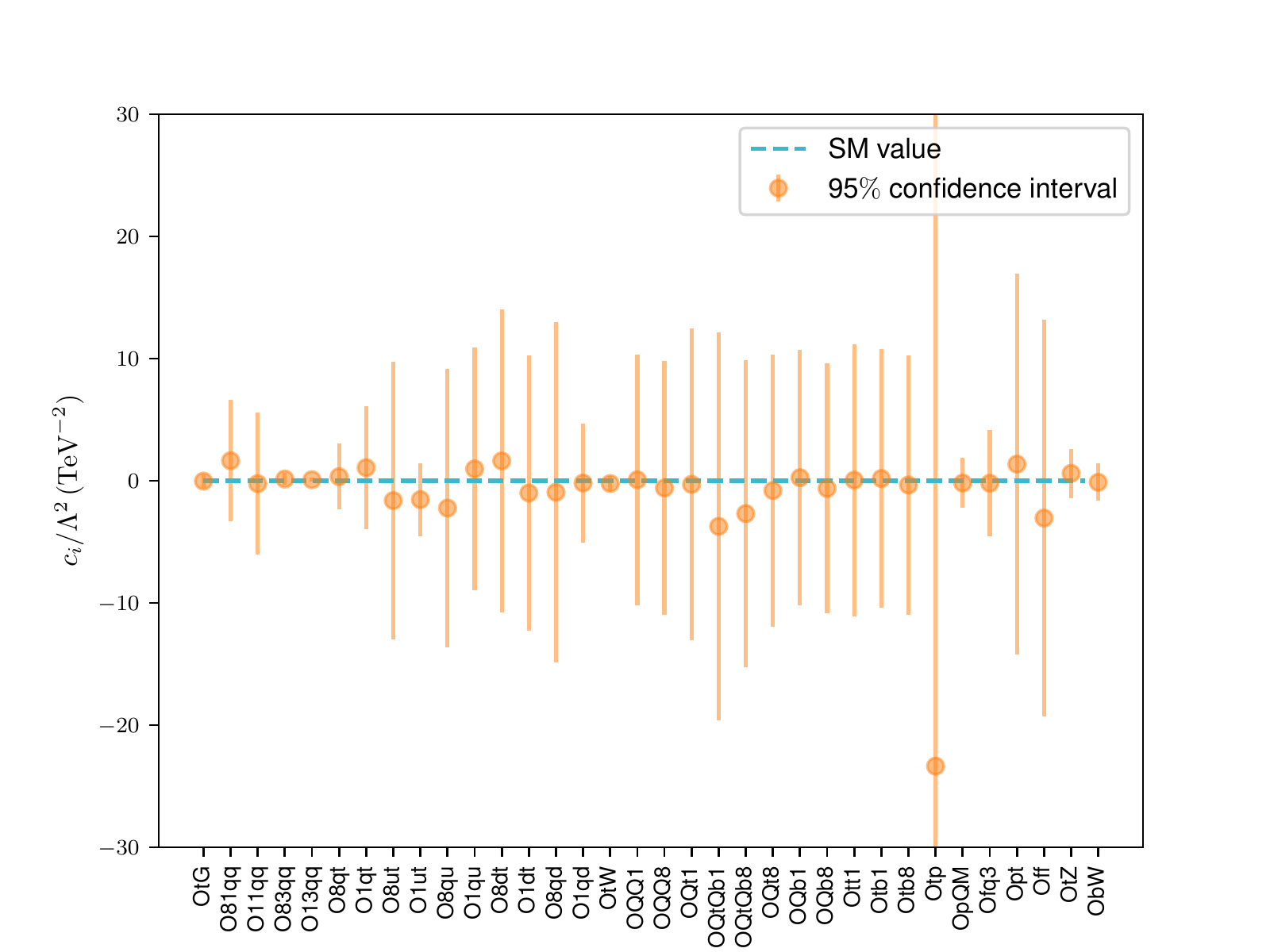}
\includegraphics[width=0.49\linewidth]{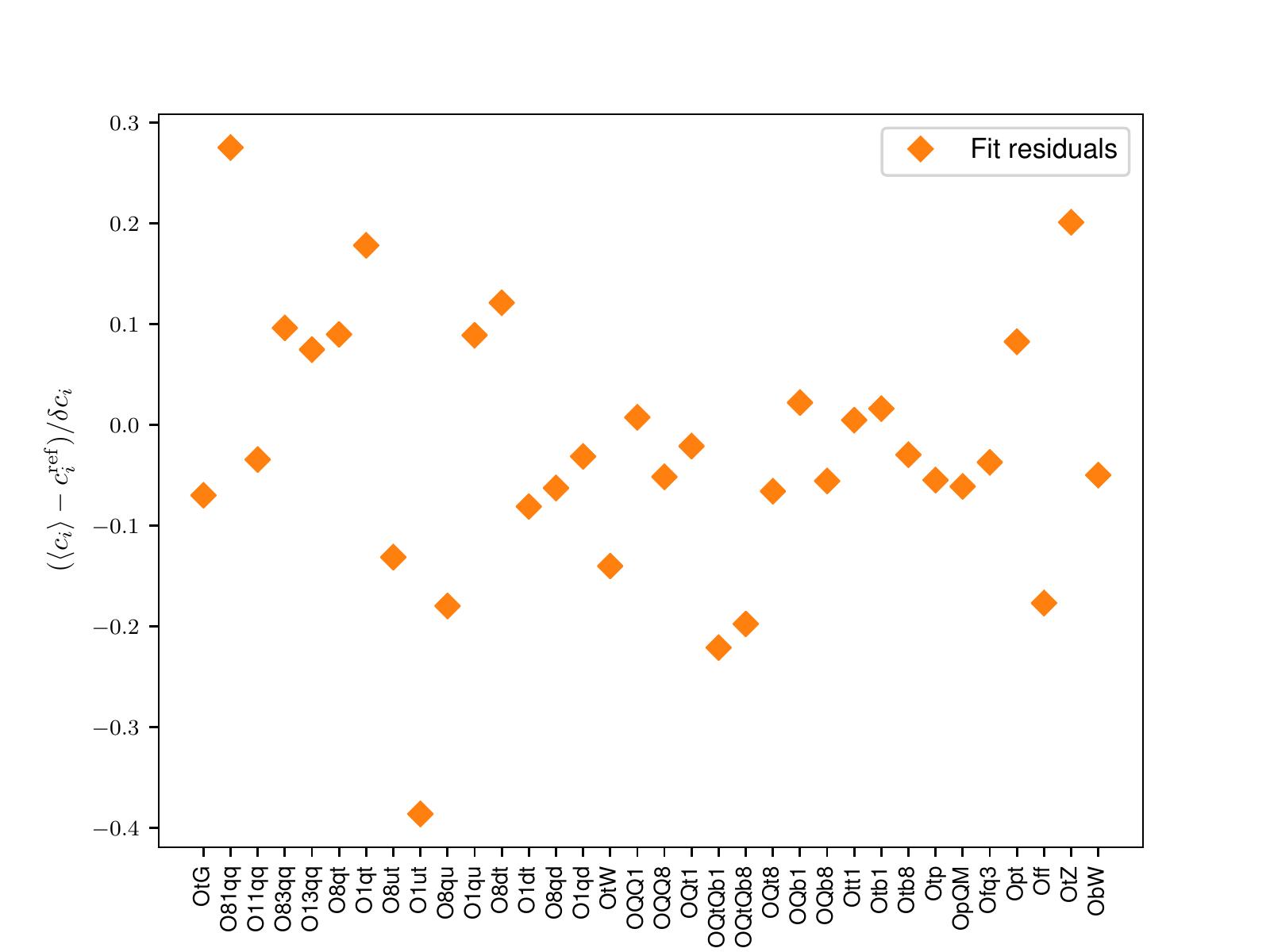}
\caption{\small Left: the best-fit values, $\la c_i \ra$,
  with the corresponding 95\% confidence intervals, $\delta c_i$,
  for the $N_{\rm op}=34$ SMEFT degrees of freedom considered in this
  analysis.
  The dashed blue line indicates the SM, $c_i=0$.
  Right plot: the associated fit residuals, which measure the deviation
  of the fit results with respect to the SM value $c_i^{(\rm ref)}=0$ in units
  of the 95\% CL uncertainties $\delta c_i$.
     \label{fig:SMEFT_std_dev_crossvalON_NLO_HO} }
  \end{center}
\end{figure}

In Fig.~\ref{fig:SMEFT_std_dev_crossvalON_NLO_HO} we display
the best-fit values of the $N_{\rm op}=34$
degrees of freedom, $\la c_i\ra$, together with the corresponding
95\% confidence levels $\delta c_i$.
The dashed line indicates the SM prediction as reference.
As with elsewhere in this work, we show the values
of the degrees of freedom $c_i/\Lambda^{-2}$ in units of TeV$^{-2}$,
which coincide with $c_i$ for $\Lambda=1$ TeV.
In the right panel of the same
figure, we show the associated fit residuals $r_i$, Eq.~(\ref{eq:fitresiduals}),
which measure the deviation
of the fit results with respect to the SM in units
of the 95\% CL $\delta c_i$ uncertainties.

From this comparison, we find that
the fit results are in good agreement with the SM within
uncertainties, the fit residuals satisfying $|r_i|\le 0.4$
for all operators.
Note that the correlations between degrees of freedom imply that
the fluctuations around the best-fit results are in general smaller
as compared to the case in which all operators are completely independent.

From Fig.~\ref{fig:SMEFT_std_dev_crossvalON_NLO_HO}, we also observe that
there is a rather wide range of values for the fit uncertainties $\delta c_i$
obtained for the different degrees of freedom.
For example, a very small uncertainty is found for
the coefficients associated to {\tt OtG} or {\tt O83qq},
while much larger uncertainties are obtained for the fit coefficients
associated to other degrees of freedom, including all the four-heavy-quark
operators, such as {\tt OQQ1}, and for {\tt Otp}.
In most cases, the origin of these differences in the size of the $\delta c_i$ 
uncertainties can be traced back to 
Table~\ref{table:operatorprocess}: different
degrees of freedom are constrained by different processes,
and in each case the available amount of experimental information
varies widely.
For instance, the four-heavy-quark operators are constrained by only
two data points (the $b\bar{b}t\bar{t}$ and $t\bar{t}t\bar{t}$
cross-sections), hence the large uncertainties of the associated coefficients.
Likewise, {\tt Otp} is only constrained from the $t\bar{t}h$ cross-section 
measurements.

The interpretation of the 95\% CL uncertainties shown in 
Fig.~\ref{fig:SMEFT_std_dev_crossvalON_NLO_HO}, requires some care.
The reason is that the available data on
top production at the LHC, summarised in
Tables~\ref{eq:input_datasets}--\ref{eq:input_datasets2},
does not allow us to fully separate all possible independent
directions in the SMEFT parameter space.
As a consequence, as illustrated in Sect.~\ref{sec:closuretest} at the closure
test level,
there will be in general large (anti-)correlations between the fit
parameters, reflecting this degeneracy in the parameter space.
As we will show now, in general more stringent bounds
are obtained if each operator is fitted individually
and the contributions of all other operators are set to zero,
as compared to the bounds obtained in the global fit.

\begin{figure}[t]
  \begin{center}
\includegraphics[width=0.9\linewidth]{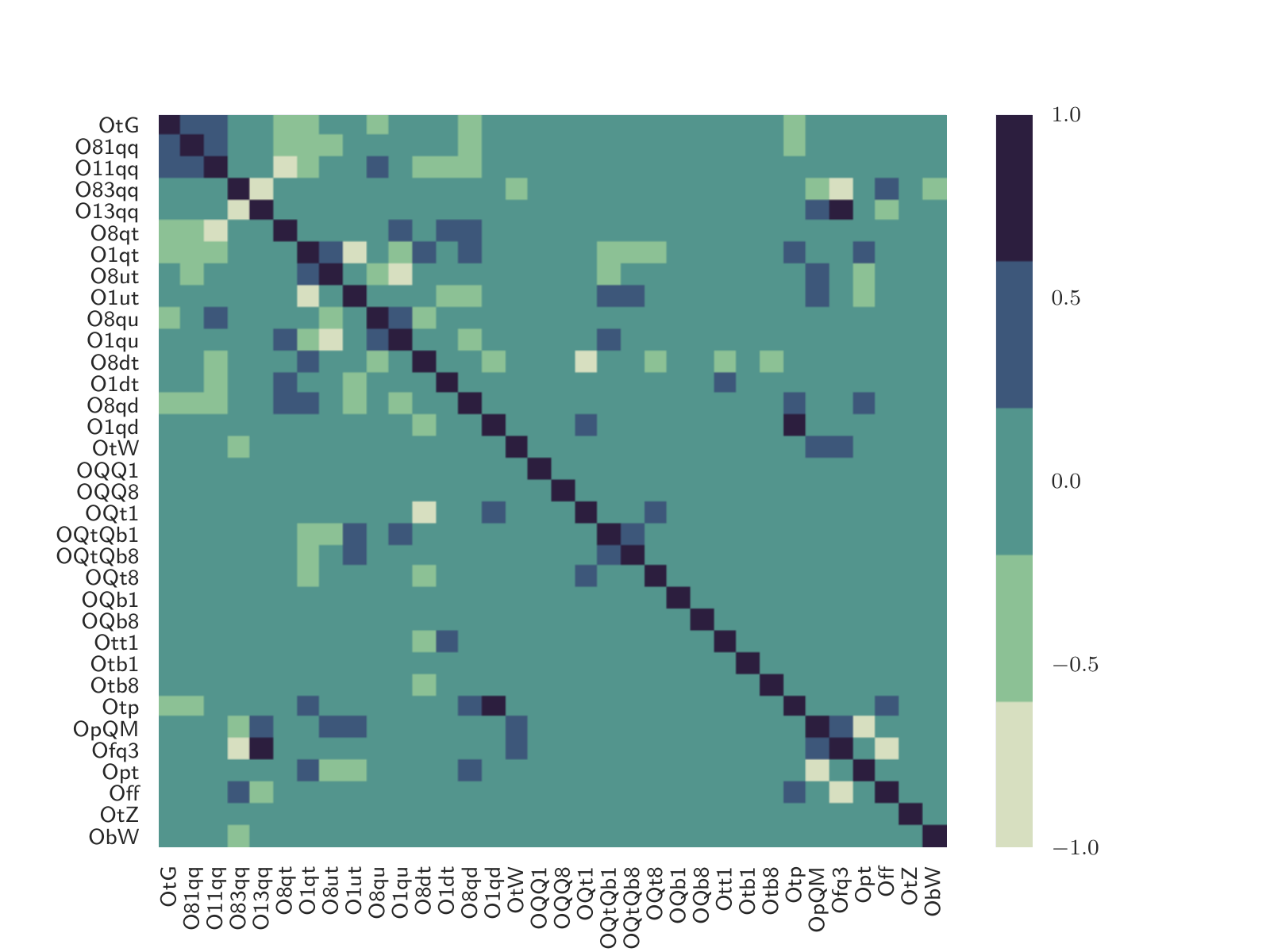}
\caption{\small Heat map indicating the values of the correlation coefficient
  $\rho(c_i,c_j)$
  between the $N_{\rm op}=34$ fitted coefficients
  shown in Fig.~\ref{fig:SMEFT_std_dev_crossvalON_NLO_HO},
  see text for more details.
     \label{fig:heatmap_crossvalON_NLO_HO} }
  \end{center}
\end{figure}

To quantify this point, in Fig.~\ref{fig:heatmap_crossvalON_NLO_HO} we show
a heat map indicating the values of the correlation coefficient,
Eq.~(\ref{eq:correlationL2CT}),
between the $34$ degrees of freedom constrained from the fit.
In this heat map, dark blue regions correspond to degrees of freedom
that are significantly correlated, while light green regions
are instead degrees of freedom that are significantly anti-correlated.
Indeed, we find that specific pairs of coefficients $c_i$ exhibit a
significant amount of (anti-)correlation, such as for instance {\tt O1qd} and
{\tt Otp}.
The effects of such correlations are ignored in fits
where these degrees of freedom are constrained individually rather
than marginalised from the global fit results, and lead
in general to artificially tighter constraints.

Given the overall agreement between the fit results and
the SM, it becomes possible
to interpret the uncertainties $\delta c_i$ as upper bounds on
the parameter space of the SMEFT degrees of freedom.
Such upper bounds provide important information for BSM
model building, since they need to be satisfied
for any UV-complete theory at high energies that has the SM as the low-energy
effective theory.
These bounds can also be compared with previous SMEFT studies of
the top quark sector reported in the literature.
While on the one hand our global SMEFT analysis is based on a wider
LHC dataset than previous analysis of top quark production,
on the other hand it explores a larger parameter space
with reduced model assumptions.
Therefore, a priori, one could either expect stronger
(from the larger dataset) or weaker (from the reduction in model assumptions)
bounds as compared to previous studies: only performing the actual fit itself
can shed light on this question.

In order to compare with previous results, we will follow here the discussion 
in Appendix~A of the Top LHC Working Group EFT 
note~\cite{AguilarSaavedra:2018nen}, to where we direct the reader for further 
details.
We note that the results quoted in~\cite{AguilarSaavedra:2018nen}
are in many cases restricted to fitting one operator at a
time, or at most marginalising over a small subset of operators,
and thus these limits might be too optimistic due to neglecting 
correlations with other directions in the SMEFT parameter space.
We will quote here both the direct limits obtained from the top-quark 
measurements, and the indirect
limits derived from non-top processes such as
low-energy observables, the decays of $B$ mesons,
electroweak precision observables, and Drell-Yan production.
See Sect.~\ref{eq:nontopops} for a related discussion of the existing
experimental constraints on SMEFT degrees of freedom that
do not involve top quarks.

In Table~\ref{eq:tableCoeffBounds} we report
the values of the 95\% confidence level bounds (in units
of TeV$^{-2}$, assuming $\Lambda=1$ TeV)
for the coefficients of the
34 SMEFT degrees of freedom derived from
the marginalisation of the results of the SMEFiT global analysis.
We compare our results with those obtained elsewhere in the literature
either
from the direct analysis of
top quark production (``direct'')  or from indirect bounds
from other processes not involving top quarks (``indirect''). 
We note
that for several degrees of freedom, such as for {\tt Off} and {\tt Otb1},
the bounds reported here have been
obtained for the first time. In Table~\ref{eq:tableCoeffBounds2} 
we additionally show the results for the differing theory settings used in the 
global fit; namely using only
$\mathcal{O}(\Lambda^{-2})$ corrections and LO QCD in the SMEFT calculations.

As recommended in~\cite{AguilarSaavedra:2018nen}, it is important
to also quote the bounds derived from fitting individual
coefficients, one at a time, in order
to compare them with the global fit results.
The results from such single-operator fits are provided
in Table~\ref{eq:tableCoeffBoundsIndividual}
using the same settings as in the baseline global fit (as well
as by varying the theory settings, see the discussion in 
Sect.~\ref{sec:perturbativeinstability}).
In the case of the individual fits of the operators
that are very loosely constrained (in particular, for most of the 
four-heavy-quark
degrees of freedom) we find that the SMEFiT approach is affected
by numerical stability issues.
Therefore, for such operators (identified in italics), it is more
reliable to quote instead the 95\% CL
bounds obtained from the analytical minimisation of the $\chi^2$, which
for these cases has a relatively simple form.

By comparing the bounds obtained in the global and individual
fits, Tables~\ref{eq:tableCoeffBounds}, \ref{eq:tableCoeffBounds2} 
and~\ref{eq:tableCoeffBoundsIndividual}
respectively, one finds that for essentially all degrees of freedom
the bounds obtained
from the individual fits are either more stringent than
or comparable to the marginalised results from the global fit.
As discussed above, the reason for this can be traced back to
the fact that within the single-operator fits one is
neglecting cross-correlations between the different
directions spanned by the fitted degrees of freedom.
For instance, the 95\% CL bound associated to {\tt OtG}
is $[-0.4,+0.4]$ in the global fit, while it is $[-0.08,+0.03]$
if the corresponding coefficient is fitted individually.
Another example is {\tt Otp}, whose bound is $[-60,+10]$ in the global fit,
and $[-5.3,+1.6]$ in the individual fit, i.e.~it is more stringent by about 
an order of magnitude.

Another important advantage of providing the results for the individual
operators is that it allows us to better assess the impact that varying the 
theory settings has on the fit results.
For instance, as we will discuss in Sect.~\ref{sec:perturbativeinstability},
accounting for the quadratic $\mathcal{O}(\Lambda^{-4})$ terms
leads to an improvement in the bounds of most operators, but assessing
this effect is more transparently done in the case of the individual
than in the global fits, where one has additional factors
to take into account in the interpretation of the results.

\begin{table}[p]
  \centering
  \small
   \renewcommand{\arraystretch}{1.33}
        \begin{tabular}{l|C{1.4cm}||C{2.3cm}||C{3.0cm}|C{3.0cm}}
    \multirow{2}{*}{Notation} &
    \multirow{2}{*}{DoF}   &  {\bf SMEFiT}  & Direct  &  Indirect  \\
     &      &  {(marginalised)}  & (other)  &  (other)  \\
    \toprule
        {\tt OQQ1} &$\ccc{1}{QQ}{}$&    $ [ -9.4 , 9.4 ] $   &    &        \\
        {\tt OQQ8} &$\ccc{8}{QQ}{}$&       $[ -10 , 9.4 ]  $    &   &  \\
        {\tt OQt1} &$\ccc{1}{Qt}{}$&       $ [ -13 , 12 ] $         & $[-5.0,4.9]$~\cite{Zhang:2017mls}   &  \\
        {\tt OQt8} &$\ccc{8}{Qt}{}$&         $ [ -12 , 10 ] $  	&  $[-10.3,9.3]$~\cite{Zhang:2017mls}  &  \\
        {\tt OQb1} &$\ccc{1}{Qb}{}$&       $ [ -9.7 , 9.7 ] $  	 &     &  \\
        {\tt OQb8} &$\ccc{8}{Qb}{}$&       $[ -9.8 , 9.3 ]  $   	  &     &  \\
        {\tt Ott1} &$\ccc{1}{tt}{}$&   $ [ -11 , 11 ]  $  	 & $[-2.9,2.8]$~\cite{Zhang:2017mls}     &  \\
        {\tt Otb1} &$\ccc{1}{tb}{}$&  $  [ -9.5 , 9.9 ] $  &      &  \\
        {\tt Otb8} &$\ccc{8}{tb}{}$&   $  [ -10 , 9.7 ]  $ 	  &      &  \\
        {\tt OQtQb1} &$\ccc{1}{QtQb}{}$&     $ [ -25 , 8.9 ]   $	 &  &  \\
        {\tt OQtQb8} &$\ccc{8}{QtQb}{}$&      $  [ -18 , 8.6 ]  $ 	&  &  \\
        \midrule
            {\tt  O81qq }  &  	$\ccc{1,8}{Qq}{}$  & $ [ -4.7 , 7.8 ]  $ 	 & $[-6.9, 4.9]$~\cite{Zhang:2017mls}     &  \\
            {\tt O11qq }  &  $\ccc{1,1}{Qq}{} $   &  $ [ -6.8 , 7.4 ] $ 	& $[-3.1, 3.2]$~\cite{Zhang:2017mls}     & \\
            {\tt O83qq }  &  $\ccc{3,8}{Qq}{} $  & $  [ -1.3 , 1.6 ]  $ 	 &  $[-6.1, 6.7]$~\cite{Zhang:2017mls}    &  \\
            {\tt  O13qq }  &  $\ccc{3,1}{Qq}{} $  & $ [ -1.1 , 1.3 ] $ 		 & $[-0.7, 1.2]$~\cite{Buckley:2015lku}     &   \\
            {\tt O8qt }  &   $\ccc{8}{tq}{}$    & $ [ -3.7 , 4.1 ] $ 	 &  $[-6.8, 3.5]$~\cite{Zhang:2017mls}     & \\
            {\tt O1qt }  & $\ccc{1}{tq}{} $    &  $ [ -5.3 , 7.5 ] $ 	&  $[-2.8, 2.8]$~\cite{Zhang:2017mls}     &  \\
            {\tt O8ut }  &  $ \ccc{8}{tu}{}$    &  $ [ -14 , 10 ]  $ 		&  $[-8.0, 4.8]$~\cite{Zhang:2017mls}     & \\
            {\tt O1ut }  &  $\ccc{1}{tu}{} $    & $ [ -5.8 , 2.6 ] $ 		&   $[-3.6, 3.5]$~\cite{Zhang:2017mls}   & \\
            {\tt O8qu }  & 	$\ccc{8}{Qu}{} $    &  $ [ -15 , 9.1 ]  $ 		& $[-8.1, 4.0]$~\cite{Zhang:2017mls}     & \\
            {\tt O1qu }  & 	$\ccc{1}{Qu}{} $    & $ [ -9.8 , 12 ]  $ 	 &  $[-3.3, 3.4]$~\cite{Zhang:2017mls}    & \\
            {\tt O8dt }  &   $	\ccc{8}{td}{} $    &  $ [ -9.5 , 17 ]   $ 		 &  $[-12, 9.3]$~\cite{Zhang:2017mls}    &  \\
            {\tt O1dt }  &  $	\ccc{1}{td}{}$   &  $ [ -13 , 10 ]   $		 &   $[-4.9, 5.0]$~\cite{Zhang:2017mls}    &  \\
            {\tt O8qd }  &  $\ccc{8}{Qd}{}  $   & $  [ -14 , 11 ]   $ 		 & $[-11.8, 9.4]$~\cite{Zhang:2017mls}      & \\
            {\tt O1qd}   & $\ccc{1}{Qd}{}  $    &  $[ -3.5 , 2.6 ]   $ 		 & $[-5.0, 5.0]$~\cite{Zhang:2017mls}      & \\
        \midrule
            {\tt OtG }    &  $c_{tG}$   	&   $ [ -0.4 , 0.4 ] $   	 &  $[-1.3,1.2]$~\cite{Buckley:2015lku}    & \\
	    {\tt  OtW}  &  $\ccc{}{tW}{}$		& $ [ -1.8 , 0.9 ] $   	& $[-4.0,3.5]$~\cite{Buckley:2015lku}    &  $[-2.8,2.0]$ (EW)   \\
  {\tt ObW} &$\ccc{}{bW}{}$ &    $ [ -2.6 , 3.1 ]  $  		  &      &  $[-15,37]$ (EW)    \\
  {\tt OtZ} &$\ccc{}{tZ}{}$ &  $ [ -2.1 , 4.0 ] $   		 &  $\ccc{}{tB}{}$: $[-4.1,4.1]$~\cite{Englert:2017dev}     &
 $\ccc{}{tB}{}$:~$[-5.8,15.4]$~(EW)  \\
  {\tt Off} &$\ccc{}{\varphi tb}{}$ &     $ [ -27 , 8.7 ] $ 		 &      &   \\
  {\tt  Ofq3}  & $\ccc{3}{\varphi Q}{}$  &  $ [ -5.5 , 5.8 ]   $  	  & $[-4.1, 2.0]$~\cite{Buckley:2015lku}     &    \\
  {\tt OpQM} &$\ccc{-}{\varphi Q}{}$ &  $ [ -3.5 , 3 ] $   		 &  $\ccc{1}{\varphi Q}{}$: $[-3.1, 3.1]$~\cite{Buckley:2015lku}
  &  $[-3.4,7.4]$ (EW)  \\
  {\tt Opt} &$\ccc{}{\varphi t}{}$ &  $ [ -13 , 18 ] $  		 &  $[-9.7,8,3]$~\cite{Buckley:2015lku}    &
  $[-2.0,5.6]$ (EW)\\
  {\tt Otp} & $\ccc{}{t\varphi}{}$&   $ [ -60 , 10 ] $ 	  &      &    \\
        \bottomrule
  \end{tabular}
  \caption{\label{eq:tableCoeffBounds}
    \small The 95\% confidence level bounds (in units
    of TeV$^{-2}$, assuming $\Lambda=1$ TeV)
    for the coefficients of the
    34 SMEFT degrees of freedom derived from the present analysis
    from the marginalisation of the global fit results.
    We also quote results obtained elsewhere 
    from the direct analysis of
    top quark production  and from indirect bounds from processes
    not involving top quarks.
}
\end{table}


\begin{table}[p]
  \centering
  \small
  \renewcommand{\arraystretch}{1.35}
  \begin{tabular}{l|C{1.4cm}||C{2.3cm}|C{2.3cm}|C{2.3cm}}
    \multicolumn{5}{c}{\bf SMEFiT global analysis (marginalised)}\\[0.2cm]
    \toprule
Notation &
DoF   &  Baseline  & $\mathcal{O}(\Lambda^{-2})$ only  &  LO QCD  \\
    \toprule
        {\tt OQQ1} &$\ccc{1}{QQ}{}$&    $ [ -9.4 , 9.4 ] $   &  $* $  &
        $[ -9.6, 9.5 ]$\\
        {\tt OQQ8} &$\ccc{8}{QQ}{}$&       $[ -10 , 9.4 ]  $    &  $*$ &
        $[ -9.9 , 9.4 ]$\\
        {\tt OQt1} &$\ccc{1}{Qt}{}$&       $ [ -13 , 12 ] $         & $* $    &
        $[ -26 , 12 ] $\\
        {\tt OQt8} &$\ccc{8}{Qt}{}$&         $ [ -12 , 10 ] $  	& $*$  &
        $ [ -20 , 10 ]$\\
        {\tt OQb1} &$\ccc{1}{Qb}{}$&       $ [ -9.7 , 9.7 ] $  	 & $*$    &
        $[ -9.7 , 9.8 ] $\\
        {\tt OQb8} &$\ccc{8}{Qb}{}$&       $[ -9.8 , 9.3 ]  $   	  &
        $* $& $ [ -9.7 , 9.2 ]$  \\
        {\tt Ott1} &$\ccc{1}{tt}{}$&   $ [ -11 , 11 ]  $  	 &
        $*$& $ [ -17 , 12 ]$ \\
        {\tt Otb1} &$\ccc{1}{tb}{}$&  $  [ -9.5 , 9.9 ] $  &
        $* $& $ [ -9.6 , 10 ]$ \\
        {\tt Otb8} &$\ccc{8}{tb}{}$&   $  [ -10 , 9.7 ]  $ 	  &  $* $    &
        $[ -11 , 9.8 ] $\\
        {\tt OQtQb1} &$\ccc{1}{QtQb}{}$&     $ [ -25 , 8.9 ]   $	 &
        $* $& $[ -19 , 9.2 ] $ \\
        {\tt OQtQb8} &$\ccc{8}{QtQb}{}$&      $  [ -18 , 8.6 ]  $ 	&
        $*$& $ [ -15 , 9.0 ] $ \\
        \midrule
            {\tt  O81qq }  &  	$\ccc{1,8}{Qq}{}$  & $ [ -4.7 , 7.8 ]  $ 	 &
            $[ -7.0 , 12 ]$& $ [ -7.5 , 7.6 ]  $ \\
            {\tt O11qq }  &  $\ccc{1,1}{Qq}{} $   &  $ [ -6.8 , 7.4 ] $ 	&
            $* $& $[ -17 , 7.5 ]$ \\
            {\tt O83qq }  &  $\ccc{3,8}{Qq}{} $  & $  [ -1.3 , 1.6 ]  $ 	 &
            $[ -7.7 , 9.1 ] $& $[ -0.8 , 1.3 ] $  \\
            {\tt  O13qq }  &  $\ccc{3,1}{Qq}{} $  & $ [ -1.1 , 1.3 ] $ 		 &
            $[ -0.5 , 0.6 ] $& $ [ -1 , 0.8 ]$  \\
            {\tt O8qt }  &   $\ccc{8}{tq}{}$    & $ [ -3.7 , 4.1 ] $ 	 &
            $[ -10 , 8.1 ]$& $[ -3.2 , 6.8 ]$ \\
            {\tt O1qt }  & $\ccc{1}{tq}{} $    &  $ [ -5.3 , 7.5 ] $ 	&
            $*$& $[ -6.1 , 15 ] $ \\
            {\tt O8ut }  &  $ \ccc{8}{tu}{}$    &  $ [ -14 , 10 ]  $ 		&
            $ [ -13 , 9.3 ] $&  $[ -22 , 9.4 ] $\\
            {\tt O1ut }  &  $\ccc{1}{tu}{} $    & $ [ -5.8 , 2.6 ] $ 		&
            $* $& $[ -8.8 , 17 ]$ \\
            {\tt O8qu }  & 	$\ccc{8}{Qu}{} $    &  $ [ -15 , 9.1 ]  $ 		&
            $[ -15 , 8.5 ]$& $[ -13 , 16 ]$ \\
            {\tt O1qu }  & 	$\ccc{1}{Qu}{} $    & $ [ -9.8 , 12 ]  $ 	 &
            $*$& $[ -14 , 8.8 ] $ \\
            {\tt O8dt }  &   $	\ccc{8}{td}{} $    &  $ [ -9.5 , 17 ]   $ 	&
            $[ -9.5 , 13 ]$&  $[ -9.5 , 45 ] $ \\
            {\tt O1dt }  &  $	\ccc{1}{td}{}$   &  $ [ -13 , 10 ]   $		 &
            $* $& $[ -18 , 9.0 ]$ \\
            {\tt O8qd }  &  $\ccc{8}{Qd}{}  $   & $  [ -14 , 11 ]   $ 		 &
            $[ -15 , 11 ]$& $[ -38 , 16 ] $ \\
            {\tt O1qd}   & $\ccc{1}{Qd}{}  $    &  $[ -3.5 , 2.6 ]   $ 		 &
            $*$& $ [ -1.9 , 5.0 ] $ \\
        \midrule
            {\tt OtG }    &  $c_{tG}$   	&   $ [ -0.4 , 0.4 ] $   	 &
            $[ -0.4 , 0.4 ] $& $[ -0.4 , 0.4 ]$\\
            {\tt  OtW}  &  $\ccc{}{tW}{}$		& $ [ -1.8 , 0.9 ] $   	&
            $[ -0.8 , 1.1 ] $&  $[ -2.0 , 1.0 ] $   \\
            {\tt ObW} &$\ccc{}{bW}{}$ &    $ [ -2.6 , 3.1 ]  $  		  & *    &
            $[ -1.3 , 4.8 ]$\\
  {\tt OtZ} &$\ccc{}{tZ}{}$ &  $ [ -2.1 , 4.0 ] $   		 &    $[ -14 , 8.0 ]$   &
 $[ -4.6 , 5.9 ]$ \\
    {\tt Off} &$\ccc{}{\varphi tb}{}$ &     $ [ -27 , 8.7 ] $ 		 &
   * & $[ -8.9 , 7.0 ] $  \\
    {\tt  Ofq3}  & $\ccc{3}{\varphi Q}{}$  &  $ [ -5.5 , 5.8 ]   $  	  &
    $[ -2.7 , 2.2 ]$& $[ -8.9 , 7.0 ]$    \\
    {\tt OpQM} &$\ccc{-}{\varphi Q}{}$ &  $ [ -3.5 , 3 ] $   		 &
    $ [ -6.8 , 11 ]$&  $[ -2.6 , 3.3 ] $   \\
    {\tt Opt} &$\ccc{}{\varphi t}{}$ &  $ [ -13 , 18 ] $  		 &
    $ [ -9.7 , 20 ] $& $[ -23 , 7.3 ]$        \\
    {\tt Otp} & $\ccc{}{t\varphi}{}$&   $ [ -60 , 10 ] $ 	  &
    $[ -8.1 , 5.0 ] $ &  $[ -9.5 , 11 ]$   \\
        \bottomrule
  \end{tabular}
  \caption{\label{eq:tableCoeffBounds2}
    \small Same as Table~\ref{eq:tableCoeffBounds} for the
    95\% CL bounds obtained in the global fit, now comparing
    the results obtained using the baseline theory settings with
    those obtained when  only the linear $\mathcal{O}\lp \Lambda^{-2}\rp$
    terms are included and when only LO QCD calculations are used
    for the SMEFT contribution.
}
\end{table}


\begin{table}[p]
  \centering
  \small
  \renewcommand{\arraystretch}{1.35}
  \begin{tabular}{l|C{1.4cm}||C{2.3cm}|C{3.5cm}|C{2.3cm}}
    \multicolumn{5}{c}{\bf SMEFiT individual bounds (single-operator fits)}\\
      \toprule
Notation &
DoF   &  Baseline  & $\mathcal{O}(\Lambda^{-2})$ only  &  LO QCD  \\
    \toprule
        {\tt OQQ1} &$\ccc{1}{QQ}{}$&  $\mathit{[ -5.2,4.9]}$    & $\mathit{[-54,83]}$   &   $\mathit{[-5.4,5.2]}$     \\
        {\tt OQQ8} &$\ccc{8}{QQ}{}$&  $\mathit{[ -14,12]}$         &  $\mathit{[-200,18]}$ & $\mathit{[-21,16]}$   \\
        {\tt OQt1} &$\ccc{1}{Qt}{}$&     $\mathit{[ 4.5,4.5 ]}$         &   $\mathit{[-610,210]}$ &  $\mathit{[-4.9,4.9]}$  \\
        {\tt OQt8} &$\ccc{8}{Qt}{}$&    $\mathit{[-10,8.1]}$    	&  $\mathit{[-69,28]}$ & $\mathit{[-11,8.7]}$   \\
        {\tt OQb1} &$\ccc{1}{Qb}{}$&   $\mathit{[6.9,6.7]}$    	 &  $\mathit{[-1.9\,10^3,-110]}$   & $\mathit{[-6.1,6.0]}$   \\
        {\tt OQb8} &$\ccc{8}{Qb}{}$&    $\mathit{[-16,12]}$    	  & $\mathit{[-260,-14]}$    & $\mathit{[-15,11]}$   \\
        {\tt Ott1} &$\ccc{1}{tt}{}$&   $\mathit{[-2.9,2.7]}$	 &  $\mathit{[-26,41]}$  &$\mathit{[-3.4,3.2]}$  \\
        {\tt Otb1} &$\ccc{1}{tb}{}$& $\mathit{[-6.8,6.8]}$  & $\mathit{[-2.1\,10^4,-1.4\,10^3]}$     & $\mathit{[-6.1,6.1]}$   \\
        {\tt Otb8} &$\ccc{8}{tb}{}$&  $\mathit{[-17,12]}$	  & $\mathit{[-270,-15]}$     & $\mathit{[-15,11]}$   \\
        {\tt OQtQb1} &$\ccc{1}{QtQb}{}$&   $\mathit{[-5.4,5.5]}$ 	 &  $\mathit{[160,2.8\,10^3]}$&  $\mathit{[-4.8,4.9]}$  \\
        {\tt OQtQb8} &$\ccc{8}{QtQb}{}$&  $\mathit{[-14,14]}$   	&  $\mathit{[910,1.6\,10^4]}$&   $\mathit{[-13,13]}$ \\
        \midrule
            {\tt  O81qq }  &  	$\ccc{1,8}{Qq}{}$  & $[ -0.6 , 0.1 ]$	 & $ [ -1.2 , 0.3 ]$ & $[ -0.6 , 0.07 ]$ \\
            {\tt O11qq }  &  $\ccc{1,1}{Qq}{} $   &  $[ -0.2 , 0.02 ]$	&  $*$   &$ [ -0.2 , 0.03 ]$ \\
            {\tt O83qq }  &  $\ccc{3,8}{Qq}{} $  & $[ -0.5 , 0.4 ]$	 &  $[ -3.3 , -0.08 ]$  &$ [ -0.7 , 0.2 ]$ \\
            {\tt  O13qq }  &  $\ccc{3,1}{Qq}{} $  &	$ [ -0.1 , 0.09 ]$ &  $[ -0.1 , 0.2 ]$  & $ [ -0.1 , 0.09]$  \\
            {\tt O8qt }  &   $\ccc{8}{tq}{}$    & $[ -1.3 , 0.4 ]$ &    $[ -2.1 , 1.5 ]$   &$[ -0.7 , 0.09 ]$ \\
            {\tt O1qt }  & $\ccc{1}{tq}{} $    &  $[ -0.3 , 0.02 ]$	&   $*$  & $ [ -0.3 , 0.03 ]$ \\
            {\tt O8ut }  &  $ \ccc{8}{tu}{}$    & 	$[ -1.1 , 0.04 ]$	&  $[ -2.0 , 0.09]$   &$[ -0.9 , 0.03 ]$ \\
            {\tt O1ut }  &  $\ccc{1}{tu}{} $    & 	$ [ -0.2 , 0 ]$	&  $*$    &$ [ -0.4 , 0.03 ]$ \\
            {\tt O8qu }  & 	$\ccc{8}{Qu}{} $    & 	$[ -2.6 , 0.2 ]$	& $[ -4.4 , 0.3 ]$    & $[-2.6,0.1]$\\
            {\tt O1qu }  & 	$\ccc{1}{Qu}{} $    &$ [ -0.5 , 0.02 ]$ 	 &  $*$   & $	 [ -0.4 , 0.03 ]$\\
            {\tt O8dt }  &   $	\ccc{8}{td}{} $    &  	$[ -2.5 , -0.01 ]$ &   $[ -4.6 , -0.2 ]$  & $ [ -1.6 , 0.02 ]$ \\
            {\tt O1dt }  &  $	\ccc{1}{td}{}$   & $[ -0.8 , 0 ]$	 &     $*$  &$[ -0.6 , 0.03 ]$  \\
            {\tt O8qd }  &  $\ccc{8}{Qd}{}  $   & $[ -2.7 , 0.3 ]$		 &       $ [ -3.7 , 0.9 ]$& $ [ -1.9 , 0.07 ]$\\
            {\tt O1qd}   & $\ccc{1}{Qd}{}  $    &  	$[ -0.9 , -0.01]$	 &   $*$  &$[ -0.9 , 0.05 ]$ \\
        \midrule
            {\tt OtG }    &  $c_{tG}$   	&   $[ -0.08 , 0.03 ]$ 	 &   $[ -0.08 , 0.03 ]$  &$	 [ -0.1 , 0.04 ]$ \\
  {\tt  OtW}  &  $\ccc{}{tW}{}$		&$[ -0.4 , 0.2 ]$	&$[ -0.3 , 0.1 ]$ &  $ [ -0.4 , 0.2]$ \\
  {\tt ObW} &$\ccc{}{bW}{}$ &   $[ -0.6 , 0.2 ]$		  &  $* $   &  $ [ -0.7 , 0.2 ]$  \\
  {\tt OtZ} &$\ccc{}{tZ}{}$ &  	$[ -2.8, 4.5  ]$	 &   $[ -17 , 4.6 ]$  & $[ -6.3 , 7.4 ]$   \\
  {\tt Off} &$\ccc{}{\varphi tb}{}$ &  $[ -9.4 , 9.5 ]$   	 & $*$    & $[ -9.7, 9.8 ]$  \\
  {\tt  Ofq3}  & $\ccc{3}{\varphi Q}{}$  &   $[ -0.9 , 0.6 ]$	  &$ [ -1.0 , 0.6 ]$    &   $[ -1.0 , 0.6 ]$ \\
  {\tt OpQM} &$\ccc{-}{\varphi Q}{}$ &$[ -4.2 , 3.9 ]$		 &$[ -4.2 , 3.8 ]$  &$[ -5.1 , 4.6 ]$   \\
  {\tt Opt} &$\ccc{}{\varphi t}{}$ &  $ [ -6.4 , 7.3]$		 &  $ [ -6.9, 7.8 ]$  &  $[ -7., 8.0 ]$ \\
  {\tt Otp} & $\ccc{}{t\varphi}{}$&   $[ -5.3 , 1.6 ]$	  &    $ [ -5.1, 1.6 ]$  &  $[ -5.4 , 1.6 ]	$   \\
        \bottomrule
  \end{tabular}
  \caption{\label{eq:tableCoeffBoundsIndividual}
    \small Same as Table~\ref{eq:tableCoeffBounds2}, now for the results of
    individual fits when only one operator is constrained at a time.
    The bounds in italics have been obtained from the analytical minimisation
    of the $\chi^2$ rather than
    using the SMEFiT numerical approach, see text for more details.}
    \end{table}

\begin{figure}[t]
  \begin{center}
\includegraphics[width=0.99\linewidth]{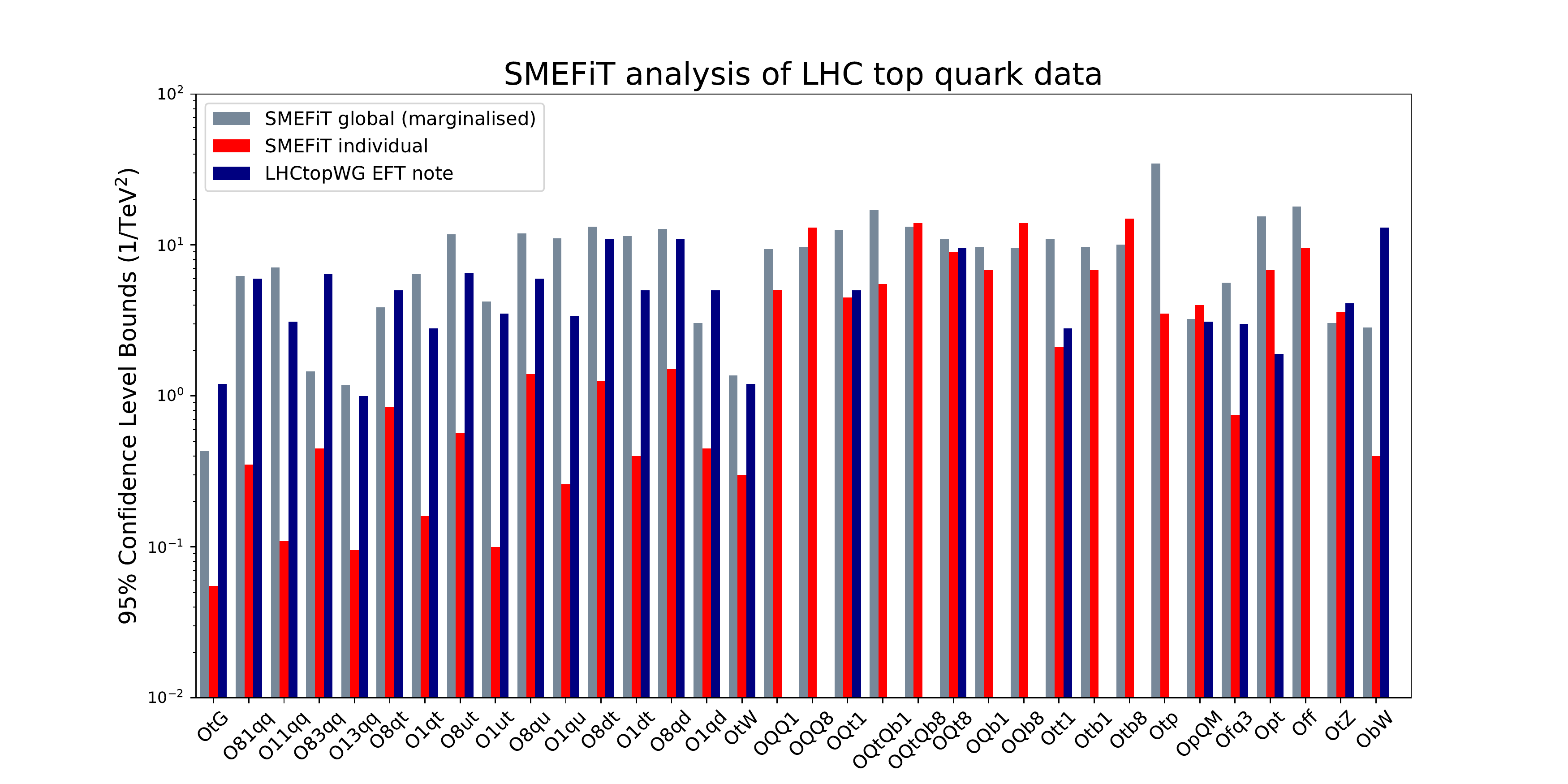}
    \caption{\small Graphical representation of the results of 
      Tables~\ref{eq:tableCoeffBounds}
      and~\ref{eq:tableCoeffBoundsIndividual},
      where we compare the 95\% CL bounds on the 34 degrees of freedom
      included the present analysis, both in the marginalised (global) 
      and in the individual fit cases, with the
      bounds reported in the LHC Top WG EFT note~\cite{AguilarSaavedra:2018nen}.
      \label{fig:SMEFiT-bounds}
  }
  \end{center}
\end{figure}

The graphical representation of the comparison between
the global fit results and the bounds reported in
the LHC top WG EFT note (Table~\ref{eq:tableCoeffBounds}), as well as with
the individual fit results (Table~\ref{eq:tableCoeffBoundsIndividual}),
is shown in Fig.~\ref{fig:SMEFiT-bounds}.
For the purposes of visualisation only, we have symmetrised the bounds
reported there, that is, if a given operator has a 95\% CL bound of
$\lc \delta c_i^{(\rm min)},\delta c_i^{(\rm max)}\rc$, then we show
\be
\overline{\delta c_i}
\equiv
\lp\delta c_i^{(\rm max)} - \delta c_i^{(\rm min)}\rp/2 \, .
\ee
We find that for some of the fitted degrees of freedom our bounds are stronger 
than those reported in previous studies, in some cases such as for {\tt ObW} by 
nearly one order of magnitude.
Another example is provided by the chromomagnetic operator {\tt OtG},
for which the bound found in this work,
$[-0.4,+ 0.4]$, is improved by a factor of three
as compared to the bound quoted in the Top WG EFT note, $[-1.3, +1.2]$.
From this comparison one can also appreciate how the individual bounds
are in general rather tighter than the marginalised ones, except
for some of the four-heavy-quark operators (and for {\tt OtZ}) where
they are instead comparable.

\begin{figure}[t]
  \begin{center}
    \includegraphics[scale=0.55,angle=-90]{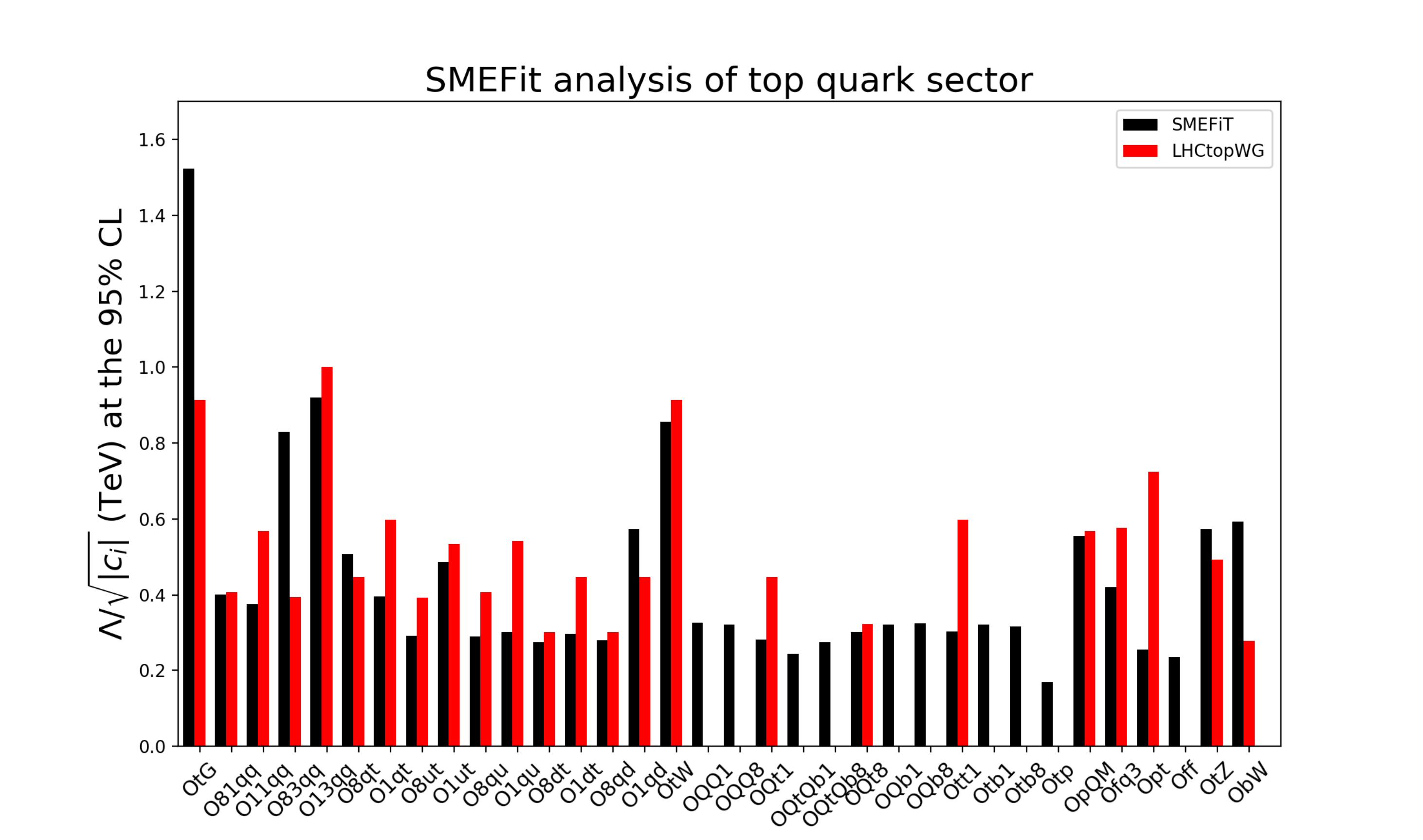}
    \caption{\small Same as Fig.~\ref{fig:SMEFiT-bounds}, now representing
      the marginalised bounds obtained from the global fit as bounds on
      $\Lambda/\sqrt{|c_i|}$.
     \label{fig:SMEFiT-bounds2}
  }
  \end{center}
\end{figure}

Another useful way to present our results is by representing the bounds
on $\Lambda/\sqrt{|c_i|}$ that are derived from the fit.
This is interesting because, assuming UV completions where the values of the 
fitted degrees of freedom $c_i$ are $\mathcal{O}(1)$, plotting the
results this way indicates the approximate reach in energy
that is being achieved by the SMEFT global analysis.
This comparison is shown in Fig.~\ref{fig:SMEFiT-bounds2}, which
is the analogous plot as Fig.~\ref{fig:SMEFiT-bounds} now representing
the same bounds as bounds on the ratio $\Lambda/\sqrt{|c_i|}$ (now
only for the marginalised bounds from the global fit).
We find that for the degrees of freedom that are better constrained we achieve
sensitivity up to scales as high as $\Lambda\simeq 1.5$ TeV, in particular
thanks to the chromomagnetic operator {\tt OtG} which is well determined
from the differential measurements of top quark pair production.
Future measurements based on larger statistics should allow
us to prove even higher scales, in particular by means of the high-luminosity 
LHC datasets.

\subsection{The impact of the NLO QCD and $\mathcal{O}\lp \Lambda^{-4}\rp$ corrections}
\label{sec:perturbativeinstability}

The baseline fit results presented above are based on theory calculations
that account both for the NLO QCD corrections to the SMEFT contributions
and for the quadratic $\mathcal{O}\lp \Lambda^{-4}\rp$
terms in Eq.~(\ref{eq:smeftXsecInt}), see also the discussion
in Sect.~\ref{sec:smefttop}.
Here we aim to assess the robustness and stability of our results by
comparing the baseline fit results with those of fits based on two alternative 
theory settings.
Firstly we compare with a fit where only LO QCD effects are included for the 
SMEFT contributions, and then with a fit that includes only the linear 
$\mathcal{O}\lp \Lambda^{-2}\rp$
terms in the effective theory expansion (but still based on NLO QCD 
for the SMEFT contributions).

These comparisons have been carried out in the case of both the marginalised
results obtained from the global fit and
of the fits to individual degrees of freedom.
In Table~\ref{eq:tableCoeffBounds2} we show the
95\% CL bounds on the fitted degrees of freedom
obtained in the global analysis, and compare
the results obtained using the baseline theory settings with
those obtained either when only the linear $\mathcal{O}\lp \Lambda^{-2}\rp$
terms are included or when only LO QCD calculations are used
for the SMEFT contribution.
In Table~\ref{eq:tableCoeffBoundsIndividual} we show
the corresponding comparison in the case of individual fits.
Recall that, as mentioned above, some of the individual
bounds reported in Table~\ref{eq:tableCoeffBoundsIndividual}
have been evaluated from the analytical minimisation of the $\chi^2$,
which for those cases is more robust than the numerical
minimisation.

As can be seen from Table~\ref{eq:tableCoeffBoundsIndividual}, the individual
bounds that one obtains at $\mathcal{O}\lp \Lambda^{-2}\rp$ are very loose
for most of the four-heavy-quark operators.
This indicates that, using only the linear SMEFT contribution, one has very
limited sensitivity to these degrees of freedom.
For this reason, we do not attempt to quote any bounds for the four-heavy-quark
operators in the global fit
based on $\mathcal{O}\lp \Lambda^{-2}\rp$ theory in Table~\ref{eq:tableCoeffBounds2}:
this small sensitivity might hinder the reliability of numerical approaches
such as the ones we adopt here.
This problem goes away once we include the $\mathcal{O}\lp \Lambda^{-4}\rp$
contributions, due to the additional sensitivity provided by the quadratic 
terms.
In this case, we can reliably quote 95\% CL bounds
for both global and individual fits.

\begin{figure}[t]
  \begin{center}
   \includegraphics[width=0.99\linewidth]{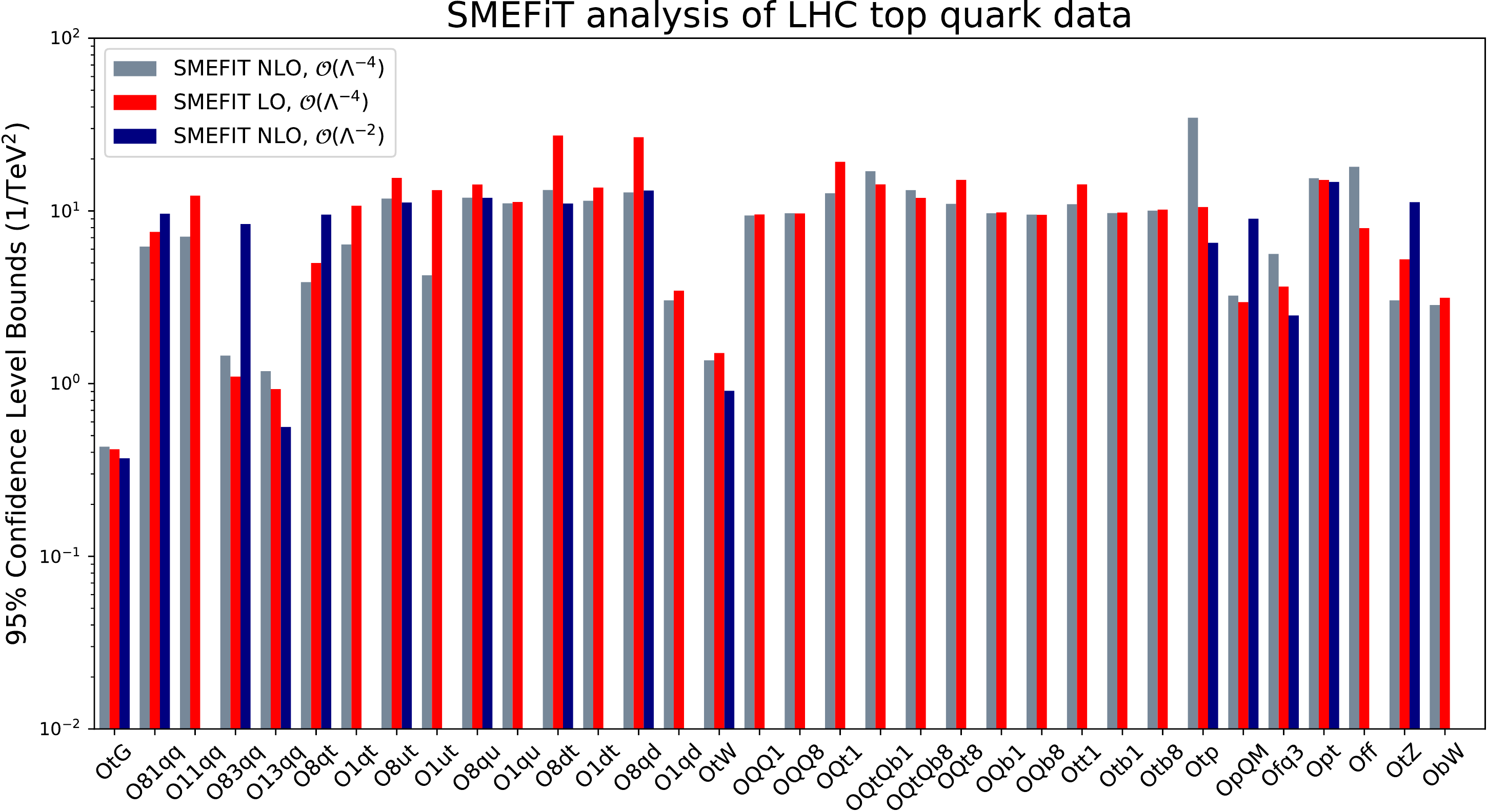}
    \caption{\small
      Same as Fig.~\ref{fig:SMEFiT-bounds}, now 
      comparing the baseline fit results reported in 
      Table~\ref{eq:tableCoeffBounds} with those from
      Table~\ref{eq:tableCoeffBounds2} based on
      the corresponding fits where (i) only LO QCD effects are included in the 
      SMEFT contributions and (ii) only the linear 
      $\mathcal{O}\lp \Lambda^{-2}\rp$
      terms are accounted for (and thus neglecting the quadratic terms).
     Note that, at the level of the $\mathcal{O}\lp \Lambda^{-2}\rp$ fit,
     several degrees of freedom are absent and we quote no bounds
     for the four-heavy-quark operators.
\label{fig:pertvariations}
}
\end{center}
\end{figure}

In Fig.~\ref{fig:pertvariations} we show
the graphical representation of the bounds
reported in Table~\ref{eq:tableCoeffBounds2} for the global fit
results with different theory settings.
Note that, for the  $\mathcal{O}\lp \Lambda^{-2}\rp$ fit, several
degrees of freedom are absent and we quote no bounds for the four-heavy-quark 
operators for the reasons mentioned above.
For the case of the two global fits with theory variations, we do not
show the comparison with the SM predictions or
the fit residuals, as was done for the case of the fit with baseline
theory settings in Fig.~\ref{fig:SMEFT_std_dev_crossvalON_NLO_HO}.
The reason is that, in both cases, the agreement with the SM is as good as 
in the case of the global fit with baseline theory settings within 
uncertainties.

From these comparisons, one finds
that the impact of the NLO QCD corrections varies depending on
the specific operators considered.
For the majority of operators, such as the two-light-two-heavy operators,
the bounds derived from the data are either comparable or
moderately improved once the
NLO QCD corrections to the SMEFT contributions are accounted for.
For example, the bound on the coefficient of the {\tt O11qq}
operator worsens
from $[ -6.8 , +7.4 ]$, with baseline theory settings, to
$[ -17 , +7.5 ]$, with LO QCD theory in the SMEFT contribution.
Note, however, that this improvement does not necessarily mean that
the NLO QCD corrections associated to the contributions of {\tt O11qq} 
itself are important.
Indeed, at the level of single-operator fits, the bounds
at NLO and LO are quite similar, $[ -0.2 , 0.02 ]$
and $[-0.2, 0.03]$, respectively.

Similar considerations apply to those operators
whose bounds in the global fit worsen when the NLO QCD corrections
to the SMEFT contributions are missing.
First, for these three operators, namely {\tt Otp}, {\tt Ofq3}, and {\tt Off}, 
the bounds are relatively loose due to the limited fit sensitivity, so they
are potentially affected by larger statistical fluctuations.
Second, at the level of individual fits, one finds that including
or not NLO QCD effects has essentially no impact
on the resulting bounds.
Therefore, the observed effect is most likely a consequence of the fact that 
adding NLO QCD corrections rearranges the weight of the different degrees of 
freedom in the global fit, leading to an overall modification of the bounds.

Concerning the impact of the quadratic $\mathcal{O}\lp \Lambda^{-4}\rp$
terms, from the comparisons in Fig.~\ref{fig:pertvariations},
we find that for most degrees of freedom the bounds are
similar regardless of whether or not these quadratic terms
are included in the fit.
This is the expected behaviour for those operators
for which the dominant sensitivity in the fit arises
already at $\mathcal{O}\lp \Lambda^{-2}\rp$, as indicated in
Table~\ref{table:operatorprocess}.
For other operators, including $\mathcal{O}\lp \Lambda^{-4}\rp$ corrections
leads to more stringent bounds.
For instance, for the coefficient associated to the {\tt O8qt} degree
of freedom the baseline bounds of $\lc -3.7,+4.1\rc$ are degraded
to $\lc -10,+8.1\rc$ if only the linear $\mathcal{O}\lp \Lambda^{-2}\rp$
corrections are taken into account.
Another example is provided by {\tt OtZ}, whose baseline bound
of $\lc -2.1,+4.0\rc$ becomes rather looser in the linear
approximation, $\lc -14,+8.0\rc$.

As already mentioned several times,
within a global fit it is in general not possible
to precisely pinpoint how a variation of the theory settings translates
into a difference in the resulting constraints on the fitted
degrees of freedom, with obvious exceptions
such as for those operators
whose contributions vanish at $\mathcal{O}\lp \Lambda^{-2}\rp$.
For such assessment, the results of the single-operator fits reported in
Table~\ref{eq:tableCoeffBoundsIndividual} are more suitable.
For example, from the results obtained in the single operator
fits we can confirm that the improvement in the bounds
obtained for the {\tt O8qt} and {\tt OtZ} degrees of freedom
upon the inclusion of the quadratic $\mathcal{O}\lp \Lambda^{-4}\rc$
corrections is genuine, rather than an artefact of the global fit.
The impact of including the quadratic terms is particularly manifest
for the four-heavy-quark degrees of freedom, where one finds
improvements of up to several orders of magnitude.
For instance, while for {\tt Otb1} the linear bounds
are almost non-existent, $\lc -2\, \cdot10^{4},-1.4\,\cdot 10^{3}\rc$,
they are improved down to $\lc -6.8,+6.8\rc$ once the $\mathcal{O}\lp \Lambda^{-4}\rp$
contributions are taken into account.

To conclude this assessment of the impact of the
NLO QCD and $\mathcal{O}\lp \Lambda^{-4}\rp$ corrections on the
fit results,  in Table~\ref{tab:chi2table-variations} we show
the comparison of the
$\chi^2/n_{\rm dat}$ values of the fit obtained
with the baseline theory settings
(NLO QCD and $\mathcal{O}\lp \Lambda^{-4}\rp$ corrections)
with the corresponding values obtained in the fits where
either only LO QCD effects or $\mathcal{O}\lp \Lambda^{-2}\rp$
corrections are included in the SMEFT corrections.
The corresponding comparison with the prior (SM) theory calculations
was reported in  Table~\ref{tab:chi2table}.

One finds that, for all three theory settings, the total
$\chi^2/n_{\rm dat}$ is similar ($\simeq 1$).
The lowest value
is found when using LO QCD theory for the SMEFT corrections.
In that case, we find $\chi^2/n_{\rm dat}= 0.84$, with the abundant
data on
$t\bar{t}$ differential distributions driving the improvement
as compared to the baseline settings.
We note that this effect is not statistically significant,
so it could also be explained by a fluctuation.
It will be interesting to revisit this comparison once more
precise top production measurements become available,
and assess whether or not there
is evidence for the need of NLO QCD corrections to achieve the optimal
description of the experimental data.

\begin{table}[p]
  \centering
  \footnotesize
   \renewcommand{\arraystretch}{1.34}
   \begin{tabular}{l|c|c|c|c}
      \toprule
      Dataset   &  \multicolumn{3}{c|}{$\chi^2/n_{\rm dat}$}  &  $n_{\rm dat}$  \\
      &  NLO + $\mathcal{O}\lp \Lambda^{-4}\rp$  & NLO + $\mathcal{O}\lp \Lambda^{-2}\rp$   &
      LO + $\mathcal{O}\lp \Lambda^{-4}\rp$&  \\
\midrule
{\tt ATLAS\_tt\_8TeV\_ljets} [ $m_{t\bar{t}}$ ]  &    1.25     & 1.67  & 1.68  &    7     \\
{\tt CMS\_tt\_8TeV\_ljets} [ $y_{t\bar{t}}$ ]   &     1.17     &  1.04  &0.82   &   10    \\
{\tt CMS\_tt2D\_8TeV\_dilep } [ $\lp m_{t\bar{t}},y_t\rp$ ] &    1.38     & 1.18   & 1.38 &    16   \\
{\tt CMS\_tt\_13TeV\_ljets2 } [ $m_{t\bar{t}}$ ]    &     1.28      & 0.67   &  0.67  &    8    \\
{\tt CMS\_tt\_13TeV\_dilep } [ $m_{t\bar{t}}$ ]  &     1.42      & 1.28 &    1.17 & 6    \\
{\tt CMS\_tt\_13TeV\_ljets\_2016} [ $m_{t\bar{t}}$ ]	       &   1.87    & 1.17 & 0.57  &    10   \\
{\tt ATLAS\_WhelF\_8TeV}    &      0.27      & 0.77  & 1.97  &  3    \\
{\tt CMS\_WhelF\_8TeV}     &     1.18     &  0.25 & 1.43 &    3   \\
\midrule
%
{\tt CMS\_ttbb\_13TeV}    &        1.29     &  3.02 & 1.14 &   1    \\
{\tt CMS\_tttt\_13TeV}    &      0.02     & 0.12 & 0.00  &    1    \\
{\tt ATLAS\_tth\_13TeV}      &     0.55      & 1.98 & 0.14 &    1    \\
{\tt CMS\_tth\_13TeV}    &      0.01     & 2.30  & 0.31 &     1    \\
{\tt ATLAS\_ttZ\_8TeV}    &      5.29    & 0.01 & 0.85  &    1    \\
{\tt ATLAS\_ttZ\_13TeV}    &     1.06     & 1.74 & 0.19 &    1    \\
{\tt CMS\_ttZ\_8TeV}    &     0.06      & 1.23 & 0.16  &    1     \\
{\tt CMS\_ttZ\_13TeV}   &       0.67     & 1.44 & 0.06 &    1    \\
{\tt ATLAS\_ttW\_8TeV}    &       0.27     & 2.08 & 0.12  &    1    \\
{\tt ATLAS\_ttW\_13TeV}    &       0.65      & 1.89 & 0.54  &    1    \\
{\tt CMS\_ttW\_8TeV}   &       0.54    & 1.96 & 0.35 &    1    \\
{\tt CMS\_ttW\_13TeV}&        0.09     & 1.96 & 0.39 &    1    \\
\midrule
{\tt CMS\_t\_tch\_8TeV\_dif}         &  0.32      & 0.46 & 0.24 &   6   \\
{\tt ATLAS\_t\_tch\_8TeV} [ $y_t$ ]          &  0.43     & 3.01 & 0.57 &    4    \\
{\tt ATLAS\_t\_tch\_8TeV} [ $y_{\bar{t}}$ ]         &   0.45     & 0.35 & 0.34  &    4   \\
{\tt ATLAS\_t\_sch\_8TeV}        &   1.92    & 1.41 & 1.77  &   1  \\
{\tt ATLAS\_t\_tch\_13TeV}          & 0.09     & 0.02 & 0.03 &  2   \\
{\tt CMS\_t\_tch\_13TeV\_dif} [ $y_t$ ]          & 0.49       &0.43   & 0.47 &  4    \\
{\tt CMS\_t\_sch\_8TeV}          & 0.76      & 0.07  &  0.78 &     1   \\
{\tt ATLAS\_tW\_inc\_8TeV}         &  0.06     & 0.00 & 0.14  &   1   \\
{\tt CMS\_tW\_inc\_8TeV}       &  0.07     & 0.00 & 0.12  &   1   \\
{\tt ATLAS\_tW\_inc\_13TeV}       &   0.82     & 0.57 & 0.91  &    1    \\
{\tt CMS\_tW\_inc\_13TeV}         &  1.68     & 0.82 &  1.23 &   1    \\
{\tt ATLAS\_tZ\_inc\_13TeV}          &  0.00      & 0.93 & 0.00  &   1    \\
{\tt CMS\_tZ\_inc\_13TeV}        &    0.34    & 0.02 & 0.07 &    1    \\
\midrule
Total &	     	  {\bf 1.06}     & {\bf 1.07} & {\bf 0.84 }  &  {\bf 103}     \\
\bottomrule
  \end{tabular}
   \caption{\small  Same as
     Table~\ref{tab:chi2table}, now comparing the
     $\chi^2/n_{\rm dat}$ values of the fit obtained
   with the baseline theory settings
   (NLO QCD and $\mathcal{O}\lp \Lambda^{-4}\rp$ corrections)
   with the corresponding values obtained in the fits where
   either only LO QCD effects or $\mathcal{O}\lp \Lambda^{-2}\rp$
   corrections are included.
     \label{tab:chi2table-variations}
  }
\end{table}


\subsection{Dataset dependence and high-energy behaviour}
\label{sec:highenergy}

\begin{figure}[t]
  \begin{center}
   \includegraphics[width=0.99\linewidth]{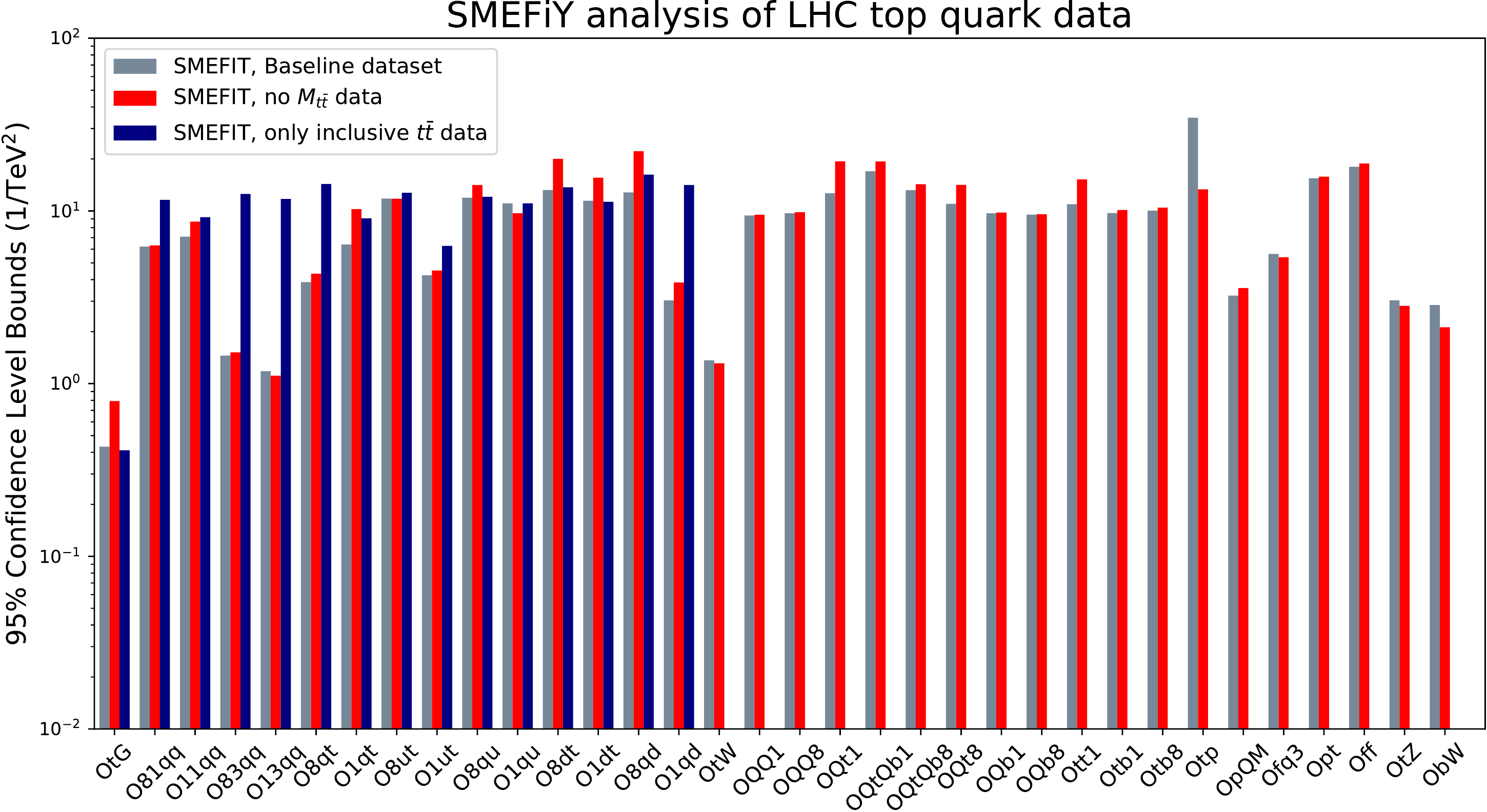}
    \caption{\small
      Same as Fig.~\ref{fig:SMEFiT-bounds}, now 
      comparing the fit results based on the baseline dataset with those 
      obtained from two different datasets.
      In the first case, for the $t\bar{t}$ production measurements,
      instead of the $m_{t\bar{t}}$ distributions as indicated in 
      Table~\ref{tab:chi2table}
      we use the corresponding $y_{t\bar{t}}$ distributions.
      In the second case, the fit is performed only using inclusive
      $t\bar{t}$ production measurements as input.
\label{fig:SMEFiT_DatasetVariations}
}
\end{center}
\end{figure}

Within the SMEFiT framework it is straightforward to repeat
the analysis with arbitrary variations of the input dataset.
To investigate the dependence of our results with respect
to this choice of input dataset, in Fig.~\ref{fig:SMEFiT_DatasetVariations}
we show a similar comparison as that of Fig.~\ref{fig:SMEFiT-bounds}, now 
assessing how the baseline fit results vary if a different
input dataset is used.
In the first case, instead of the $m_{t\bar{t}}$ distributions indicated in
Table~\ref{tab:chi2table},
we use the corresponding $y_{t\bar{t}}$ distributions
for the inclusive $t\bar{t}$ production measurements.
In the second case, the fit is performed only using inclusive
$t\bar{t}$ production measurements as input, and excluding
all other processes.
Note that in the latter case the fit has sensitivity to only
a subset of 15 degrees of freedom.

The rationale behind performing a fit replacing the $m_{t\bar{t}}$
distributions in inclusive
top-quark pair production with the corresponding  $y_{t\bar{t}}$  ones is to 
gauge the sensitivity of our results to the high-energy
region, since the $m_{t\bar{t}}$ distribution is the one more directly sensitive
to it.
This was also illustrated by the large values of the SMEFT-induced shifts 
$\delta_{\rm th}$ found in the comparisons with experimental data at large 
$m_{t\bar{t}}$ in Fig.~\ref{fig:datath_2}.
Although high-energy measurements enhance the sensitivity to SMEFT effects,
one should avoid being dominated by the highest energy bins 
since this could jeopardise the effective theory interpretation.
Therefore, one would ideally like to see that the bounds do
not become markedly worse once the $m_{t\bar{t}}$
distributions are replaced by the $y_{t\bar{t}}$ ones, since that would 
otherwise indicate that fit results are determined by high-energy events.

Concerning the fit based only on inclusive $t\bar{t}$ measurements,
one would like to find that the bounds obtained from a SMEFT fit
to a partial dataset are comparable to or looser than those
from the baseline global dataset.
Note that this is a non-trivial consistency check of the whole methodology;
when additional experimental constraints are included
in the analysis, then the bounds on the fitted coefficients
must by necessity be either unchanged or smaller.
If this were not the case, it would imply that fit results are driven not by 
the experimental data but by biased methodological choices.

From the comparison in Fig.~\ref{fig:SMEFiT_DatasetVariations} between
the fits with either the baseline dataset or the $t\bar{t}$-only dataset, 
we find that the constraints on {\tt OtG} are unchanged.
This result is not unexpected, since it is well-known that the information on 
the chromomagnetic operator is dominated by inclusive $t\bar{t}$ production.
We also observe that the bounds for some of the 2-light-2-heavy degrees of
freedom such as {\tt O83qq} and {\tt O81qq} worsen, presumably as a consequence
of the missing constraints provided by other processes, such as
$t\bar{t}$ production in association with $W$ or $Z$ bosons.
Indeed, for all the degrees of freedom directly constrained by the
inclusive $t\bar{t}$ measurements,
the bounds found in the global fit are comparable or superior to those
obtained in the $t\bar{t}$-only fit.

The other comparison shown in Fig.~\ref{fig:SMEFiT_DatasetVariations}
is that between the fit with the baseline dataset and with the same dataset
where we have replaced the $m_{t\bar{t}}$ distributions with
the corresponding $y_{t\bar{t}}$ ones.
In this case, we find that the results are qualitatively stable, 
and do not display large differences.
For a subset of the degrees of freedom, in particular those
constrained by inclusive $t\bar{t}$ data, we find that somewhat
more stringent bounds are obtained in the fits based on the
$m_{t\bar{t}}$, rather than the $y_{t\bar{t}}$, distributions.
For instance the bounds on the coefficient of {\tt OtG} are found to be
$ [ -0.4 , +0.4 ]$ when fitting $m_{t\bar{t}}$ and $[-0.8, + 0.8]$ when
fitting instead $y_{t\bar{t}}$.
These results suggest that indeed
the fit benefits from the high-energy reach of the $m_{t\bar{t}}$ distributions, 
although only slightly.

Another way to study the impact that the SMEFT
corrections have on the description of the experimental
data at high energies is to focus on the constraints provided by the
tail of the  invariant mass distribution $m_{t\bar{t}}$ in top quark
pair production, where energy-growing effects enhance the sensitivity to
SMEFT corrections~\cite{Farina:2018lqo}.
In order to highlight the impact that these energy-growing effects have
on the description of the $m_{t\bar{t}}$ tails, it is useful to compute the shift 
induced by the SMEFT corrections to the SM calculation separated into the 
contributions from different degrees of freedom.
For simplicity, in the following we restrict ourselves
to the linear $\mathcal{O}\lp \Lambda^{-2}\rp$ corrections.
In this case, following the notation of Eq.~(\ref{eq:smeftXsecInt}),
we want to compare the size of the individual corrections defined as
\be
\label{eq:smeftshifts}
\Delta_i^{\rm (smeft)} \equiv \kappa_i \frac{\delta c_i}{\Lambda^2} \, ,\qquad
\widetilde{\Delta}_i^{\rm (smeft)} \equiv \widetilde{\kappa}_{ii} \frac{\lp \delta c_i\rp^2}{\Lambda^4}\,,
\qquad 
i=1,\ldots, N_{\rm op}  \, ,
\ee
for the different bins of the $m_{t\bar{t}}$ distribution, and identify which 
degrees of freedom dominate at high energy.
Note that, as discussed in Sect.~\ref{sec:generaldiscussion},
in general there are several reasons why a given operator
might or might not lead to
energy-growing effects.
In Eq.~(\ref{eq:smeftshifts}), we will use as $\delta c_i$ the 95\% CL 
bounds for the baseline fit reported in Table~\ref{eq:tableCoeffBounds}.

\begin{figure}[t]
  \begin{center}
    \includegraphics[scale=0.38]{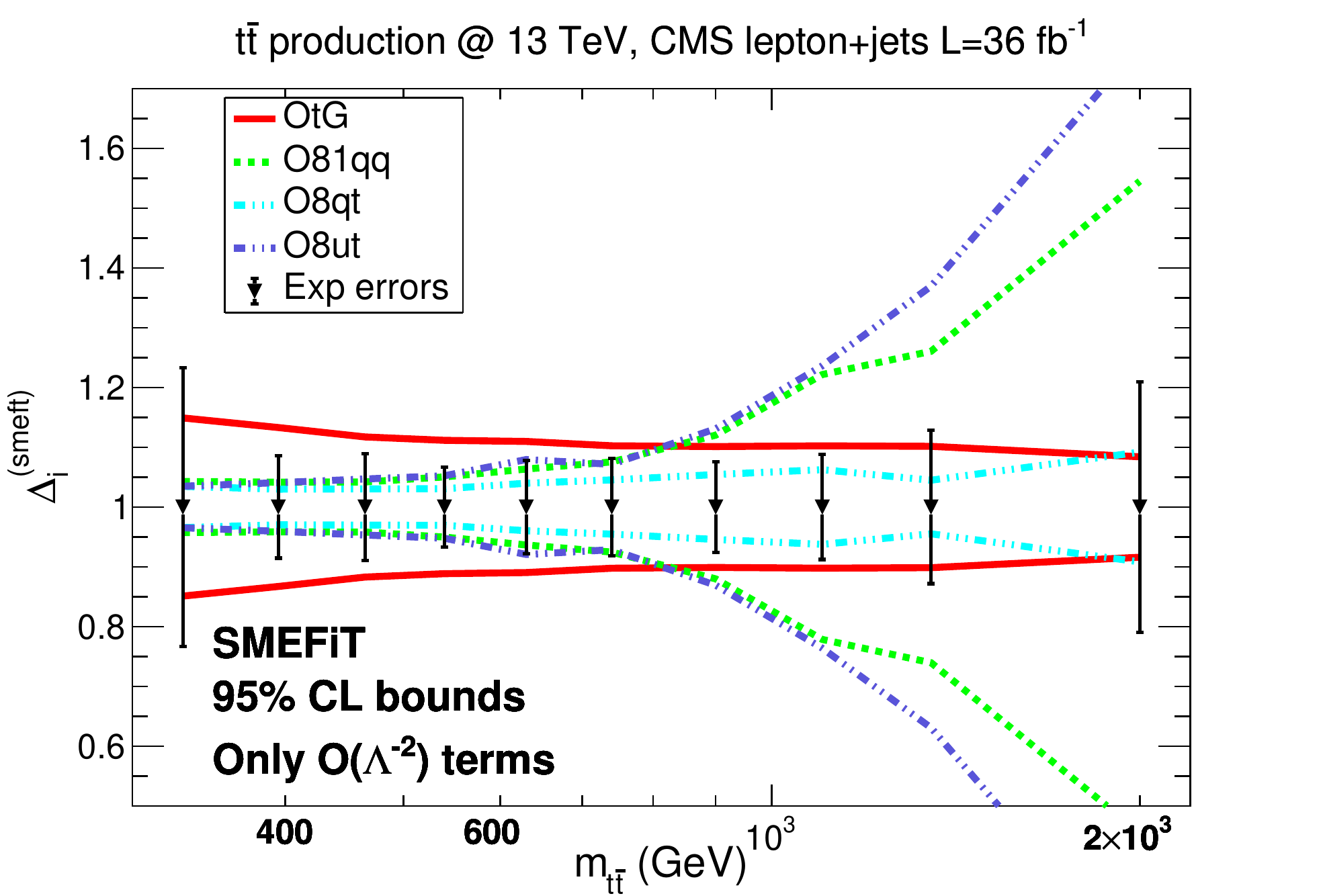}
    \includegraphics[scale=0.38]{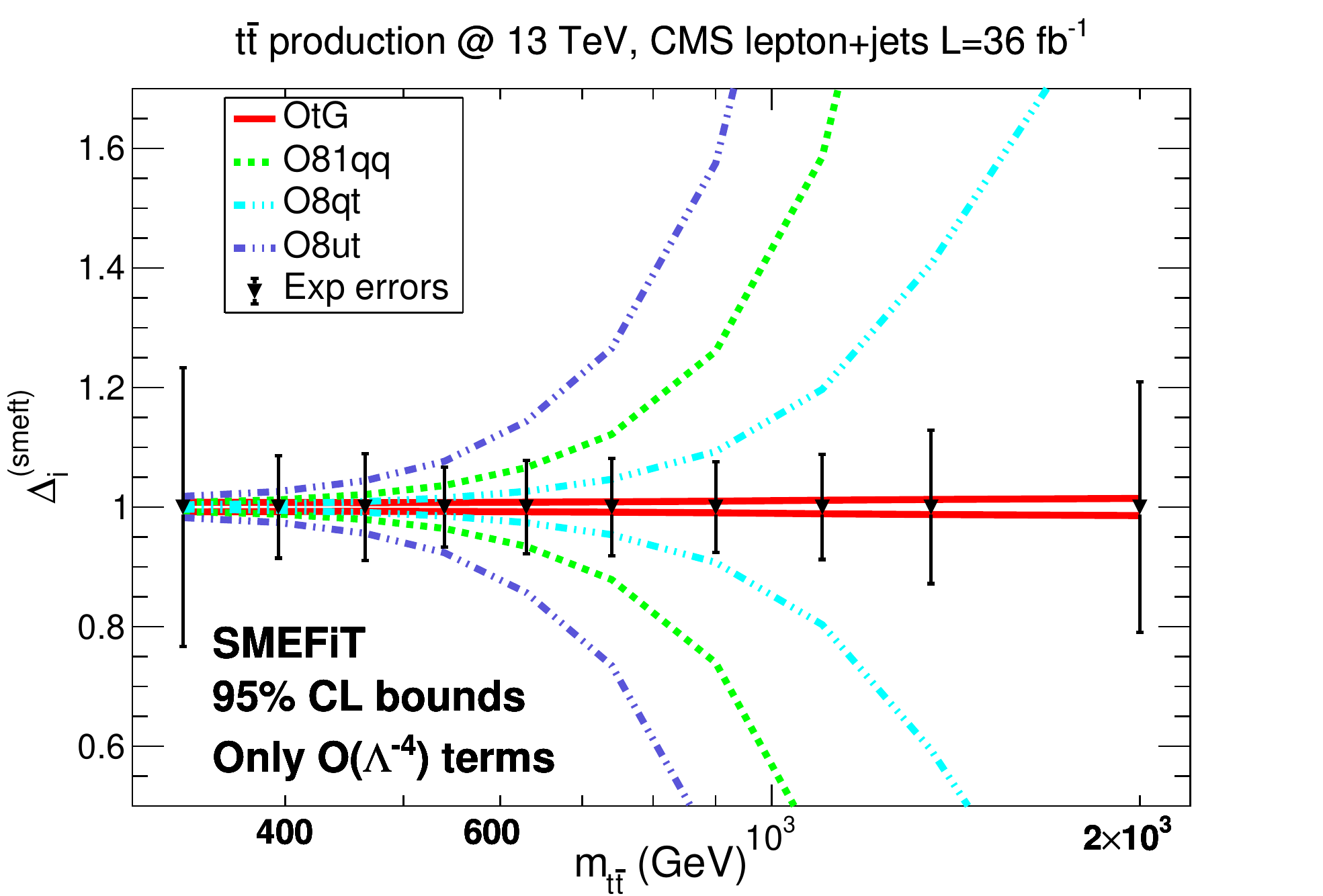}
    \caption{\small The shifts induced by representative
      SMEFT degrees of freedom to the SM cross-sections,  
      Eq.~(\ref{eq:smeftshifts}),
      for the $m_{t\bar{t}}$ distribution in the top quark pair
      production measurements at $\sqrt{s}=13$ TeV from CMS, based
      on $\mathcal{L}=36$ fb$^{-1}$ and the lepton+jets 
      channel~\cite{Sirunyan:2018wem}.
      We show the shifts arising from the linear (left) and from 
      the purely quadratic (right) terms.
      The shifts $\Delta_i^{(\rm smeft)}$ have
      been computed assuming the 95\% CL bounds $\delta c_i$ of the baseline 
      fit reported in Table~\ref{eq:tableCoeffBounds}.
      For reference, we also indicate the size of the corresponding 
      experimental uncertainties.
     \label{fig:highenergycomparison}
  }
  \end{center}
\end{figure}

In Fig.~\ref{fig:highenergycomparison} we show the values
of the SMEFT-induced shifts, Eq.~(\ref{eq:smeftshifts}),
for the different bins of the $m_{t\bar{t}}$ distribution
from the CMS measurement at 13 TeV in the lepton+jets final state,
based on
an integrated luminosity of $\mathcal{L}=36$ fb$^{-1}$~\cite{Sirunyan:2018wem},
which has the best coverage of the TeV region.
To facilitate the visualisation, we restrict ourselves to the contributions
associated to four representative
degrees of freedom: {\tt OtG}, {\tt O81qq}, {\tt O8qt}, and {\tt O8ut}. 
For reference, we also show the corresponding total experimental uncertainty
for each of the $m_{t\bar{t}}$ bins.

We observe that several operators
lead to effects that grow with the energy.
The steepest growth is found for the {\tt O8ut} degree of freedom, but
other operators that lead to energy-growing effects are 
{\tt O81qq} and {\tt O8qt}.
Other operators are less sensitive to the high-energy region.
This is illustrated by the case of {\tt OtG}, whose
sensitivity is concentrated in the $t\bar{t}$ threshold production region.
It is therefore clear that pushing the reach of the experimental
measurements deep into the TeV region will further
increase the sensitivity to these energy-growing
degrees of freedom.
In this respect, a major concern will be to appropriately disentangle
potential SMEFT signatures from the information used to constrain
the proton structure in global fits, in particular the large-$x$ gluon.

It should be emphasised that the individual shifts in
Fig.~\ref{fig:highenergycomparison} cannot be directly combined
to construct the actual shift to the SM prediction in each cross-section
bins, due to the replica-by-replica correlations between the various degrees
of freedom.
With this caveat, it is clear that the SMEFT-induced shifts
could not be much larger than the bounds derived in this analysis without
degrading the agreement between theory predictions and experimental data,
a similar conclusion that what was derived from the comparisons with
experimental data shown in Figs.~\ref{fig:datath_1}--\ref{fig:datath_4}

\begin{figure}[t]
  \begin{center}
  \includegraphics[width=0.99\linewidth]{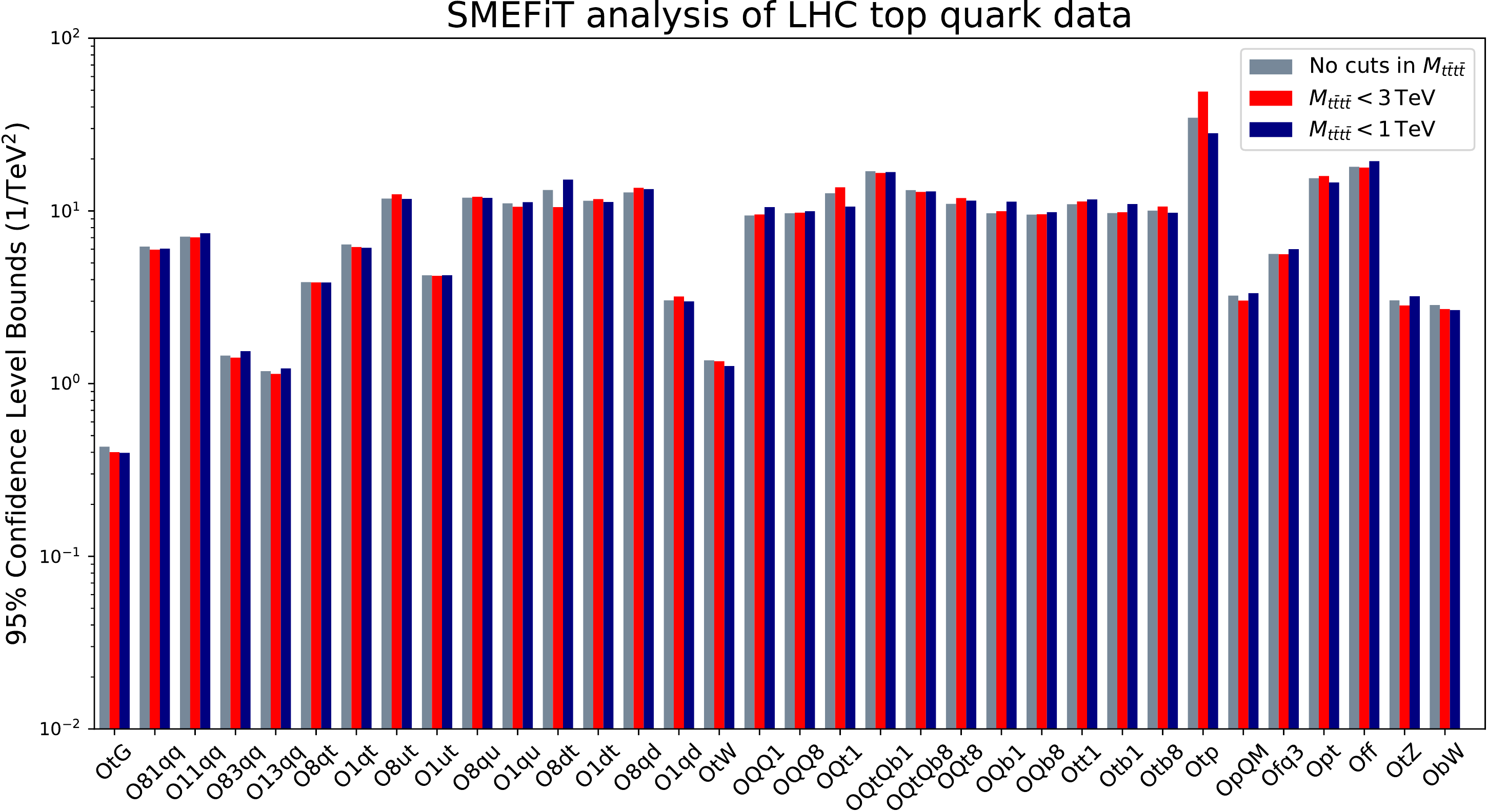}
    \caption{\small Same as Fig.~\ref{fig:SMEFiT-bounds}, now 
      comparing the baseline fit results with those obtained
      from two fits with differing cuts on $m_{t\bar{t}t\bar{t}}$
      used in the theoretical calculation of the $t\bar{t}t\bar{t}$
      production cross-section.
     \label{fig:mttcuts}
  }
  \end{center}
\end{figure}

To conclude this discussion about high-energy effects, 
another of the input processes in the fit that in principle is sensitive
to the high energy region is $t\bar{t}t\bar{t}$ production, where
the invariant mass of the 4-top final state $m_{t\bar{t}t\bar{t}}$ can reach
values of up to several TeV.
In order to further assess the stability of our results with respect to the
high-energy region, we have repeated the baseline
fit imposing different cuts on the value of the
4-top invariant mass, from a loose cut requesting
$m_{t\bar{t}t\bar{t}} \le 3$~TeV to a more stringent cut
with $m_{t\bar{t}t\bar{t}} \le 1$~TeV.
The results of these fits are displayed in Fig.~\ref{fig:mttcuts},
and do not show any sensitivity to the value of 
$ m_{t\bar{t}t\bar{t}}$ adopted in the theory calculation.
We recall that in the current analysis
a single  $t\bar{t}t\bar{t}$ cross-section has been included; 
future measurements of this process, including
possibly in differential form, could then become more
sensitive to the high-energy region.

\section{Summary and outlook}
\label{sec:summary}

In this work we have presented a novel approach to carry out
global analyses of the SMEFT.
This new framework, which we have denoted by SMEFiT, is flexible,
modular, robust upon enlarging the fitted parameter space,
and resilient with respect to problems that arise frequently
in SMEFT fits such as degeneracies and flat directions.
Its main ingredients are the MC replica method to construct
a representation of the probability distribution in the space of dimension-6 
SMEFT degrees of freedom, and cross-validation, which prevents of over-fitting.
Our results are provided as a sample of $N_{\rm rep}$ MC replicas, which
can be used to derive predictions for related cross-sections and combined
with other constraints on the SMEFT parameter space.

As a proof-of-concept of the SMEFiT framework, we have presented a detailed 
analysis of top quark production measurements at the LHC 8 TeV and 13 TeV.
We have included a wide range of top quark data, in terms of total rates
and differential distributions.
The theoretical SM and SMEFT cross-sections have been evaluated including
NLO QCD corrections by default; in the SM case, we have also considered 
NNLO effects for the most accurately measured processes, namely differential
distributions in $t\bar{t}$ and single top $t$-channel production.
This combination of state-of-the art calculations with
precision LHC measurements has allowed us to provide constraints
on $N_{\rm op}=34$ independent operators from the dimension-six Lagrangian in 
the Warsaw basis. 

Our results are in good agreement with the SM expectations: we find that
all the $N_{\rm op}=34$ fitted SMEFT degrees of freedom are consistent with
the SM result within uncertainties at the 95\% CL.
We have compared our results with existing bounds on the same operators 
presented in the literature, and have provided individual constraints 
on the operators in the SMEFiT framework.
We have also studied the robustness of our results
with respect to the inclusion of higher-order NLO QCD corrections,
or $\mathcal{O}\lp \Lambda^{-4}\rp$ effects, and variations of
the input dataset.
We have found that including either NLO QCD corrections to the SMEFT 
contributions or the quadratic $\mathcal{O}\lp \Lambda^{-4}\rp$ terms
leads to stronger bounds for most of the degrees of freedom in the fit.

The results of this analysis are available upon request
as a sample of $N_{\rm rep}=1000$ MC replicas
representing the probability distribution in the space
of Wilson coefficients for the $N_{\rm op}=34$ SMEFT operators
considered here.
These replicas can be used to compute statistical properties of
the distribution such as variances, correlations, and higher moments, 
and can be combined with other processes that provide
complementary information on the SMEFT parameter space.

The study presented in this work is the first proof-of-principle application of  the SMEFiT framework. 
Further studies and extensions can be envisioned.
The next steps will be to consider a larger basis of fitted SMEFT operators
by including other types of LHC processes beyond top
quark production in the input dataset.
These new measurements should include total rates and differential 
distributions in Higgs production, single and pair production of electroweak 
vector bosons, and also other processes directly
sensitive to the TeV region, such as di-jet and multi-jet production.
Eventually, one also might need to account for measurements from previous 
colliders such as LEP and from lower energy experiments.
In this respect, our results pave the way towards a truly global fit of
the SMEFT at dimension-six where direct constraints are simultaneously provided
for the majority of the operators.

\paragraph{Acknowledgements.}
J.~R. would like to thank Gerhard Raven and Wouter Verkerke for illuminating
discussions.
F.~M. has received fundings from the European Union's Horizon 2020 research and innovation programme as part of the Marie Sk\l{}odowska-Curie Innovative Training Network MCnetITN3 (grant agreement no. 722104) and by the F.R.S.-FNRS under the `Excellence of Science` EOS be.h project n. 30820817.  
Computational resources have been provided by the supercomputing facilities of the Universit\'e catholique de Louvain (CISM/UCL) and the Consortium des \'Equipements de Calcul Intensif en F\'ed\'eration Wallonie Bruxelles (C\'ECI). 
N.~H., J.~R., and E.~S. are supported by the European Research Council Starting
Grant ``PDF4BSM''.
J.~R. is also partially supported by the Netherlands Organization for Scientific
Research (NWO).
E.~R.~N. is supported by the European Commission through the Marie 
Sk\l{}odowska-Curie Action ParDHonS FFs.TMDs (grant number 752748), and was
supported by the UK Science and Technology Facility Council through grant 
ST/P000630/1.
E.~V. is supported by a Marie Sk\l{}odowska-Curie Individual
Fellowship of the European Commission's Horizon 2020 Programme under
contract number 704187.
C.~Z.~is supported by IHEP under Contract No.~Y7515540U1.

\FloatBarrier

\phantomsection
\addcontentsline{toc}{section}{References}

\appendix
\section{Notation and conventions}
\label{app:notation}

In this appendix we summarise the notation and conventions that are adopted
in this work concerning the relevant dimension-6 SMEFT operators.
We adopt the notation of~\cite{Grzadkowski:2010es}, where
flavour indices are labelled by $i,j,k$ and $l$; left-handed fermion SU(2) 
doublets are denoted by $q$, $l$; right-handed fermion singlets by $u$, $d$, 
$e$; the Higgs doublet by $\varphi$; the antisymmetric $SU(2)$ tensor by 
$\varepsilon\equiv i\tau^2$; $\tilde{\varphi}=\varepsilon\varphi^*$; 
$\FDF\equiv \varphi^\dagger(iD_\mu \varphi) - (iD_\mu\varphi^\dagger) \varphi$; 
$\FDFI\equiv \varphi^\dagger\tau^I(iD_\mu \varphi) - (iD_\mu\varphi^\dagger) \tau^I\varphi$ where $\tau^I$ are the Pauli matrices; 
$T^A\equiv \lambda^A/2$ where $\lambda^A$ are Gell-Mann matrices.

With these considerations, the dimension-6 SMEFT four-quark operators
relevant for the interpretation of top quark measurements at the LHC are the 
following:
\begin{align}
	\qq{1}{qq}{ijkl}
	&= (\bar q_i \gamma^\mu q_j)(\bar q_k\gamma_\mu q_l)
	 \nonumber
	,\\
	\qq{3}{qq}{ijkl}
	&= (\bar q_i \gamma^\mu \tau^I q_j)(\bar q_k\gamma_\mu \tau^I q_l)
 \nonumber
	,\\
	\qq{1}{qu}{ijkl}
	&= (\bar q_i \gamma^\mu q_j)(\bar u_k\gamma_\mu u_l)
         \nonumber
	,\\
	\qq{8}{qu}{ijkl}
	&= (\bar q_i \gamma^\mu T^A q_j)(\bar u_k\gamma_\mu T^A u_l)
         \nonumber
	,\\
	\qq{1}{qd}{ijkl}
	&= (\bar q_i \gamma^\mu q_j)(\bar d_k\gamma_\mu d_l)
         \nonumber
	,\\
	\qq{8}{qd}{ijkl}
	&= (\bar q_i \gamma^\mu T^A q_j)(\bar d_k\gamma_\mu T^A d_l)
        \label{eq:FourQuarkOp} 
	,\\
	\qq{}{uu}{ijkl}
	&=(\bar u_i\gamma^\mu u_j)(\bar u_k\gamma_\mu u_l)
         \nonumber
	,\\
	\qq{1}{ud}{ijkl}
	&=(\bar u_i\gamma^\mu u_j)(\bar d_k\gamma_\mu d_l)
         \nonumber
	,\\
	\qq{8}{ud}{ijkl}
	&=(\bar u_i\gamma^\mu T^A u_j)(\bar d_k\gamma_\mu T^A d_l)
         \nonumber
	,\\
	\hc{\qq{1}{quqd}{ijkl}}
	&=(\bar q_i u_j)\:\varepsilon\;
	(\bar q_k d_l)
         \nonumber
	,\\
	\hc{\qq{8}{quqd}{ijkl}}
	&=(\bar q_iT^A u_j)\;\varepsilon\;
	(\bar q_kT^A d_l)
         \nonumber
	 .
\end{align}
Recall that these operators satisfy all the symmetries of the SM, in particular
gauge symmetry before electroweak symmetry breaking.
Another class of relevant SMEFT operators are those that contain two quarks
coupled to Higgs fields or gauge boson fields; the ones relevant for
top quark measurements are given by:
\begin{align}
	\hc{\qq{}{u\varphi}{ij}}
	&=\bar{q}_i u_j\tilde\varphi\: (\varphi^{\dagger}\varphi)
         \nonumber
	,\\
	\qq{1}{\varphi q}{ij}
	&=\FDF (\bar{q}_i\gamma^\mu q_j)
         \nonumber
	,\\
	\qq{3}{\varphi q}{ij}
	&=\FDFI (\bar{q}_i\gamma^\mu\tau^I q_j)
         \nonumber
	,\\
	\qq{}{\varphi u}{ij}
	&=\FDF (\bar{u}_i\gamma^\mu u_j)
         \nonumber
	,\\
	\hc{\qq{}{\varphi ud}{ij}}
	&=(\tilde\varphi^\dagger iD_\mu\varphi)
	(\bar{u}_i\gamma^\mu d_j)
         \label{eq:TwoQuarkOp} 
	,\\
	\hc{\qq{}{uW}{ij}}
	&=(\bar{q}_i\sigma^{\mu\nu}\tau^Iu_j)\:\tilde{\varphi}W_{\mu\nu}^I
         \nonumber
	,\\
	\hc{\qq{}{dW}{ij}}
	&=(\bar{q}_i\sigma^{\mu\nu}\tau^Id_j)\:{\varphi} W_{\mu\nu}^I
         \nonumber
	,\\
	\hc{\qq{}{uB}{ij}}
	&=(\bar{q}_i\sigma^{\mu\nu} u_j)\quad\:\tilde{\varphi}B_{\mu\nu}
         \nonumber
	,\\
	\hc{\qq{}{uG}{ij}}
	&=(\bar{q}_i\sigma^{\mu\nu}T^Au_j)\:\tilde{\varphi}G_{\mu\nu}^A \, ,
         \nonumber
\end{align}
where $W_{\mu\nu}^I$ and $B_{\mu\nu}$ are the field-strength tensors
of the electroweak interaction and $G_{\mu\nu}^A$ is the QCD one.

In Eqns.~(\ref{eq:FourQuarkOp}) and~(\ref{eq:TwoQuarkOp}),
non-Hermitian operators are indicated with a double dagger symbol.
In the case of Hermitian operators involving vector Lorentz bilinears, complex conjugation
is the same as the transposition of generation indices: $\qq{}{}{ij}{}^* = \qq{}{}{ji}$ and by extension, for four-fermion operators, $\qq{}{}{ijkl}{}^* = \qq{}{}{jilk}$.
In addition, it is understood in the notation above
that the implicit sum over flavour indices only includes independent combinations.

\providecommand{\href}[2]{#2}\begingroup\raggedright\endgroup

\end{document}